\documentclass{aa} 
\usepackage[flushleft]{threeparttable}
\usepackage{txfonts}
\usepackage{xcolor}
\usepackage{lscape}
\usepackage{xcolor}

\usepackage{graphicx}
\usepackage{amssymb}
\usepackage{txfonts}
\usepackage[colorlinks=true,allcolors=blue]{hyperref}

\newcommand{\oiii}{\hbox{[O\,{\sc iii}]}\xspace}
\newcommand{\ha}{\hbox{H$\alpha$}\xspace}
\newcommand{\hb}{\hbox{H$\beta$}\xspace}
\newcommand{\feii}{\hbox{Fe\,{\sc ii}}\xspace}
\newcommand{\cii}{\hbox{[C\,{\sc ii}]}\xspace}
\newcommand{\sii}{\hbox{[S\,{\sc ii}]}\xspace}
\newcommand{\nii}{\hbox{[N\,{\sc ii}]}\xspace}
\newcommand{\oi}{\hbox{[O\,{\sc i}]}\xspace}
\newcommand{\hei}{\hbox{He\,{\sc i}}\xspace}
\newcommand{\heii}{\hbox{He\,{\sc ii}}\xspace}
\newcommand{\nv}{\hbox{N\,{\sc v}}\xspace}
\newcommand{\siiv}{\hbox{Si\,{\sc iv}}\xspace}
\newcommand{\civ}{\hbox{C\,{\sc iv}}\xspace}
\newcommand{\mgii}{\hbox{Mg\,{\sc ii}}\xspace}
\newcommand{\oii}{\hbox{[O\,{\sc ii}]}\xspace}
\newcommand{\neiii}{\hbox{[Ne\,{\sc iii}]}\xspace}

\newcommand{\kms}{\hbox{km~s$^{-1}$}\xspace}
\begin{document}

   \title{GA-NIFS: The highly overdense system BR1202-0725 at $z \sim 4.7$}

   \titlerunning{The highly overdense system BR1202-0725 at $z \sim 4.7$}
    
   \subtitle{A double AGN with fast outflows plus eight companion galaxies}

   \authorrunning{S. Zamora et al.}

   \author{S. Zamora \inst{\ref{inst:SNS}}\fnmsep\thanks{E-mail: \href{mailto:sandra.zamoraarenal@sns.it}{sandra.zamoraarenal@sns.it}} 
    \and  
    Giacomo Venturi \inst{\ref{inst:SNS}}
    \and
    Stefano Carniani  \inst{\ref{inst:SNS}}
    \and 
    Elena Bertola\inst{\ref{inst:arcetri}}
    \and
    Eleonora Parlanti \inst{\ref{inst:SNS}}
    \and
    Michele Perna \inst{\ref{inst:cab}}
    \and
    Santiago Arribas \inst{\ref{inst:cab}}
    \and
    Torsten Böker \inst{\ref{inst:esa_baltimore}}
    \and
    Andrew J. Bunker \inst{\ref{inst:oxford}}
    \and
    Stéphane Charlot \inst{\ref{inst:sorbonne}}
    \and
    Francesco D'Eugenio \inst{\ref{inst:cambridge1},\ref{inst:cambridge2}}
    \and
    Roberto Maiolino \inst{\ref{inst:cambridge1},\ref{inst:cambridge2}}
    \and
    Bruno Rodr\'iguez Del Pino \inst{\ref{inst:cab}}
    \and
    Hannah \"Ubler \inst{\ref{inst:max_planck}}
    \and 
    Giovanni Cresci \inst{\ref{inst:arcetri}}
    \and
    Gareth C. Jones \inst{\ref{inst:cambridge1},\ref{inst:cambridge2}}
    \and
    Isabella Lamperti \inst{\ref{inst:firence},\ref{inst:arcetri}}
    }
    
   \institute{Scuola Normale Superiore, Piazza dei Cavalieri 7, I-56126 Pisa, Italy\label{inst:SNS}
        \and 
        INAF - Osservatorio Astrofisco di Arcetri, largo E. Fermi 5, 50127 Firenze, Italy\label{inst:arcetri}
        \and
        Centro de Astrobiolog\'{\i}a (CAB), CSIC-INTA, Ctra. de Ajalvir km 4, Torrej\'on de Ardoz, E-28850, Madrid, Spain\label{inst:cab}
        \and
        European Space Agency, c/o STScI, 3700 San Martin Drive, Baltimore, MD 21218, USA \label{inst:esa_baltimore}
        \and
        University of Oxford, Department of Physics, Denys Wilkinson Building, Keble Road, Oxford OX13RH, United Kingdom \label{inst:oxford}
        \and
        Sorbonne Universit\'e, CNRS, UMR 7095, Institut d'Astrophysique de Paris, 98 bis bd Arago, 75014 Paris, France \label{inst:sorbonne}
        \and
        Kavli Institute for Cosmology, University of Cambridge, Madingley Road, Cambridge, CB3 0HA, UK \label{inst:cambridge1}
        \and
        Cavendish Laboratory - Astrophysics Group, University of Cambridge, 19 JJ Thomson Avenue, Cambridge, CB3 0HE, UK 
        \label{inst:cambridge2}
        \and
        $^{}$Max-Planck-Institut f\"ur extraterrestrische Physik (MPE), Gie{\ss}enbachstra{\ss}e 1, 85748 Garching, Germany \label{inst:max_planck}
        \and
        Dipartimento di Fisica e Astronomia, Università di Firenze, Via G. Sansone 1, 50019, Sesto F.no (Firenze), Italy 
        \label{inst:firence}
}
         
   \date{ }

% \abstract{}{}{}{}{} 
% 5 {} token are mandatory
 
  \abstract
{Distant quasars (QSOs) in galaxy overdensities are considered key actors in the evolution of the early Universe.  
In this work, we performed an analysis of the kinematic and physical properties of the BR1202-0725 system at  $z\sim4.7$, one of the most overdense fields known in the early Universe, consisting of a QSO, a submillimeter galaxy (SMG), and three Lyman-$\alpha$ emitters. We used data from the JWST/NIRSpec Integral Field Unit (IFU) to analyze the rest-frame optical emission of each source in the system. 
Based on the continuum and Balmer line emission, we estimated a bolometric luminosity of $\log(L_{\rm bol}/[ {\rm erg~s^{-1}}])$ = 47.2 $\pm$ 0.4 and a black hole mass of $\log(M_{\rm BH}/M_\odot) = 10.1\pm0.5$ for the QSO, which are consistent with previous measurements obtained with ground-based observations.
The NIRSpec spectra of the SMG revealed instead unexpected \oiii and H$\alpha$+\nii profiles. The overall \oiii line profile is blue-shifted by more than 700~\kms relative to the systemic velocity of the galaxy. Additionally, both the \oiii and H$\alpha$+\nii lines show prominent broad ($\sim$1300~\kms), blueshifted  wings associated with outflowing ionized gas. The analysis of NIRSpec and X-ray observations indicates that the SMG likely hosts an accreting supermassive black hole as supported by the following results: (i) the excitation diagnostic diagram is consistent with ionization from an active galactic nucleus (AGN); (ii) the X-ray luminosity is higher than $10^{44}\rm~erg~s^{-1}$; and (iii) it hosts a fast outflow ($v_{\rm out}$ $\sim$ 5000~\kms), comparable to those observed in luminous QSOs. 
Therefore, the QSO-SMG pair of BR1202-0725 represents one of the highest-redshift double AGN to date, with a projected separation of 24 kpc. Finally, we investigated the environment of this system and found four new galaxies, detected in both \oiii and \ha emission, at the same redshift of the QSO and within a projected distance of 5 kpc from it. This overdense system includes at least ten galaxies in a field of view of only 980~kpc$^{2}$.

}


   \maketitle
% \usepackage{multirow}
% \usepackage{multirow}
% \usepackage{multirow}

\section{Introduction}
Cosmological numerical simulations expect luminous active galactic nuclei (AGN) to reside in the most overdense regions of the Universe \citep{ni2020,habouzit2019,valentini2021,weinberger2018,costa2014,dimatteo2012,sijacki2009,zana2022, Zana:2023, barai2018}. In such fields, both the growth of supermassive black holes (BHs) and the formation of new stars are thought to be triggered by galaxy mergers and interactions. 

However, whether distant quasars (QSOs) form and evolve in galaxy overdensities is still an open question \citep[see e.g.,][and references therein]{villforth2023}. Rest-frame UV observations with ground-based telescopes and \textit{Hubble} Space Telescope (HST) have led to contrasting results. Some studies have found a large number of galaxies within a few Megaparsecs of $z\sim6$ QSOs \citep{bosman2020,mignoli2020,meyer2022,overzier2022,zheng2006} while others have reported no sources or only a few galaxies in the vicinity of the QSO \citep[e.g.,][]{willott2005,morselli2014,simpson2014}.
%(e.g., Kim et al. 2009)
Even observational programs conducted with the Atacama Large Millimeter/submillimeter Array (ALMA) have produced differing and sometimes contradictory results in the identification of companion galaxies near high-redshift quasars. Some studies have identified up to three \cii emitters in the $\sim30\arcsec\times30\arcsec$ ALMA field of view (FoV; corresponding to a physical scale of $\sim$ 180 $\times$ 180 kpc$^2$ at $z \sim 6$), while others have only reported the emission of the QSOs \citep{decarli2017, Decarli:2018, Willott:2017, Neeleman:2019, Venemans:2019, Venemans:2020, Bischetti:2024}. However, \cite{zana2022, Zana:2023} show that the dust obscuration, angular resolution, and data sensitivity may explain the different results obtained with HST. Also the small FoV and frequency coverage of ALMA can explain the scatter in the number of detected satellite galaxies among different observations.

The launch of the \textit{James Webb} Space Telescope (JWST) has opened a new window on the study of galaxy overdensities at high redshift through rest-frame optical emission lines. NIRCam Wide Field Slitless Spectroscopy (WFSS) observations have revealed more than ten \oiii emitters within a few Megaparsecs from the $z\sim6$ QSOs J0305–3150 and J0100+2802 \citep{wang2023, kashino2023}. Only a few of these sources had been previously detected with HST and ALMA. Additionally, NIRSpec observations in integral field spectroscopy (IFS) mode have enabled the detailed investigations of the close environment of high-$z$ sources, finding a high prevalence of overdensities \citep[e.g.,][]{jones2024, delpino2024, arribas2024, lamperti2024}. Many AGN and QSOs at $z>3$ followed-up with NIRSpec IFS show the presence of at least one companion galaxy up to 30 kpc and $\pm$1000~\kms of the main target \citep{perna2023b, marshall2023, marshall2024, loiacono2024, uebler2023,uebler2024}. Some of these companions show spectroscopic features associated with AGN ionization, indicating that the system may host a double or even a triple AGN. Analyzing a sample of AGN at $z\sim3$, \cite{perna2023} conclude that dual AGN with projected separations below 30 kpc may be more common (by at least a factor of 2 to 3) than what is predicted by the most recent cosmological simulations \citep{derosa2019, volonteri2022}. Specifically, the companion galaxies of five out of 16 AGN (i.e., $\sim$20--30\%) have optical emission line flux ratios consistent with AGN excitation. In comparison, the above theoretical models predict a dual AGN fraction $<1\%$.

\begin{figure*}
\includegraphics[width=\textwidth]{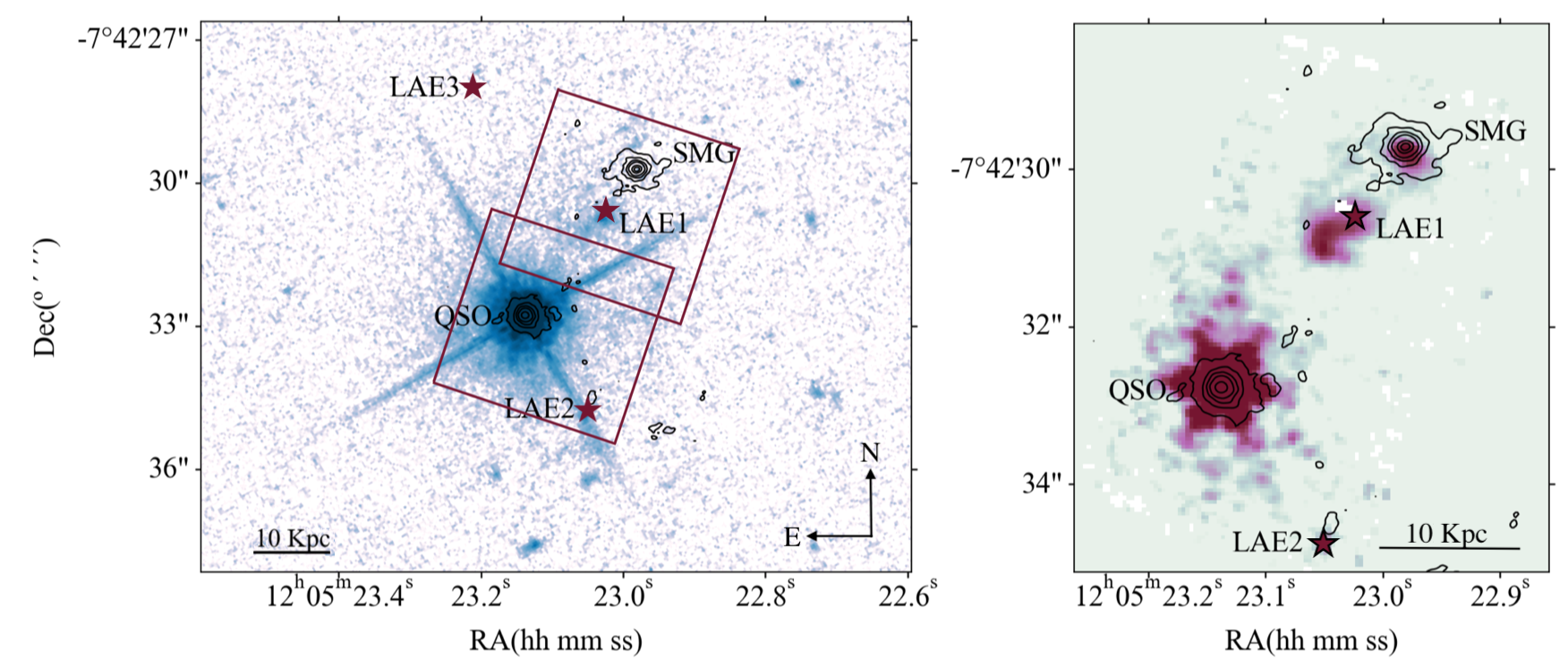}
\caption{HST ACS F775W filter ($i$-band; left) and JWST NIRSpec IFU \ha ($3.74~{\rm \mu m}<\lambda<3.75~{\rm \mu m}$; right) image  of BR1202-0725. Dust continuum emission from ALMA in Band 7 is shown as black contours. Star symbols illustrate the position of the three Lyman-$\alpha$ emitters. North is up, east is left. A logarithmic color scale is used.}
\label{fig:aper_ha}
\end{figure*}

BR1202-0725 ($z\sim4.7$) is one of the most studied overdense fields at high redshift, acting as a laboratory for studying different galaxy types, all evolving in the same environment. The system (see Fig. \ref{fig:aper_ha}) is composed of a submillimeter galaxy (SMG), a QSO - with a SMG-QSO projected separation of 24 kpc - and three Lyman-alpha emitters \citep[LAE1, LAE2, and LAE3;][]{drake2020,carniani2013,carilli2013}. 

The QSO was observed for the first time by \citet{irwin1991} and was the first object at $z > 4$ to be detected in CO emission \citep{ohta1996}. The optically obscured SMG was also found in these observations, at $\sim $ 4\arcsec\ to the northwest of the QSO \citep[see also ][]{omont1996b,iono2006}. Both the QSO and the SMG have far infrared (FIR; $\sim$40–500 $\mu$m) luminosities higher than 10$^{13}$ $L_\odot$ \citep{omont1996,iono2006,carilli2002}, which suggest very strong star formation activity (SFR $>  1000{\rm ~M_{\odot}~yr^{-1}}$; \citealt{carilli2013}) and, which according to the commonly accepted paradigm is likely triggered and sustained by merger events.

The QSO exhibits a large Ly$\alpha$ halo \citep[$\sim$55 kpc, surface brightness > 1$\times$ 10$^{-17}$ erg~s$^{-1}$~cm$^{-2}$~arcsec$^{-2}$;][]{drake2020} with high velocity widths (FWHM$_\mathrm{Ly\alpha}$ $\sim$ 1000 \kms), probably coupled with the bridge of gas located between the QSO and the SMG observed in \cii158 $\mu$m \citep{wagg2012,carilli2013,carniani2013}. The QSO contains a mass of $\sim$ 8 $\times$ 10$^{10}$ $M_\odot$ of cold molecular gas \citep[derived from the CO(J=1–0) line luminosity;][]{riechers2006}. The CO emission of the QSO can be separated into at least two components: the main one (FWHM$_\mathrm{CO}$= 700 \kms), with kinematics compatible with a rotating disk, and another one separated by +0.4\arcsec\ in declination, having a velocity shift of 180 \kms, which is likely another fainter galaxy in the merger \citep{salome2012}. The QSO also shows an excess of emission in the red wing of the line profile of \cii, which suggested an outflow of atomic gas from the QSO expelling mass at a rate of $\dot{M}_\textrm{out}$ $\sim$ 80 $M_\odot$~yr$^{-1}$ \citep{carilli2013}.

The SMG  was detected in \cii line emission with $L_{[\textrm{C}~\textsc{ii}]}$ = 10$^{10}$ $L_\odot$ and its ratio $L_{[\textrm{C}\,\textsc{ii}]}$/$L_\mathrm{FIR}$ = (8.3 $\pm$ 1.2) $\times$ 10$^{-4}$ is in agreement with previous high-redshift studies. The analysis of the CO(5--4) observations indicates the presence of a companion galaxy at 0.6\arcsec\ to the west-northwest of the SMG with a relative velocity of +600~\kms \citep[see Fig. 4 of][]{salome2012}.

LAE1 ($z_{\rm Ly\alpha}$ = 4.703) is located northwest of the QSO in the direction of the \cii bridge and LAE2 ($z_{\rm Ly\alpha}$ = 4.698) toward the southwest (see left panel of Fig~\ref{fig:aper_ha}), both at the same redshift as the QSO. These objects were spectroscopically confirmed by the detection of Ly$\alpha$ emission in \citet{hu1996} and \citet{williams2014}, respectively. In LAE1, the \nii122 $\mu$m  emission \citep{decarli2017,pavesi2016,lee2019} and the absence of high-ionization emission lines in the UV spectrum obtained with the FORS2 on the VLT \citep[\nv $\lambda$1240 \AA, \siiv $\lambda$1396 \AA, \civ $\lambda$1549 \AA, and \heii $\lambda$1640 \AA;][]{williams2014} suggest that this source is not photoionized by the QSO. Similarly, also LAE2 lacks high-ionization emission lines, and together with the \cii158 $\mu$m / \nii122 $\mu$m ratio for LAE2 indicates that its emission is dominated by H II regions and may be driven by star formation ($L_{[\textrm{C}\,\textsc{ii}]}$/$L_{[\textrm{N}\,\textsc{ii}]}$ = 2.3$^{+2.7}_{-0.9}$) and the CO(2-1) observations suggest that it has a considerable reservoir of molecular gas \citep[$\sim4\times10^{10}\,M_{\odot}$, ][]{jones2016}. These observations led to the conclusion that star formation is the primary source of emission for all the LAEs in the system.  An additional companion (LAE3; $z_{\rm Ly\alpha}$ = 4.7019) discovered by \citet{drake2020} and located 5\arcsec\ north of the QSO, shows an SFR(Ly$\alpha$) of 5 $M_\odot$~yr$^{-1}$ and an EW(Ly$\alpha$) > 34.05 \AA, consistent with star formation as the primary driver of Ly$\alpha$ emission.

In this work, we present new JWST NIRSpec IFS observations of BR1202-0725. We analyze the kinematic and physical properties of the main objects presented above, and we also study four newly identified galaxy companions that are part of the system. The observations on which the work is based are presented in Sect. \ref{sec:observations} together with an overview of the data reduction. Section \ref{sec:data_analysis} provides a detailed description of the system based on these observations and illustrates the data analysis. Sections \ref{sec:QSO} to \ref{sec:companions} focus on the results obtained for QSO, SMG and LAEs, respectively. In the latter, we also present the four newly detected galaxy companions of the QSO in addition to the LAEs and the ones found by \citet{salome2012}. The discussion is given in Sect. \ref{sec:discussion} and a summary of our results and final conclusions are presented in Sect. \ref{sec:conclusions}. 

Throughout this paper, we adopt the standard cosmological parameters $H_0$ = 70 \kms~Mpc$^{-1}$, $\Omega_{\rm M}$ = 0.30, and $\Omega_\Lambda$ = 0.70, giving a cosmology corrected scale of 6.48 kpc~arcsec$^{-1}$ at $z$ = 4.7.

\section{Observations and data reduction}\label{sec:observations}

We used JWST NIRSpec IFS \citep{jakobsen2022,boker2023, boker2022} observations obtained as part of the NIRSpec IFS GTO program ``Galaxy Assembly with NIRSpec IFS'' %\textbf{(\href{https://ga-nifs.github.io/}{GA-NIFS})}
(GA-NIFS)\footnote{Proposal ID 1220, cycle 1 (PI: Nora Luetzgendorf). GA-NIFS website: \url{https://ga-nifs.github.io/}.}. These data cover the QSO, the SMG, and two Ly$\alpha$ emitters (LAE1 and LAE2) with two different pointings (covering 3.6 $\times$ 6.5 arcsec$^2$) that overlap by 10\% because the main sources are spatially separated by a projected distance larger than 3 arcsec (see Fig. \ref{fig:aper_ha}).

The observations were acquired on 1st July 2023. Two disperser/filter combinations were used: G235H/G170LP and G395H/F290LP, with a spectral resolution $R$ $\sim$ 2700 and covering the wavelength ranges 1.66 -- 3.05 $\mu$m and  2.87 -- 5.14 $\mu$m, respectively (corresponding to the rest-frame range 2910 -- 9020 \AA\ at a redshift $\sim$ 4.7). The selected spectral configuration covers the main rest-frame optical emission lines at the redshift of the system, including the Balmer lines (H$\alpha$, H$\beta$), \oiii$\lambda \lambda$4959,5007 \AA, \nii$\lambda \lambda$6548,84 \AA, and \sii$\lambda \lambda$6717,31 \AA. Each grating was observed for 3559 s, in two different exposures to target the QSO and the SMG.

We retrieved the raw data from the MAST (Barbara A. Mikulski Archive for Space Telescopes) archive. Then, we reduced them with a customized version of the JWST pipeline (version 1.14.0), using the Calibration Reference Data System (CRDS) context jwst\_1225.pmap. Some modifications to the pipeline allow us to improve the data quality and they are described in detail in \citet{perna2023}, but here we summarize the major changes. First, we applied the ``calwebb\_detector1'' step of the pipeline to account for detector level correction. Before calibrating the count-rate images through the ``calwebb\_spec2'' module of the pipeline, we corrected them by subtracting the 1/f noise through a polynomial fitting.  We identified and removed outliers directly in the calibrated 2-d images, by applying an algorithm similar to ``lacosmic'' \citep{Vandokkum:2001}. Briefly, we computed the derivative along the dispersion direction, and we rejected the 98th percentile of the distribution \citep[see][for details]{DEugenio2024a}. The final data cube was created by combining the individual calibrated 2-D exposures by using the ‘drizzle’ weighting obtaining a cube with a spaxel size of 0.05\arcsec.

In this work, we also used \cii 158 $\mu$m and adjacent continuum observations from ALMA in Band 7 to compare the shape of the \cii spectrum of the SMG with those of \oiii and \ha from JWST and to compare the location of the dust continuum with that of H$\alpha$. We retrieved the raw visibilities of the program 2019.1.01587.S (PI: F. Lelli) from the ALMA archive and calibrated them with the Common Astronomy Software Application (CASA) by using the script included in the dataset. By using the task ``\texttt{tclean}'', we cleaned the calibrated visibilities using a ``Briggs'' weighting scale and a robust parameter equal to 0.5 to obtain the final data cube. The data cube has a beam size of 0.17\arcsec $\times$ 0.12\arcsec, roughly matching the NIRSpec IFS resolution \citep[$\sim$ 0.09\arcsec -- 0.15\arcsec;][]{DEugenio2024a}.

We used HST observations to calculate the astrometric offsets between the NIRSpec IFU and the ALMA data (see Sect. \ref{sec:astrometry}). They were acquired on 24th May 2005 with the Advanced Camera for Surveys (ACS) and the F775W filter. These data are part of proposal ID 10417 (PI: X. Fan) and are organized in four exposures and a total exposure time of 9445 s. The data were downloaded from the Hubble Legacy Archive (HLA) and reduced by the Space Telescope Science Institute (STScI) using available calibration files taken for this observation and accounting for the different dithering positions.

Finally, we also employed \textit{Chandra} archival data to investigate whether the SMG emits X-rays. BR1202-0725 was observed twice with ACIS-S, once during Cycle 3 for 10~ks (ObsID 3025, presented in \citealt{iono2006}) and a second time in Cycle 9 for 30~ks (ObsID 9232, presented in \citealt{li2021_chandra2}). The results of the X-ray data analysis are presented in Sect.~\ref{sec:chandra}.

\section{Data Analysis}\label{sec:data_analysis}
\subsection{Astrometry correction}\label{sec:astrometry}
Since astrometric offsets between the NIRSpec IFU data and the position of the sources on the sky have been reported in the literature \citep[e.g.,][]{perna2023,jones2024,jones2024-2}, we registered the JWST observations to the HST image (ACS-F775W). We verified the astrometric accuracy of the HST image by comparing the position of the star in the oV with the coordinates reported in Gaia DR3 catalog \citep[][]{gaia_collaboration2023}. We used the position of the QSO to register the JWST data to HST ones and found a shift of $\Delta$RA = +0.199 arcsec and $\Delta$Dec = --0.144 arcsec.% \textbf{($\sim$ +4 and $\sim$ --3 pixels respectively)}. 

In this work, we assumed that the ALMA astrometry is sufficiently precise for our purpose, considering that the ALMA positional accuracy varies from 5\% to 20\% of the beam depending on the S/N, which corresponds to an accuracy of 0.03 arcsec for a beam of 0.17\arcsec\ $\times$ 0.12\arcsec. The high S/N ensures that the ALMA astrometry is correct within 1 NIRSpec spaxel.

\subsection{BR1202-0725} \label{sec:obj}

The well-studied BR1202-0725 system has been observed in a wide range of molecular and atomic lines \citep[e.g.,][]{omont1996,salome2012,carilli2013,benford1999,lu2017}. However, this is the first time it was observed in rest-frame optical emission lines. The left panel of Fig. \ref{fig:aper_ha} shows the HST ACS image in the F775W filter, covering the spectral range 6801 -- 8630 \AA\ 
(corresponding to the UV rest frame emission between 1193\AA\ and 1514\AA\ at $z=4.7$) with the two JWST pointings used in this study superimposed. The JWST map at the wavelength of the H$\alpha$ emission line (3.74 $\mu$m < $\lambda$ < 3.75 $\mu$m) is shown in the right panel. The ALMA dust continuum emission is shown as black contours.

The QSO is the most prominent source in \ha in the FoV of the NIRSpec IFS data, appearing in the bottom-left part of the map, and it is dominated by the point spread function (PSF) emission. However, LAE1 is also very prominent, showing a spatially resolved structure that could be connected with the bridge between the QSO and the SMG observed in \cii \citep{carniani2013}. The SMG is located in the top-right part of the map, showing an elliptical appearance in the ionized gas. This object has been studied in other FIR and mm/submm emission lines, such as \nii at 122 and 205 $\mu$m (rest-frame), with ALMA and the IRAM Plateau de Bure Interferometer (PdBI) \citep{decarli2014,drake2020}. Despite some attempts to study the optical emission with SINFONI, that results in a non-detection of the \oii$\lambda\lambda$3726, 29 \AA\ doublet \citep{Couto:2016}, and an integrated spectrum from the AKARI space telescope encompassing the \ha emission of the QSO and SMG \citep{jun2015}, this is the first time that this optically obscured, submillimeter-bright object can be spatially resolved at rest-frame optical wavelengths, thanks to the high resolution and sensitivity of JWST.

\subsection{Point spread function modeling} \label{sec:psf_mod}
One of the main goals of this work is identifying the presence of companion galaxies close to the QSO and studying their properties. Therefore, it is fundamental to remove the point-like emission of the QSO that outshines a large fraction of the IFS FoV.

%We have thus generated images of the PSF of the instrument at different wavelengths by exploiting the emission of the QSO itself \citep[see][]{decarli2024}.
Following the approach of \cite{decarli2024}, we generated a set of 17 images of the PSF by integrating the emission of the QSO in different wavelength bands with a width varying between 0.01 $\mu$m and 0.13 $\mu$m depending on either the presence of mission line or cosmic rays in the field (see Fig.~\ref{fig:psf1} in Appendix \ref{ap:psf_mod}). The selected bands exclude the brightest forbidden emission lines, such as \oiii, \sii and \nii, as well as narrow Balmer lines, because they are not associated with the spatially unresolved emission of the accretion disk, and are therefore likely to be spatially extended.

\begin{figure}
\centering
\includegraphics[width=\linewidth]{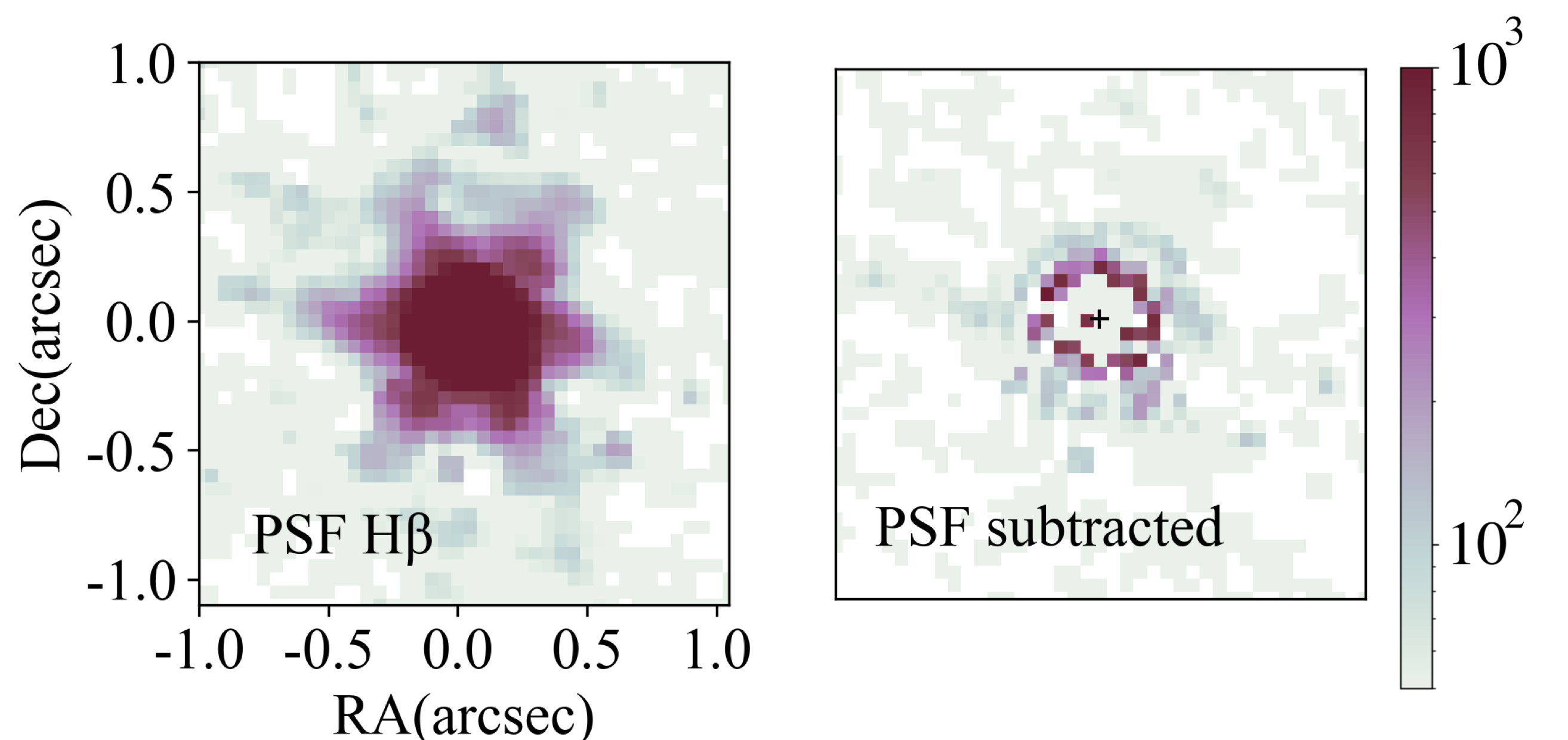}
\caption{Original (left) and PSF-subtracted (right) maps of the QSO at the wavelength corresponding to the H$\beta$ emission line.}
\label{fig:psf2}
\end{figure}

We interpolated these PSF images at each wavelength of the data cube, and we obtained a data cube with the normalized image of the PSF at each wavelength. Then, we scaled the flux of each one using an aperture of 0.25\arcsec\ centered on the peak of the QSO \citep[see][]{decarli2024}.  Finally, we subtracted the PSF models from the JWST data and Fig.~\ref{fig:psf2} shows the emission of the QSO at the wavelength of the H$\beta$ emission line (${\rm 2.75~ \mu m<\lambda<2.77~ \mu m}$, observed frame) before and after subtraction of our modeled PSF.  We notice that our procedure slightly over-subtracts the QSO emission at the center, hence the flux measurements in this region are not reliable. Therefore, we masked a region of radius of 0.2\arcsec\ from the centroid of the QSO in the PSF-subtracted data cube and limited our analysis to the emission at larger distances from the bright source.

\subsection{Aperture correction}\label{sec:ap_cor}
The aperture correction is necessary to determine the total emission flux of the targets since the PSF of the instrument depends on the wavelength.
Following \citet{loiacono2024}, we determined the flux losses from 1.7 $\mu$m to 5.2 $\mu$m for four different apertures by using the PSF models generated in Sect. \ref{sec:psf_mod}. Figure \ref{fig:ap_cor} reports the fraction of flux recovered from the four apertures with respect to an aperture with a radius of 1 arcsec, which is large enough to recover all the emission of point-like sources. All spectra and fluxes reported hereafter are corrected for the flux losses depending on the selected aperture and wavelength. We present the procedure and the results in Appendix \ref{ap:cor_aper}.

\section{QSO}\label{sec:QSO}

\begin{figure*}[ht!]
\includegraphics[width=\textwidth]{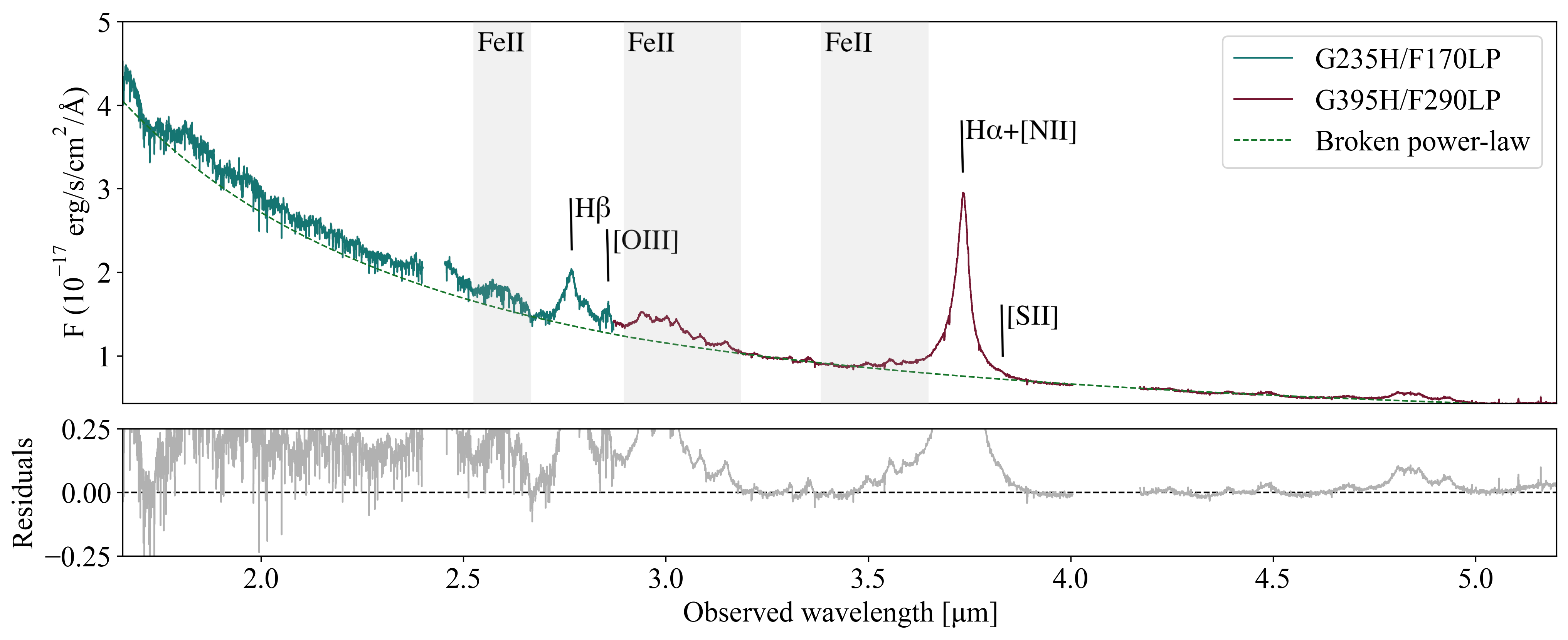}
\caption{JWST/NIRSpec rest-frame optical
spectrum of the QSO in the two gratings/filters G235H/F170LP and G395H/F290LP (green and purple lines, respectively), extracted from an aperture of $R = 0.5\arcsec$. The brightest nebular lines are marked and the \feii multiplet bands are highlighted in gray.
The green dashed line shows the broken power-law continuum model fitted to reproduce the featureless continuum emission. The gray line in the bottom panel shows the residuals of the fitting.}
\label{fig:cont_pw}
\end{figure*}

Figure \ref{fig:cont_pw} illustrates the spectrum of the QSO extracted from an aperture of $R$ = 0.5 arcsec and located at the peak of the emission (RA = 181.34634 deg, Dec = --7.70906 deg). We have calculated the errors on the integrated spectrum from the error extension of the data cube in the same spaxels of the selected region propagating them in quadrature (see Appendix \ref{ap:err}). The spectrum clearly reveals broad emission lines of \ha and \hb, and \feii emission lines arising from the broad line region (BLR).

We initially modeled the featureless continuum emission of the QSO. A simple power-law function is usually adopted to reproduce the multi-blackbody emission of the accretion disk \citep[e.g.,][]{loiacono2024}, but given the extended wavelength range covered by the NIRSpec data, we adopted two power-law functions as follows:
\begin{equation}\label{eq:cont}
\centering
f(\lambda) = 
\left\{\begin{aligned}
 A\left( \frac{\lambda}{\lambda_0}\right)^{\alpha_1} \quad \lambda <\lambda_0\\
 A\left (\frac{\lambda}{\lambda_0}\right)^{\alpha_2} \quad \lambda \geq\lambda_0\\
\end{aligned}\right.
\end{equation}
where $A$ is the normalization, $\lambda_0$ the break wavelength and $\alpha_1$ and $\alpha_2$ the power-law indexes. 
This profile has been widely used to reproduce the continuum emission of local QSOs \citep{Vanden2001} which change slope at the rest-frame wavelength of $\sim$5000 \AA\ due to the contribution of the host galaxy.

We fitted the line-free spectral regions of the spectrum with the function in Eq. \ref{eq:cont}. Figure \ref{fig:cont_pw} shows the QSO spectrum together with the continuum model, while the fitted parameters are shown in Table \ref{tab:QSO_cont}.
Considering a redshift of $z$ = 4.6943 \citep{carniani2013}, we found a rest-frame value for the break wavelength of 5107 $\pm$ 108 \AA, comparable to others found in the literature for QSOs at a similar redshift \citep[e.g.,][]{loiacono2024}.

\begin{table}
\caption{Fitted continuum parameters for the QSO.}\label{tab:QSO_cont}
\centering
\begin{tabular}{cc}
\hline \hline
Parameter & Value\\ \hline \hline
$A$ & (1.22 $\pm$ 0.05) $\times$ 10$^{-17}$erg~s$^{-1}$~cm$^{-2}$~\AA$^{-1}$ \\
$\alpha_1$ & --2.130 $\pm$ 0.009\\
$\alpha_2$ & --1.931 $\pm$ 0.005\\
$\lambda_0^{\rm obs}$ & 2.91 $\pm$ 0.06 $\mu$m\\
$\lambda_0^{\rm rest, ~ \it a}$ & 5107 $\pm$ 108 \AA \\
\hline
\end{tabular}
\begin{tablenotes}
\item $^a$ Relative to $z$ = 4.6943 \citep{carniani2013}.
\end{tablenotes}  
\end{table}

\begin{figure}[h!]
\includegraphics[width=\columnwidth]{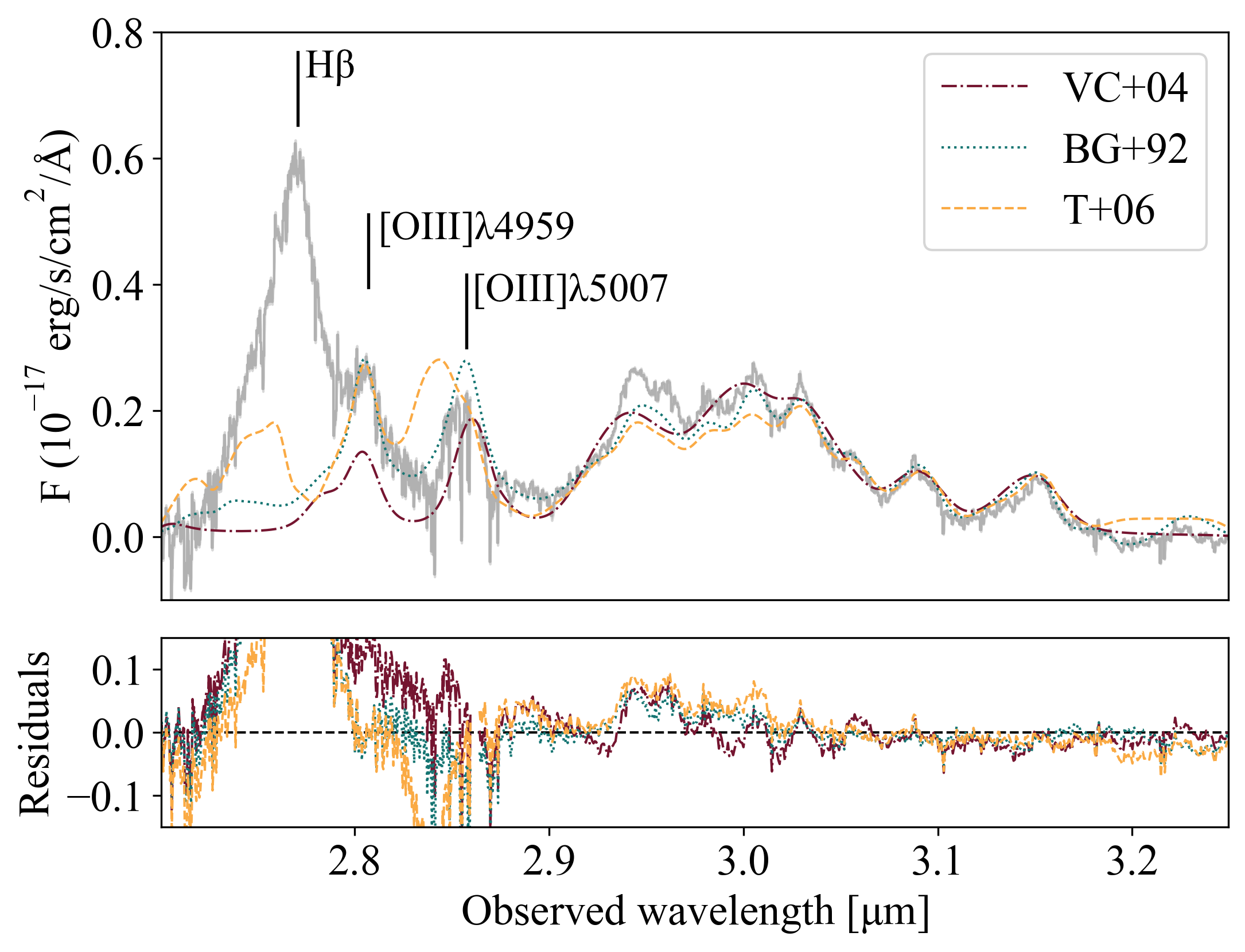}
\caption{QSO rest-frame optical continuum-subtracted spectrum (gray line) in the spectral region 4700 -- 5400 \AA. The emission lines H$\beta$ and \oiii$\lambda \lambda$4959,5007 \AA\ (marked with vertical lines) are blended with the \feii emission, which can be clearly seen at $\sim$3 $\mu$m. The following fitted \feii templates are shown: \citet[][VC+04, red line]{veron-cetty2004}, \citet[][BG+92, blue line]{boroson1992} and \citet[][T+06, yellow line]{tsuzuki2006}. The lines in the bottom panel shows the different residuals of the fitting.}
\label{fig:fit_Fe}
\end{figure}

Typical type 1 QSO spectra such as the one reported show the presence of a forest of \feii emission bumps. % stemming from the BLR.
The analysis of the \feii multiplets is not among the aims of this paper. However, it is important to obtain a continuum- and \feii-free spectrum in order to fit the optical emission lines to derive QSO properties such as the BH mass. In particular, appropriately modeling the prominent \feii multiplets in the rest-frame wavelength region between 4700 and 5400 \AA\ is critical, since they are blended with H$\beta$ and \oiii$\lambda \lambda$4959,5007 \AA\ emission lines.

We tried to reproduce the \feii emission with the following three different templates: \citet[][T+06]{tsuzuki2006}, \citet[][VC+04]{veron-cetty2004}, and \citet[][BG+92]{boroson1992}. The results are shown in Fig. \ref{fig:fit_Fe}. None of them was able to accurately reproduce all the \feii emission features in the spectrum. However, we discarded the results obtained with \citet{tsuzuki2006} and \citet{boroson1992} templates because they overpredict the flux emission at the wavelength range of the \oiii \citep[see][]{park2022}. We have thus assumed as fiducial model the results obtained with the \citet{veron-cetty2004} \feii template.

Finally, we fitted the brightest nebular lines present in the integrated continuum-subtracted spectrum, H$\beta$, \oiii$\lambda\lambda$ 4959,5007 \AA, \nii$\lambda\lambda$6548,84 \AA, H$\alpha$, and \sii$\lambda\lambda$6717,31 \AA, using a combination of Gaussian profiles convolved with the line spread function of the instrument. We have used two Gaussian components: (i) a \textit{Narrow} component with $\sigma$ < 600 \kms, and (ii) a broad one (dubbed \textit{Outflow}) with 400 \kms < $\sigma$ < 2000 \kms. For the Balmer recombination lines, H$\alpha$ and H$\beta$, we added an additional Gaussian component with a velocity dispersion spanning a range  1000~${\rm \kms} < \sigma < 5000$ \kms to reproduce the emission from the BLR. We tied the kinematics of each Gaussian component for each emitting species (centroid and line width). The two emission lines of the \oiii and \nii doublets originate from the same upper level, then we fixed the intensity ratios between each line of these doublets to their theoretical values: $I$(\nii$\lambda$6584)/$I$(\nii $\lambda$6548) = 2.94 and  $I$(\oiii$\lambda$5007)/$I$(\oiii$\lambda$4959) = 3 \citep{Osterbrock2006}. We also fixed the \ha/\hb ratio of the \textit{Narrow} and \textit{Outflow} components to take values 2.3 < \ha/\hb < 10. We included the \citet{veron-cetty2004} \feii template in the fitting, by allowing its amplitude and velocity dispersion to vary.

\begin{figure*}[h!]
\includegraphics[width=\textwidth]{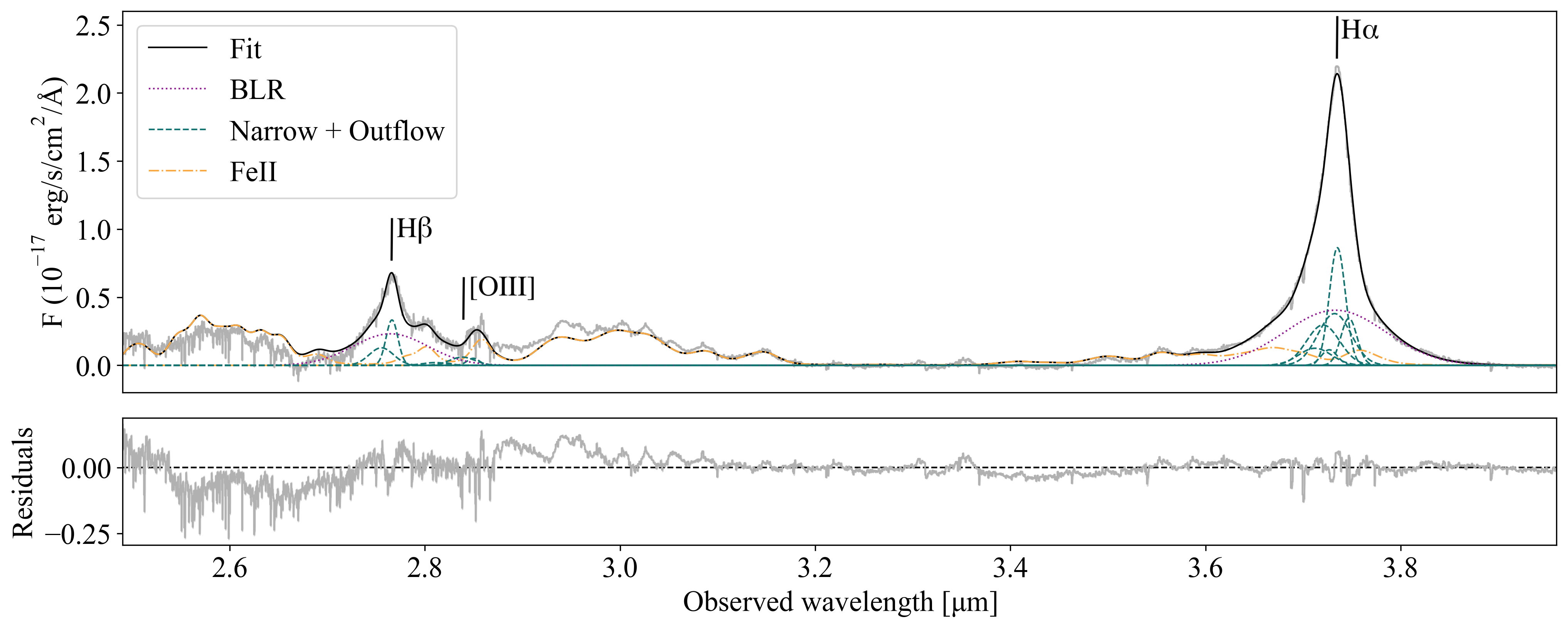}
\caption{Continuum-subtracted JWST NIRSpec spectrum of the QSO. The best-fit model is shown with a black solid line, while the results of the individual components are reported with different colors: the blue dashed line shows the \textit{Narrow} + \textit{Outflow} components, and the purple dotted line corresponds to the \textit{BLR} component. The fitted \citet[][]{veron-cetty2004} \feii template (VC+04) is shown with the orange dash-dotted line.}
\label{fig:fit}
\end{figure*}

\begin{table}[ht]
    \centering
    
    \caption{Emission line properties of the QSO as constrained from our best fit of the spectrum (Fig. \ref{fig:fit}). The top table reports the velocity shift and velocity dispersion of the emission lines, while their fluxes are reported in the bottom table.}

    \begin{tabular}{ccc}
 \multicolumn{3}{c}{Kinematics} \\
 \hline \hline
~ & $\Delta v$ [\kms\!]$^a$ & $\sigma$ [\kms\!] \\ \hline \hline
BLR &  --410 $ \pm$  40   &  4160 $ \pm$  60  \\
Narrow &  --240 $ \pm$  20   &  660 $ \pm$ 30   \\
Outflow & --1420  $ \pm$  120   & 1360 $ \pm$ 80   \\ 
\hline \\
\end{tabular}

\begin{tabular}{cccc}
 \multicolumn{4}{c}{Emission line fluxes$^b$} \\

\hline
\hline
Line & BLR & Narrow$^c$ & Outflow$^c$ \\
\hline \hline
\hb  &  220  $ \pm $  10  &  50  $ \pm $  4  & 40  $ \pm $  5 \\
\oiii  &    -    &   8 $ \pm $  3  &  20  $ \pm $  4 \\
\ha  &  520  $ \pm $  20  &  180   $ \pm $  30  &  130  $ \pm $  20 \\
\nii  &    -    &  72  $ \pm $  15  &  160  $ \pm $  50 \\

\hline
\end{tabular}

\begin{tablenotes}
\item $^a$ Relative to $z$ = 4.6943 \citep{carniani2013}.\\
\item $^b$ In units of ${\rm 10^{-17}~erg~s^{-1}~cm^{-2}}$.\\
\item $^c$ These measurements should be taken with caution due to the degeneracies in the fitting.
\end{tablenotes}  
\label{tab:QSO1}
\end{table}

In Table \ref{tab:QSO1}, we report the emission line velocities and velocity dispersions for each component and the emission-line intensities of the nebular lines after the aperture correction (see Sect. \ref{sec:ap_cor}). Figure \ref{fig:fit} shows the fitted profiles for each component. We can see that the multicomponent fitting accurately reproduces the spectrum at the wavelengths of the emission lines, with a narrow and a broad component with velocity dispersions of 660 $\pm$ 30 \kms and 1360 $\pm$ 80 \kms, respectively. The BLR is blueshifted by --410 \kms relative to $z$ = 4.6943 \citep{carniani2013} and has $\sigma$ = 4160 $\pm$ 60 \kms. 

We stress that the \textit{Narrow} and \textit{Outflow} components in the fitting are strongly affected by degeneracies, and the \feii multiplets present in the spectrum also influence the values obtained for the \oiii emission line. Therefore, these results should be interpreted with caution. For the \textbf{BLR} component, the velocity dispersion and amplitude vary depending on the fitting priors. Specifically, the velocity dispersion ranges from 3400 \kms to 4600 \kms (mean value of 3900 \kms with a standard deviation of 400 \kms), which translates into an \ha emission line flux range of 470 -- 660 ${\rm \times 10^{-17}~erg~s^{-1}~cm^{-2}}$.
However, as shown in Sect.~\ref{sec:bh_props}, the different BH masses obtained from the above range of values are well within the uncertainty on the calibration used to estimate them, therefore they are consistent among each other.

\section{SMG}\label{sec:SMG}

\begin{figure*}[h]
\includegraphics[width=\linewidth]{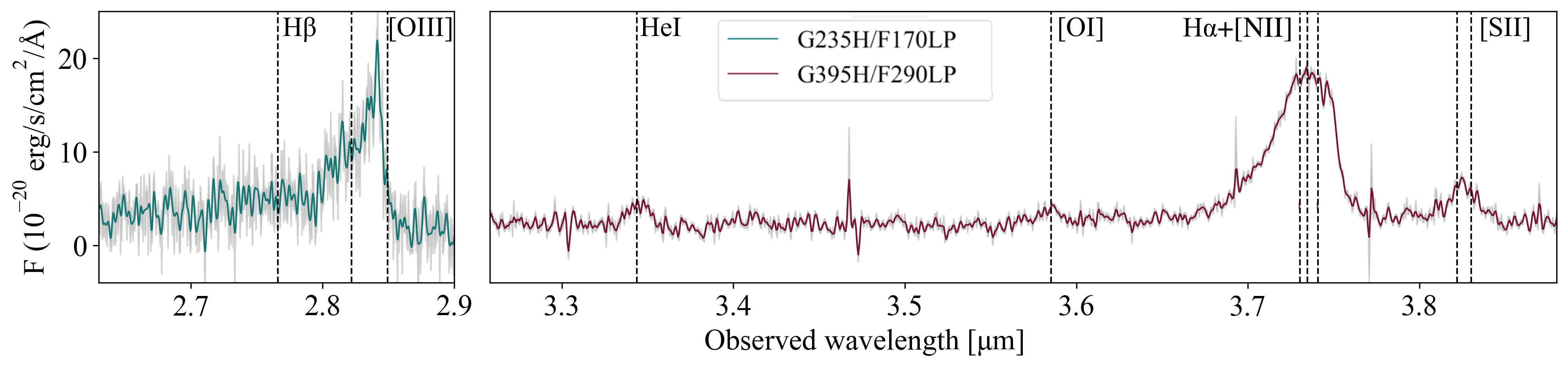}
\caption{JWST NIRSpec rest-frame optical spectrum of the SMG in the two gratings/filters G235H/F170LP and G395H/F290LP (blue and red lines, respectively), extracted from an aperture of $R = 0.3\arcsec$. The principal nebular lines are marked with vertical dashed lines and the errors of the spectrum are shown in light gray.}
\label{fig:sms_esp}
\end{figure*}

Figure \ref{fig:sms_esp} shows the integrated spectrum of the submillimeter galaxy located $\sim$24 kpc north-west of the QSO (see Fig. \ref{fig:aper_ha}). The spectrum was extracted from an aperture of $R$ = 0.3 arcsec centered on the position RA = 181.34575 deg and Dec = --7.70825 deg. In contrast to the QSO, the continuum emission from the SMG is detected with low signal-to-noise and its intensity does not depend strongly on the wavelength.

The G395H data clearly reveal blended \ha\ and \nii\ emission and the \sii doublet. \hei$\lambda$5875 \AA\ and \oi$\lambda$6300 \AA\ are also visible in the spectrum, with lower signal-to-noise ratios (S/N = 2.6). At lower wavelengths, the \oiii$\lambda$5007 \AA\ is detected with a signal-to-noise ratio higher than 21, while the \hb\ line is not detected due to the high dust attenuation of the galaxy.

\begin{figure}
\includegraphics[width=\linewidth]{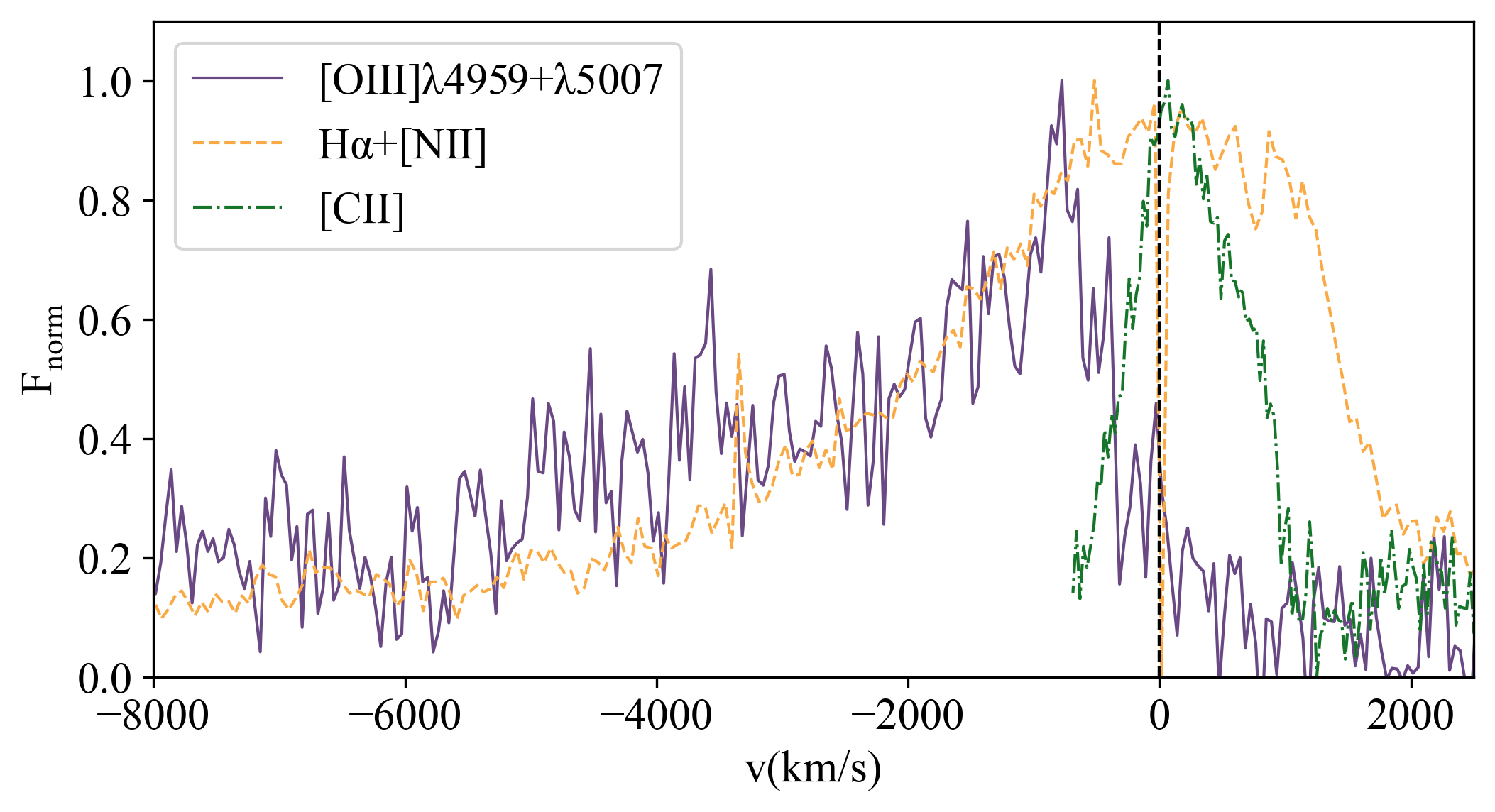}
\caption{Comparison of the SMG line profiles of the \oiii, H$\alpha$+\nii, and \cii emission lines in velocity space. All line profiles are normalized to their peak value. The vertical black dotted line indicates the zero velocity at the systemic redshift of $z = 4.6891$ for \oiii$\lambda$5007, H$\alpha$, and \cii emission lines.}
\label{fig:sms_lines}
\end{figure}

We note that the peak of \oiii\ is blueshifted with respect to the other optical lines observed in the NIRSpec data. By adopting the systemic redshift of $z = 4.6891$ determined from \cii\ and CO far-infrared emission lines  (see \citealt{carniani2013} and \citealt{salome2012}, respectively), we find that the peak of H$\alpha$+\nii is consistent with the systemic redshift of the galaxy while the peak of the \oiii\ line is shifted by --1000 \kms (Fig.~\ref{fig:sms_lines}). 
Both the profiles of the \oiii doublet and \ha\!+\nii are very broad and show an asymmetric shape with a prominent emission at shorter wavelengths that suggest the presence of galactic outflows  \cite[e.g.,][]{cresci2015, carniani2015, harrison2015, harrison2016, balmaverde2016, brusa2016, perna2017, leung2019, vayner2021, vayner2024, venturi2023, speranza2024}.

We fitted the brightest nebular lines of the SMG by using a combination of Gaussian profiles. Due to the complexity of the blended \ha and \nii profiles (see Fig. \ref{fig:sms_esp}), we started by fitting only the \oiii doublet emission to determine the kinematics of the blueshifted broad wings. 
%and decrease the degeneracies in the \nii$\lambda\lambda$6548,84 \AA\ and H$\alpha$ set. 
We used a combination of two Gaussian components: (i) a narrow component with $\sigma < 400$~\kms, and (ii) a broad component with ${\rm 400~ km~s^{-1}<\sigma < 1500~ km~s^{-1}}$. We dubbed the narrow and broad components as \textit{Narrow 2} and \textit{Outflow}. We adopted the label \textit{Narrow 2} because this component is not at the systemic redshift of the galaxy defined from \cii and CO. We used a second-order polynomial to model the continuum in this spectrum, although the results show that it is almost constant with a value of 1.6 $\times$ 10$^{-20}$ erg~s$^{-1}$~cm$^{-2}$~\AA$^{-1}$.

\begin{figure}[h]
\centering
\includegraphics[width=0.85\linewidth]{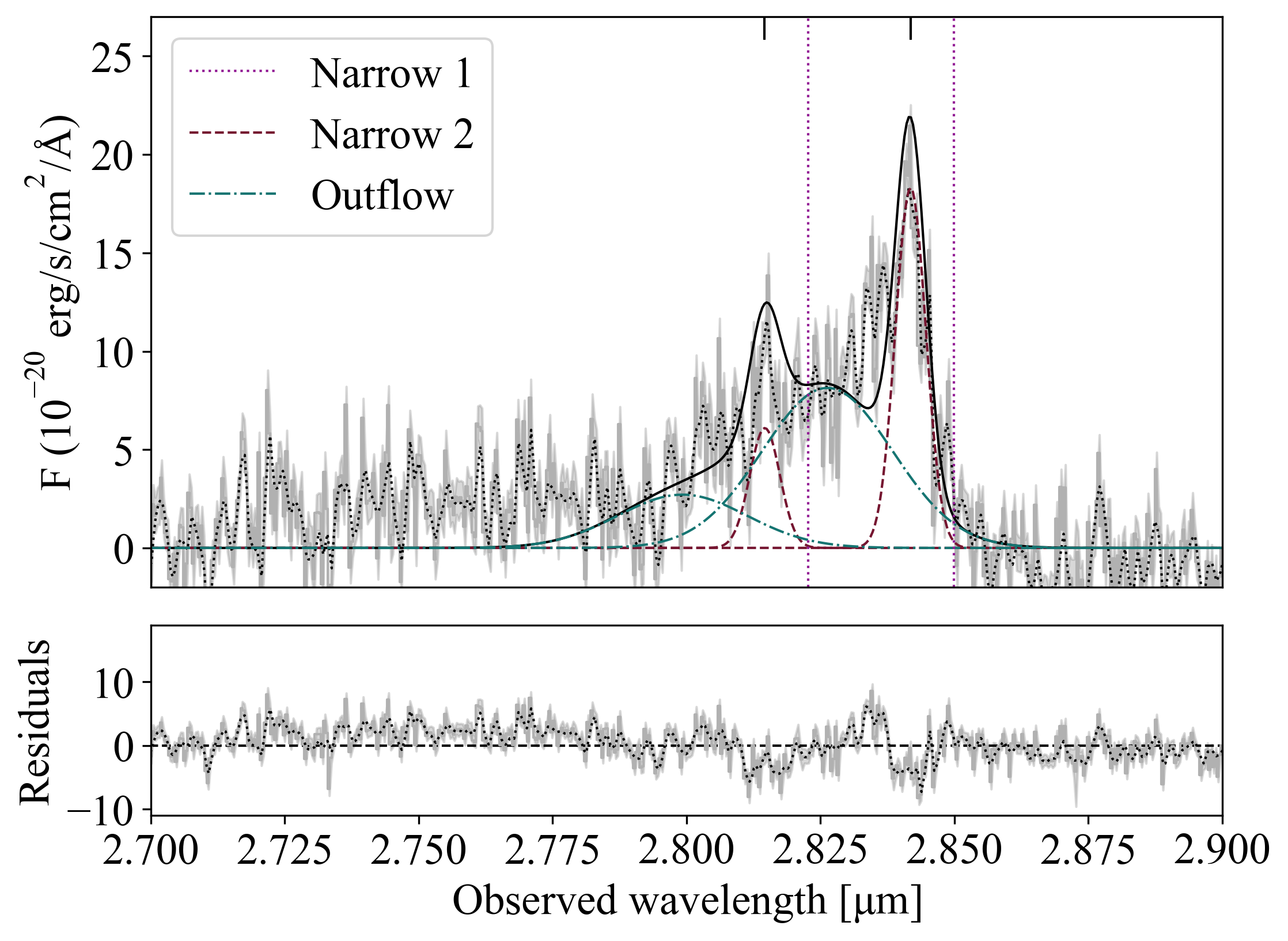}
\includegraphics[width=0.85\linewidth]{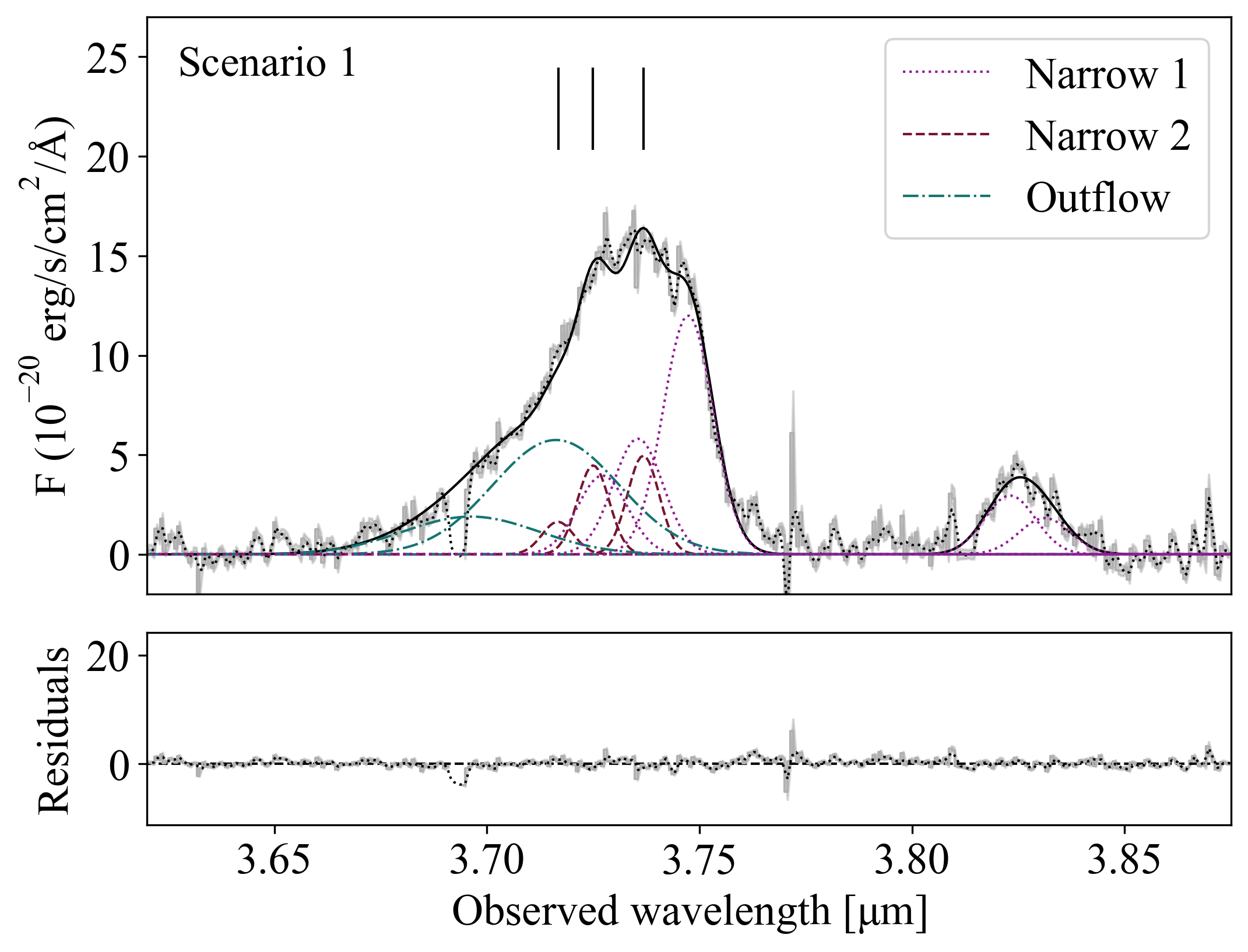}
\includegraphics[width=0.85\linewidth]{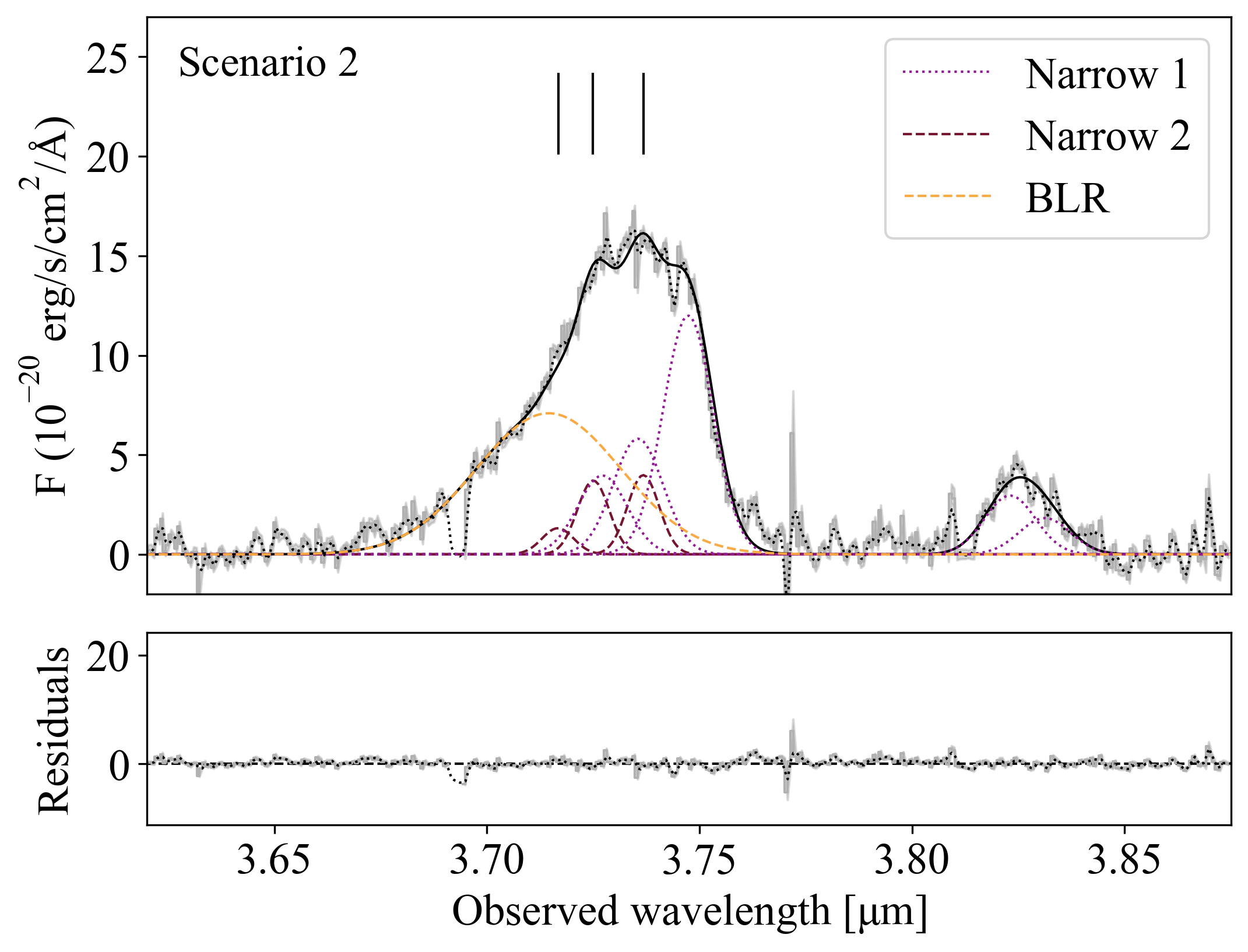}
\caption{Fitting results for the SMG spectrum, showing the \textit{Narrow 1}, \textit{Narrow 2}, \textit{BLR}, and \textit{Outflow} components (dotted purple, dashed red, dashed yellow, and dash-dotted blue lines, respectively). The errors considered in the fitting procedure are displayed in gray and the residuals of the fits are shown at the bottom of each panel. Top panel: Fit of the \oiii$\lambda\lambda$4959,5007 \AA\ emission lines. The vertical dotted lines mark the position of \oiii at the systemic redshift determined from \cii. Middle and bottom panels: Fits of the H$\alpha$, \nii$\lambda\lambda$6548,6584 \AA, and \sii$\lambda\lambda$6717,6731 \AA\ emission lines for the first (outflow) and second (BLR) scenarios described in Sect. \ref{sec:SMG}. The black vertical lines indicate the position of the emission lines for the \textit{Narrow 2} component found in \oiii.}
\label{fig:sms_fit}
\end{figure}

The upper panel of Fig. \ref{fig:sms_fit} shows the best-fit model for the \oiii\ doublet. The \textit{Outflow} component has a velocity dispersion of $1300 \pm 500$~\kms  and is blueshifted by $-2450 \pm 40$~\kms relative to the systemic velocity of the galaxy. The profile of this line is similar to the broad blueshifted wings observed in luminous QSOs, associated with fast outflowing gas. 

The \textit{Narrow 2} instead has a velocity dispersion consistent with what is expected from a rotating massive galaxy ($\sigma = 300 \pm 100 $~\kms), but 
is blue-shifted by $-760\pm 90$~\kms relative to the systemic redshift determined from \cii. This remarkable velocity shift suggests that this emission is associated with outflowing gas. A similar shift has been observed in AGN  called ``blue outliers'', in which the peak of the \oiii\ is blueshifted by more than --250~\kms with respect to the peak of \ha and \hb lines \citep{zamanov2002}. In the local and $z\sim1$ Universe, these features are observed in highly accreting AGN, and the velocity shift is thought to be due to strong nuclear outflows \citep{zamanov2002,marziani2003,lanzuisi2015,marziani2016,cracco2016,perna2021,perna2023}.  In conclusion, as similar profiles have been observed in other AGN, we believe that the entire \oiii\ emission line of the SMG (\textit{Narrow 2}+\textit{Outflow}) is tracing a ionized outflow.  

We stress that the extraction aperture of the spectrum (radius = 0.3\arcsec) encloses all the outflow emission. It does not encompass the whole galaxy emission, which is more extended than the aperture, but this does not affect our analysis, which focuses on the properties of the broad component rather than on those of the extended galaxy emission.

As discussed above, the blended \ha\!+\nii profile is complex, and the results strongly depend on the number of components used to reproduce the shape of the emission lines. Therefore, we considered the two following possible scenarios: (1)  the broad asymmetric profile of the complex line is due to the presence of galactic outflows; (2) the SMG hosts a type 1 AGN and the very broad component is associated with the emission of the broad line region. 
We discuss these alternative scenarios in the following. The combination of both is also possible; therefore, we will discuss the two extreme scenarios in the following sections.

\subsection{Scenario 1: High-velocity outflow hosted by the SMG}

The two components (\textit{Outflow} and \textit{Narrow 2}) adopted to model the \oiii line profile, significantly blueshifted relative to the systemic velocity, are not sufficient to reproduce the profile of the \ha, \nii, and \sii. Therefore, for the five emission lines of these three species we added an additional Gaussian component with $\sigma < 500$ \kms dubbed \textit{Narrow 1}.
We also fixed the kinematics (centroid and line width) of the \textit{Narrow 2} and \textit{Outflow} components to the best-fit results from the \oiii fitting process.

The middle panel of Fig.~\ref{fig:sms_fit} illustrates the best-fit model (Reduced chi-squared, $\chi$ = 13.89) and Table~\ref{tab:SMG} reports the best-fit values for the free parameters. The   H$\alpha$+\nii complex is well-reproduced with this fitting and the \textit{Narrow 1} centroid is within 110~\kms of the systemic redshift of the galaxy. As the line spread function of the instrument is about FWHM $\sim$ 100~\kms, we can conclude that \textit{Narrow 1} arises from the gas at rest in the galaxy.  This component is also consistent with the kinematics of the  \hei$\lambda$5875 \AA\ and \oi$\lambda$6300 \AA\ nebular emission lines, and also with that of \cii and CO lines. The kinematics of this component is consistent with being gas in the host, but the velocity dispersion ($\sigma$ $\sim$ 470 \kms) is larger than that of \cii and CO lines ($\sigma_{[\textrm{C}\,\textsc{ii}]}$ $\sim$ 307 \kms and $\sigma_{\rm CO}$ $\sim$ 313 \kms), which may be due to turbulent and disturbed ionized gas at large scales due to interactions with the other members of the group \citep[see][]{salome2012,lehnert2020}.

Remarkably, no H$\alpha$ \textit{Outflow} component is found in this fitting, with only \nii doublet lines being non zero. However, there could be some degeneracies in the fitting due to the complexity and broadening of the blended  H$\alpha$+\nii profile. The best-fit model indicates that \nii-to-\ha flux ratio of the \textit{Outflow} component is higher than 7 (${\rm log(\nii\,\lambda6584\,\AA/\ha) > 0.85}$), an extreme line ratio that suggests that the blueshifted H$\alpha$ component is blended with the \nii due to the degeneracy.

We determined the upper limit on the reddening calculating the standard deviation of the spectrum at the position of the H$\beta$ emission line, obtaining a value of 0.85 $\times$ 10$^{-20}$ erg~s$^{-1}$cm$^{-2}$ (1$\sigma$). By assuming that the Balmer lines has the same line width as the other nebular lines, we found that the Balmer ratios for the narrow components are H$\alpha _{Narrow\ 1}$/H$\beta _{Narrow\ 1}$ > 9.2 and H$\alpha _{Narrow\ 2}$/H$\beta _{Narrow\ 2}$ > 7.1. The extinction $A_V$ can be calculated by adopting the Galactic extinction law of \citet{reddening}, with a specific attenuation of $R_V$ = 2.97 and the theoretical ratio H$\alpha $/H$\beta $ = 2.863 from \citet[][for $n_{\rm e}$ = 100 cm$^{-3}$, $T_{\rm e}$ = 10$^4$ K, case B recombination]{Osterbrock2006}. We thus obtained $A_{V,~{Narrow\ 1}}$ > 3.5 mag and $A_{V,~{Narrow\ 2}}$ > 2.7 mag.

\subsection{Scenario 2: The SMG as a Type 1 AGN}

We present now the other more ``extreme'' scenario in which the broad blue-shifted \ha\ component might be associated with the emission of a BLR.
To test the case that the broad component in H$\alpha$+\nii is due to emission from the BLR, we  replaced the \textit{Outflow} component in fitting this line complex with another Gaussian profile dubbed \textit{BLR}, only associated to the \ha\ line, allowing its velocity dispersion to vary between 1000 and 5000~\kms, as typical of BLRs in active galactic nuclei. The results of the spectral fitting are listed in Table~\ref{tab:SMG}. We stress that, even if the broad component in H$\alpha$+\nii is due to a BLR and not an outflow, this does not change the fact that the very broad component observed in \oiii (see Sect. \ref{sec:SMG}) can only be explained by the presence of a fast outflow.

The best-fit model ($\chi$ = 13.80) illustrated in the bottom panel of Fig.~\ref{fig:sms_fit} shows that an individual \textit{BLR} Gaussian profile in \ha with velocity dispersion of 1370~\kms (Table \ref{tab:SMG}) is sufficient to reproduce the prominent wings in the spectrum of \ha\!+\nii. However, we note that the component is blueshifted by --1600~\kms with respect to the systemic velocity of the galaxy.  Velocity shifts of BLR lines of thousands of \kms have been already observed in other AGN from the Sloan Digital Sky Survey \citep[e.g.,][]{shields2009,steinhardt2012,eracleous2012,li2022,zhang2024} and several explanations have been proposed to explain the spectral displacement of the BLR Balmer lines \citep[see][]{2013ApJ...777...44J,2008ApJ...683L..21K, 2010Natur.463E...1G}:  a) dense gas clouds forming the BLR are moving on elliptical Keplerian orbits; b) a recoiling supermassive black hole; c) binary black holes. Multi-epoch spectra are necessary to investigate the origin of the blueshifted BLR among the aforementioned explanations, therefore we refrain from further speculations.

\begin{table*}[h]
\centering
\caption{Emission line properties of the SMG as constrained from our best fit of the spectrum in Fig. \ref{fig:sms_fit} (extracted from an aperture radius of 0.3\arcsec). Emission line intensities are in units of ${\rm 10^{-20}~erg~s^{-1}~cm^{-2}}$.}
\begin{tabular}{cccccccc}
        \hline \hline
~ & $\sigma$ [\kms\!] & $\Delta v$ [\kms\!]$^a$ & $F_{\rm [O\,\textsc{iii}]\lambda 5007 \AA}$ & $F_{\rm H\alpha}$ &  $F_{\rm [N\,\textsc{ii}]\lambda 6584 \AA}$ &  $F_{\rm [S\,\textsc{ii}]\lambda 6717 \AA}$ &  $F_{\rm [S\,\textsc{ii}]\lambda 6731 \AA}$ \\ \hline \hline
\multicolumn{8}{c}{Scenario 1: Outflow}  \\ \hline
Narrow 1 & 470 $\pm$ 20   &  111 $\pm$ 2    & -             & 856 $\pm $ 3  & 1772 $\pm$ 2   & 445 $\pm$ 2   & 275 $\pm$ 2 \\ 
Narrow 2 & 300 $\pm$ 100  &  --760 $\pm$ 90  & 1300 $\pm$ 30 & 417 $\pm $ 2  & 464 $\pm$ 3   &  - & -\\ 
Outflow  & 1300 $\pm$ 500 &  --2450 $\pm$ 40 & 2480 $\pm$ 50 & -             & 2142 $\pm$ 2  & -          & -         \\
\hline

\multicolumn{8}{c}{Scenario 2: BLR}  \\ \hline
Narrow 1 & 470 $\pm$ 20 &  72 $\pm$ 2 & - & 855	$\pm$ 1 & 1771 $\pm$ 1 & 445 $\pm$ 1 & 275 $\pm$ 1\\ 
Narrow 2 & 300 $\pm$ 100 &  --760 $\pm$ 90 &  1300 $\pm$ 30  & 346 $\pm$ 1  & 373 $\pm$ 2  & - & -  \\ 
Outflow &  1300 $\pm$ 500 &  --2450 $\pm$ 40 & 2480 $\pm$ 50 & - & - & -  & -\\     
BLR &  1370	 $\pm$ 50 &  --1600 $\pm$ 400 & - & 3021	$\pm$ 5 & - & - & -\\       
\hline
\end{tabular} 

\begin{tablenotes}
\item $^a$ Relative to $z$ = 4.6891 \citep{carniani2013}.\\
\end{tablenotes}

\label{tab:SMG}
\end{table*}

\section{LAEs and companions}\label{sec:companions}

\begin{figure*}
\includegraphics[width=\textwidth]{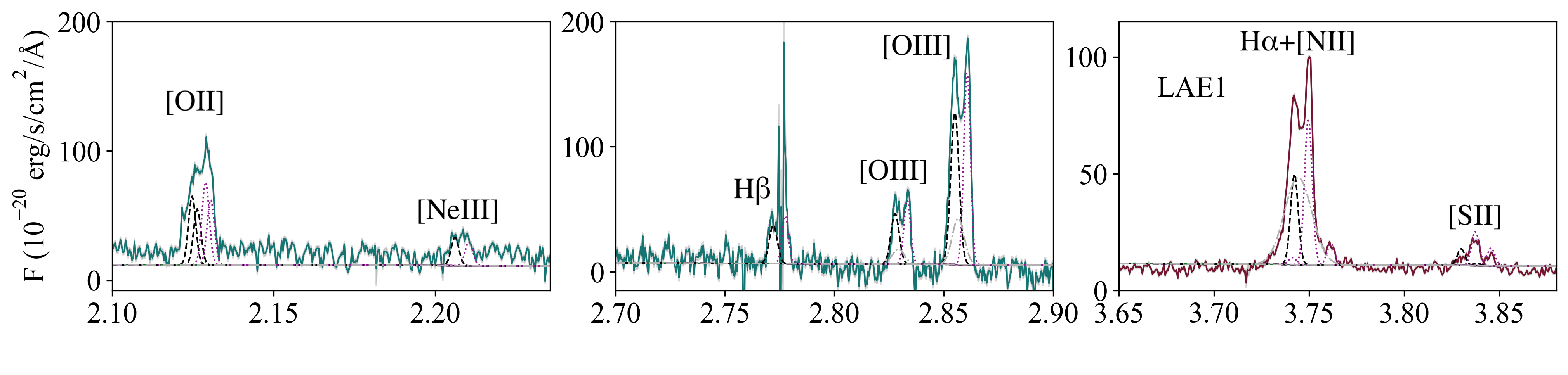}
\includegraphics[width=\textwidth]{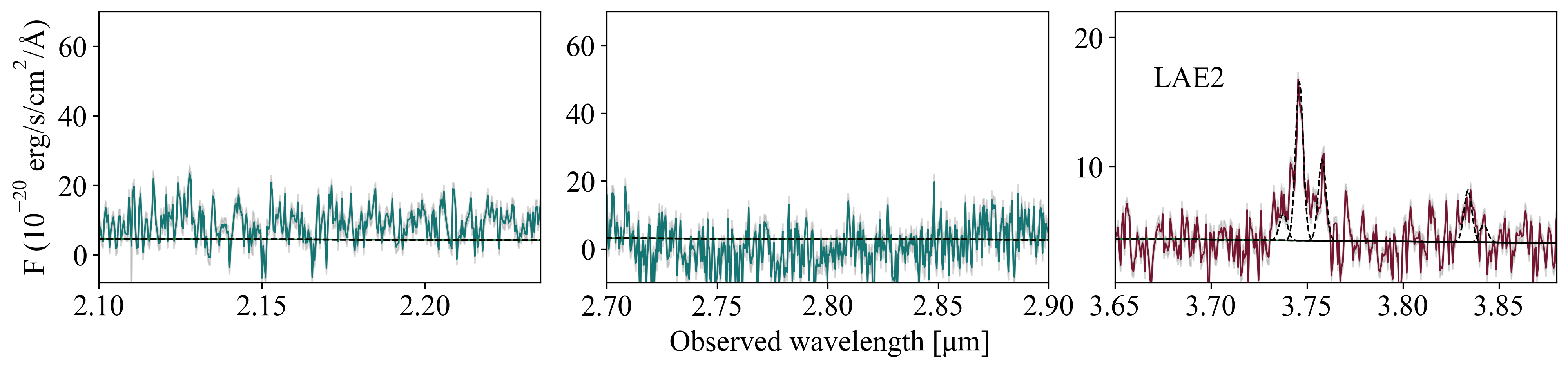}
\caption{JWST NIRSpec rest-frame optical spectra of LAE1 and LAE2 in the two gratings/filters used (G235H/F170LP and G395H/F290LP, blue and red lines, respectively), extracted from apertures of $R$ = 0.6\arcsec and 0.3\arcsec, respectively. The errors in the spectra are shown in light gray and the kinematic components used to model the emission lines in LAE1 are marked with different curve styles and colors (the two main narrow components are shown with a dashed black and a dotted purple line, respectively, while the broad component is marked shown with a dashed gray line).
The spikes that appear in the H$\beta$ emission line of LAE1 are instrumental artifacts, therefore the measurements of this emission line in LAE1 are not completely reliable.}
\label{fig:lae_spectra}
\end{figure*}

As BR1202-0725 is one of the most overdense fields known in the early Universe, we exploited the NIRSpec IFS data to study the properties of its environment. The two Lyman-$\alpha$ emitters, LAE1 and LAE2, are detected with a high level of significance in both spectral configurations of the NIRSpec data in the integrated spectra extracted from apertures of $R$ = 0.6 arcsec and 0.3 arcsec for the two sources, respectively (see Fig.~\ref{fig:lae_spectra}). 

The optical lines of LAE1 reveal a double-horned profile that might suggest the presence of a rotating disk. However, the analysis of the channel maps rather indicates a scenario of two independent galaxies that partially overlap along the line of sight. Taking into account the relative positions and velocities of these two galaxies, they could be an on-going merging system (Fig.~\ref{fig:fmapLae1}). The two sources, hereafter dubbed LAE1a and LAE1b, are separated by a projected distance of about $\sim$2.8 kpc, which is larger than the typical galaxy size at similar redshift and mass \citep[e.g., 1.5--2 kpc;][]{Ormerod2024}. The relative velocity between LAE1a and LAE1b is $\simeq$590~\kms and their emission lines have similar velocity dispersions of 160~\kms and 167~\kms, respectively. Modeling the spectral shape of the line requires an additional broad component ($\sigma$ = 570 $\pm$ 20 \kms), which may originate from tidal interactions or outflows.

The other known companion galaxy in the system, LAE2, is detected only in \ha\ and \sii, whose redshift is consistent, within the errors, with those determined from Ly$\alpha$ and \cii. 
The \ha emission map features a south-north elongation, but unfortunately, the sensitivity of current observations is not sufficient for a spatially resolved kinematic analysis of the \ha emission of the galaxy.

\begin{figure}
\centering
\includegraphics[width=0.8\linewidth]{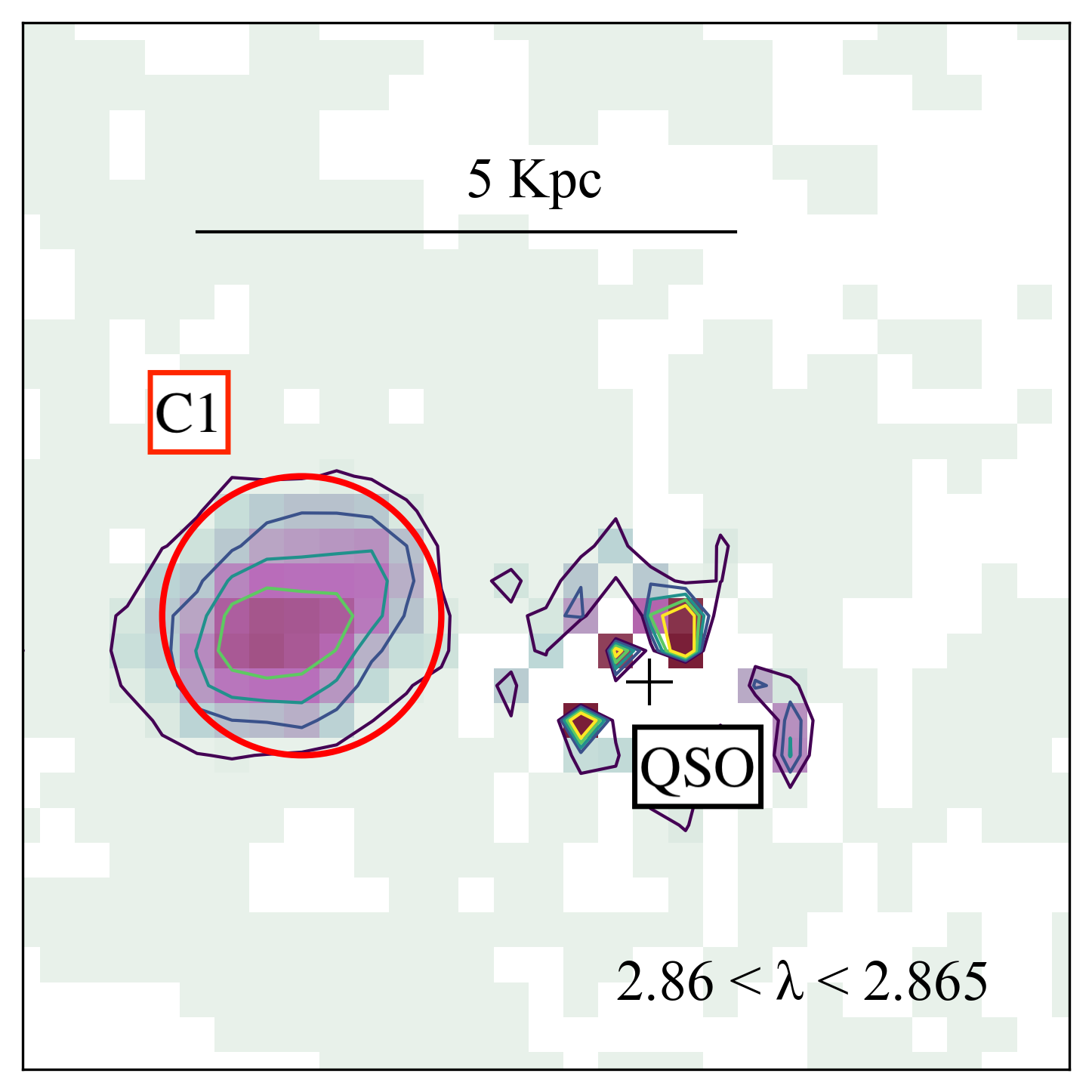}
\includegraphics[width=0.8\linewidth]{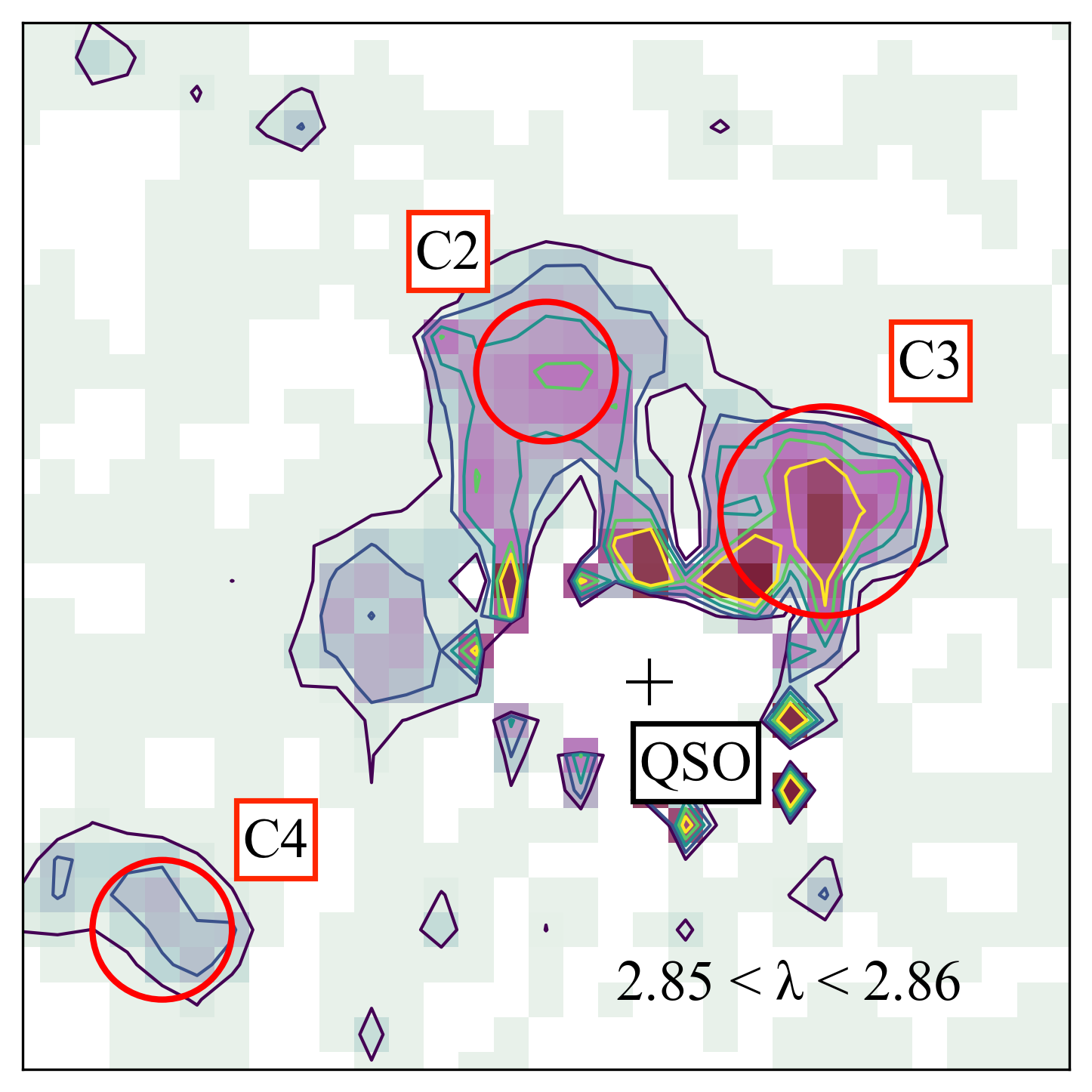}
\caption{JWST NIRSpec PSF-subtracted maps integrated over the two wavelength ranges reported in each panel, both encompassing \oiii$\lambda$5007 \AA\ emission in the same FoV. The QSO position is marked with a plus symbol. The companions found near the QSO are highlighted with red circles, which represent the apertures used to extract their integrated spectra (reported in Fig. \ref{fig:comp_spectra}).}
\label{fig:comp3}
\end{figure}

\begin{figure*}
\includegraphics[width=\textwidth]{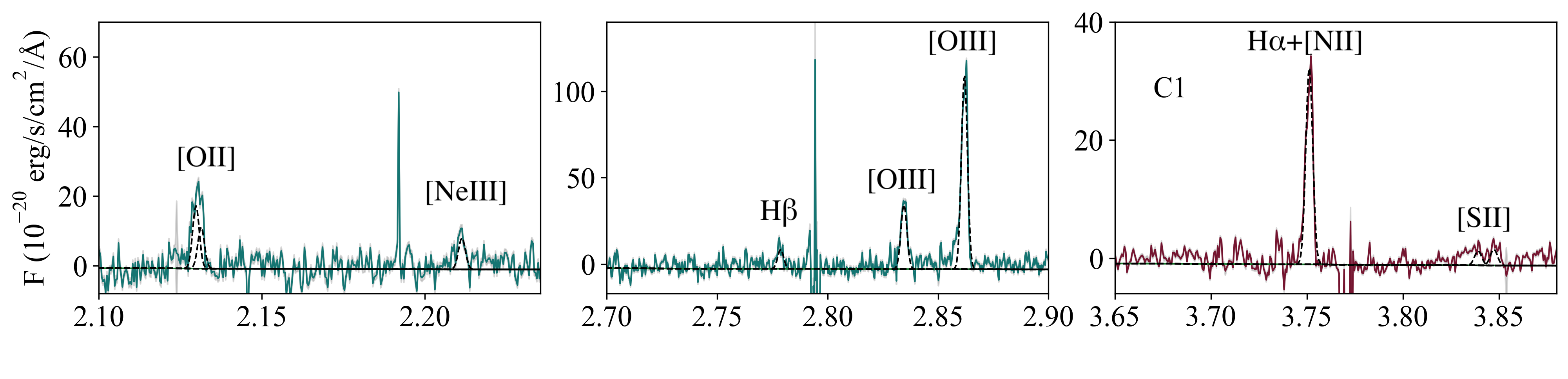}
\includegraphics[width=\textwidth]{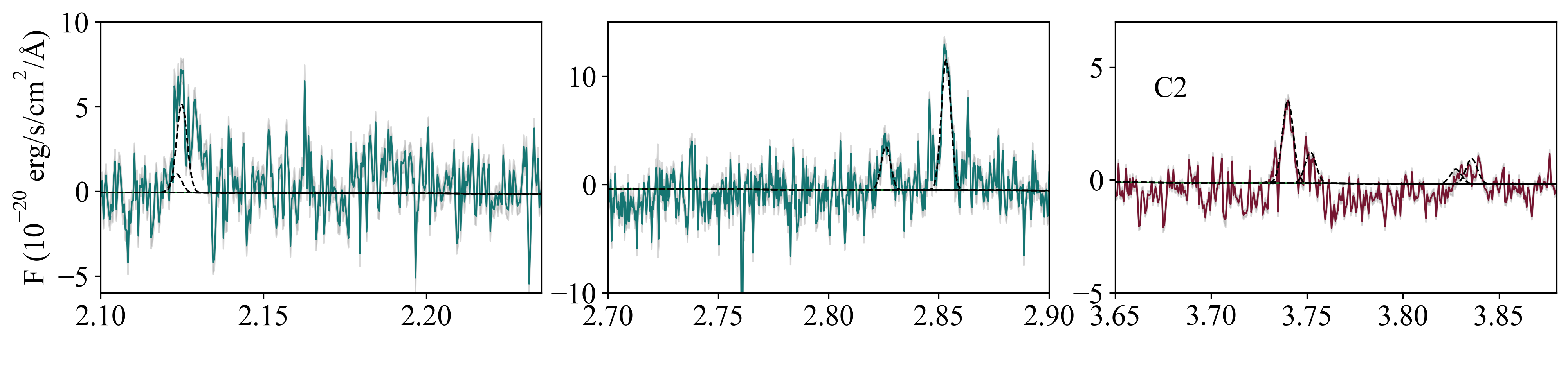}
\includegraphics[width=\textwidth]{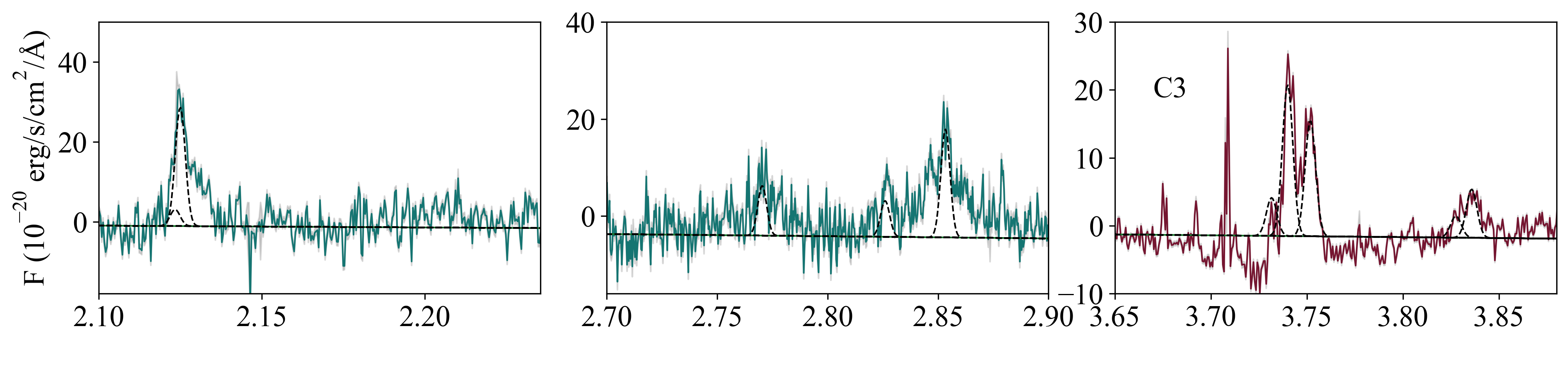}
\includegraphics[width=\textwidth]{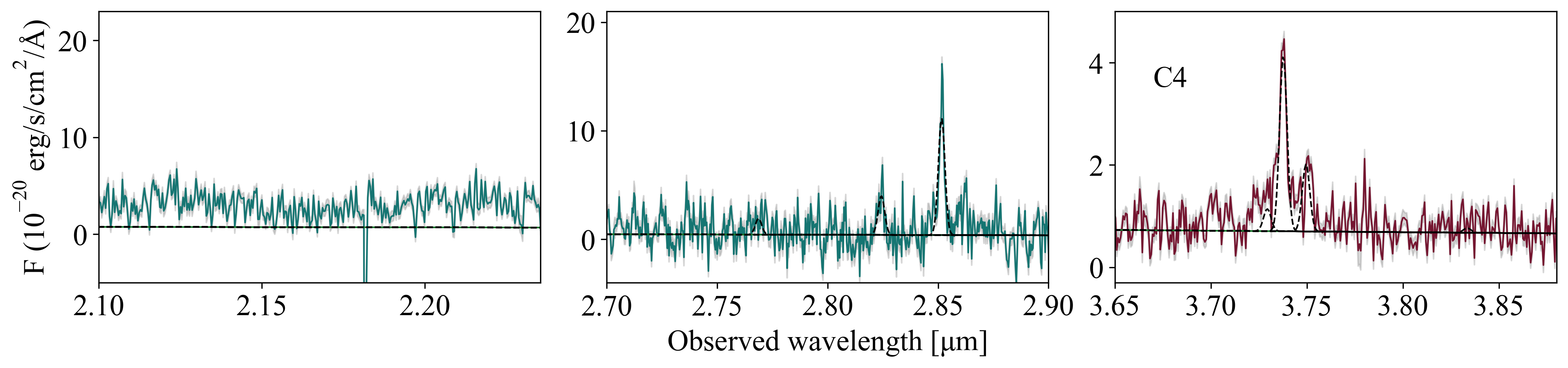}
\caption{JWST NIRSpec rest-frame optical spectra of the four detected companions, C1, C2, C3, and C4, extracted from apertures of $R$ = 0.2\arcsec, 0.1\arcsec, 0.15\arcsec, and 0.1\arcsec, respectively. The errors on the spectrum are shown in light gray. 
%\textbf{The main emission lines in each spectrum are: \oii$\lambda$3727,29 \AA\ and  \neiii$\lambda$3869 \AA (left panel); H$\beta$, and \oiii$\lambda$4959,5007 \AA\ (middle panel); and \nii$\lambda$6548 \AA, H$\alpha$, \nii$\lambda$6584, and \sii$\lambda$6716,31 \AA\ (right panel).} 
The fitting results are shown with a dashed black line.}
\label{fig:comp_spectra}
\end{figure*}

In addition to LAE1 and LAE2, we robustly identified four other companion galaxies (C1, C2, C3, C4) within 1\arcsec\ of the QSO by performing a blind line search in the PSF-subtracted data cube. The two panels of Fig. \ref{fig:comp3} show the emission maps obtained by integrating the PSF-subtracted data-cube over two wavelength ranges including \oiii$\lambda$5007 \AA\ line emission at different spectral shifts, 2.86 -- 2.865 $\mu$m and 2.85 -- 2.86 $\mu$m, respectively. The two wavelength ranges correspond to the spans of velocity 1080~--~1600~\kms and 30~--~1080~\kms\ with respect to the \oiii\ wavelength redshifted to $z=4.6943$.
One of the companions (C1) is very bright in \oiii and is located at around 3.25 kpc to the east of the QSO  with an apparent elongation to the north-west, while the two other bright sources (C2 and C3) are within a radius of $\sim$2.3 kpc from the QSO. The last one (C4) is located to the south-west of the QSO. All the sources are detected in both \oiii and \ha and have redshifts within 1200~\kms from the systemic velocity of the QSO. 
We extracted the integrated spectra of the new companion galaxies, C1, C2, C3, and C4, from the apertures shown in Fig. \ref{fig:comp3} (with radii of 0.2, 0.1, 0.15, and 0.1 arcsec for C1, C2, C3, and C4, respectively) and we report them in Fig. \ref{fig:comp_spectra}. 
A single Gaussian component per line was sufficient to fit the emission lines. The emission line observed fluxes and the main emission line ratios of these newly found sources are reported in the upper part of Table \ref{tab:companions_fluxes}, together with those of the LAEs. We stress that the errors on the Guassian parameters might be larger due to the uncertainties introduced by the PSF subtraction.
LAE3 falls outside the FoV of the NIRSpec IFS observations (Fig. \ref{fig:aper_ha}, left), therefore we cannot constrain its emission-line properties.

In conclusion, the BR1202-0725 system includes at least ten galaxies, namely the QSO, the SMG, and the companions LAE1a, LAE1b, LAE2, LAE3, C1, C2, C3, and C4, in a projected area of 980 kpc$^2$ and within a relative velocity range of $\sim1200$~\kms. A similarly large number of companions has also been recently found in SPT0311-58 at $z$ $\sim$ 6.9  thanks to NIRSpec IFU observations \citep{arribas2024}. Interestingly, this system shares several similar structural and kinematic properties with BR1202-0725. In particular, it is also formed by two main (dusty) galaxies and about ten nearby companions with relative radial velocities, and projected distances similar to what we found in BR1202-0725. Table \ref{tab:Comp} shows the coordinates, redshifts, velocity shifts and projected distances of all the galaxies found in this field.
All the spectra of the companions show H$\alpha$, H$\beta$, \oiii$\lambda \lambda$4959,5007 \AA, and \oii$\lambda \lambda$3727,29 \AA\ emission lines, except for LAE2 in which only \ha and \sii are detected, while LAE1 and C1 also show \neiii$\lambda$3869 \AA\ (see Figs. \ref{fig:lae_spectra} and \ref{fig:comp_spectra}). 
The emission line fluxes and commonly used line ratios are reported in Table~\ref{tab:companions_fluxes}.
The blind line search analysis also revealed the presence of four other candidate emitting sources in the NIRSpec FoV (see Fig. \ref{fig:map_candidates}), but the significance of the detection is too low to confirm the origin of the signals and to determine their redshifts. The spectra and coordinates of these candidate sources are reported in Fig.~\ref{fig:comp} and Table~\ref{tab:comp}, respectively (see Appendix \ref{ap:comp_spec}).

\begin{table*}
\caption{Coordinates, redshifts, projected distances from the QSO, velocities, and velocity dispersions of the companion galaxies around the QSO of BR1202-0725.}
\centering
\begin{tabular}{ccccccc}
\hline \hline
Name & \begin{tabular}[c]{@{}c@{}}RA$^a$\\ {[}deg{]}\end{tabular}
& \begin{tabular}[c]{@{}c@{}}Dec$^a$\\ {[}deg{]}\end{tabular}
& $z$  & 
\begin{tabular}[c]{@{}c@{}}Dist. QSO\\ {[}kpc{]}\end{tabular} & \begin{tabular}[c]{@{}c@{}}$\Delta v^b$\\ {[}\kms{]}\end{tabular} & \begin{tabular}[c]{@{}c@{}}$\sigma$\\ {[}\kms{]}\end{tabular} \\ \hline \hline
LAE1a & 181.34598 & --7.70853 & 4.7007  & 14.37 &  +339 $\pm$ 6 & 160 $\pm$ 7\\
LAE1b & 181.34598 & --7.70853 &  4.7120 &  19.00 & +931 $\pm$ 4 & 167	$\pm$ 4\\
LAE2 & 181.34602 & --7.70962 &  4.7066 & 14.87 & +650 $\pm$ 10 & 160 $\pm$ 10\\
C1   & 181.34654 & --7.70908 & 4.7145 & 3.28 & +1062 $\pm$ 8 & 143 $\pm$ 8 \\ 
C2   & 181.34644 & --7.70894 & 4.6973 & 3.04 & +160 $\pm$ 10 & 230 $\pm$ 10 \\ 
C3   & 181.34633 & --7.70902 & 4.6973 & 2.52 & +158 $\pm$ 3 &
226 $\pm$ 7 \\ 
C4   &  181.34659 & --7.70920  & 4.6939 & 5.07 & --23 $\pm$ 7 &
156 $\pm$ 7\\ 

 \hline   
\end{tabular}

\begin{tablenotes}
\centering
\item $^a$ Aperture position.
\item $^b$ Relative to $z$ = 4.6943 \citep{carniani2013}.\\
\end{tablenotes}
\label{tab:Comp}  
\end{table*}

\begin{table*}
\caption{Emission line observed fluxes (in units of ${\rm 10^{-17}~erg~s^{-1}~cm^{-2}}$), emission line ratios, and main properties (ionization parameter $U$, gas-phase metallicity, in terms of oxygen abundance and relative to solar, and dynamical masses) of the LAEs and confirmed companions, from their integrated spectra reported in Figs.~\ref{fig:lae_spectra} and \ref{fig:comp_spectra}.}
\small
\centering
\begin{tabular}{lccccccc}
\hline \hline
&  LAE1a & LAE1b & LAE2 & C1& C2& C3& C4		\\	
\hline \hline
H$\alpha$& 15.8	$\pm$	1.3 & 27.0	$\pm$	0.9 & 4.9	$\pm$ 0.5 &	12.0	$\pm$	1.1	&	2.1	$\pm$	0.2	&	12.3	$\pm$	0.7	&	1.35	$\pm$	0.10	\\

H$\beta$	&	16.4	$\pm$	1.9$^a$ & 21	$\pm$	3$^a$ & <0.9 & 4	$\pm$	2	&			< 0.4	&	5.7	$\pm$	1.3	&			< 0.3	\\

\neiii$\lambda$3869 \AA & 7.5	$\pm$	1.4 &	7.7	$\pm$	1.6 & <0.8 & 3.2	$\pm$	1.9 & <1.1 & <2 & <0.5 \\

\nii$\lambda$6584 \AA	& <0.4 &4.14	$\pm$	0.15 & 2.54	$\pm$	0.11 &			< 0.7	&	0.78	$\pm$	0.14	&	9.4	$\pm$	0.6	&	0.51	$\pm$	0.06	\\

%\nii$\lambda$6548 \AA	&	- &1365	$\pm$	148 & 837 	$\pm$	108 &		< 726	&	36	$\pm$	164	&	3115	$\pm$	192	&	170	$\pm$	21	\\

\oii$\lambda$3727,29 \AA	&	36	$\pm$	2 & 44	$\pm$	2 & <1.6 & 11	$\pm$	2	&	3.6	$\pm$	0.7	&	19	$\pm$	2	&	<1.0			\\

\oiii$\lambda$5007 \AA &	79	$\pm$	4 & 70	$\pm$	2 & <0.9 & 40	$\pm$	3	&	6.8	$\pm$	0.6	&	12.5	$\pm$	1.2	&	4.2	$\pm$	0.3	\\

%\oiii$\lambda$4959 \AA &	26212	$\pm$	1479 & 23455	$\pm$	781 & - & 13370	$\pm$	1048	&	2279	$\pm$	194	&	4153	$\pm$	406	&	1407	$\pm$	103	\\

%[SII]$\lambda$6717 \AA	&	3034	$\pm$	395 & 6651	$\pm$	1923 & 1612	$\pm$	291 & 872	$\pm$	523	&	367	$\pm$	140	&	1737	$\pm$	347	&			< 95	\\

%[SII]$\lambda$6731 \AA	&	< 3184 & 3306	$\pm$	393 & 542	$\pm$	252 & 973	$\pm$	529	&	638	$\pm$	115	&	4011	$\pm$	374	&			< 95	\\
\hline

H$\alpha$/H$\beta$	&	- &- &	- & 3.0	$\pm$	1.6	&	-		&	2.2	$\pm$	0.5	&	-			\\

log(\nii/H$\alpha$)	&	< --1.6 &--0.82	$\pm$	0.02 & --0.29	$\pm$	0.05 & < --1.2			& --0.43 $\pm$ 0.09			&	--0.12	$\pm$	0.04	&	--0.42	$\pm$	0.06	\\

log(\oiii/H$\beta$)	&	0.68	$\pm$	0.06 & 0.53	$\pm$	0.07 & - & 1.0	$\pm$	0.2	&	> 1.3		&	0.34	$\pm$	0.11	&	> 1.2			\\

%log([SII]/H$\alpha$)	&	- &--0.434	$\pm$	0.126 & --0.359	$\pm$	0.090 & --0.813	$\pm$	0.264	&	--0.318	$\pm$	0.171	&	--0.331	$\pm$	0.090	&	< --0.849			\\

log(\oiii/\oii)	& 0.34	$\pm$	0.04 & 0.21 $\pm$ 0.03 &	- & 0.57	$\pm$	0.11	&	0.28	$\pm$	0.10	&	--0.18	$\pm$	0.07	&	-			\\

% \nii/\oii	&	- &0.386	$\pm$	0.090 & - & -			&	0.216	$\pm$	0.060	&	0.503	$\pm$	0.067	&	-			\\

%[SII]$\lambda$6717[SII]$\lambda$6731	&	- &2.012	$\pm$	0.582 & 2.972 	$\pm$	0.537 & 0.896	$\pm$	0.538	&	0.575	$\pm$	0.220	&	0.433	$\pm$	0.086	&	-			\\
\hline
$\log(U)$	&	--2.75 $\pm$ 0.03 & --2.86 $\pm$ 0.02 & - & --2.56 $\pm$ 0.09 &	--2.80  $\pm$ 0.08	&	--3.16 $\pm$ 0.05	& - \\
12+log(O/H)	& 8.19$_{-0.22}^{+0.17}$ & 8.35$_{-0.14}^{+0.12}$ &8.77$_{-0.19}^{+0.17}$ &8.08$_{-0.25}^{+0.19}$ &8.47$_{-0.16}^{+0.13}$ & 8.58$_{-0.07}^{+0.06}$ & 8.77$_{-0.22}^{+0.17}$\\
$Z/Z_\odot$$^b$	& 0.31$_{-0.12}^{+0.15}$ & 0.45$_{-0.12}^{+0.15}$ & 1.2$_{-0.4}^{+0.6}$ & 0.25$_{-0.11}^{+0.14}$ & 0.60$_{-0.18}^{+0.21}$ & 0.77$_{-0.11}^{+0.12}$ & 1.2$_{-0.5}^{+0.6}$\\
%$M_\mathrm{dyn}$	&  &  &  &  &  &  &  \\
\hline   
\end{tabular}
\begin{tablenotes}
\centering
\item $^a$ These measurements should be taken with caution due to the presence of spikes at the wavelengths of \hb in the spectrum of LAE1 (see Fig.~\ref{fig:lae_spectra}); for this reason, we have not used the ratios involving this line to estimate the metallicity of LAE1a and LAE1b.
\item $^b$ Assuming a solar photospheric value of 12+log(O/H)$_\odot$ = 8.69 \citep{asplund2009}.
\end{tablenotes}
\label{tab:companions_fluxes}  
\end{table*}

\section{Discussion}\label{sec:discussion}

\subsection{Ionization nature of the sources in BR1202-0725}
\begin{figure*}[h!]
\centering
\includegraphics[width=0.8\textwidth]{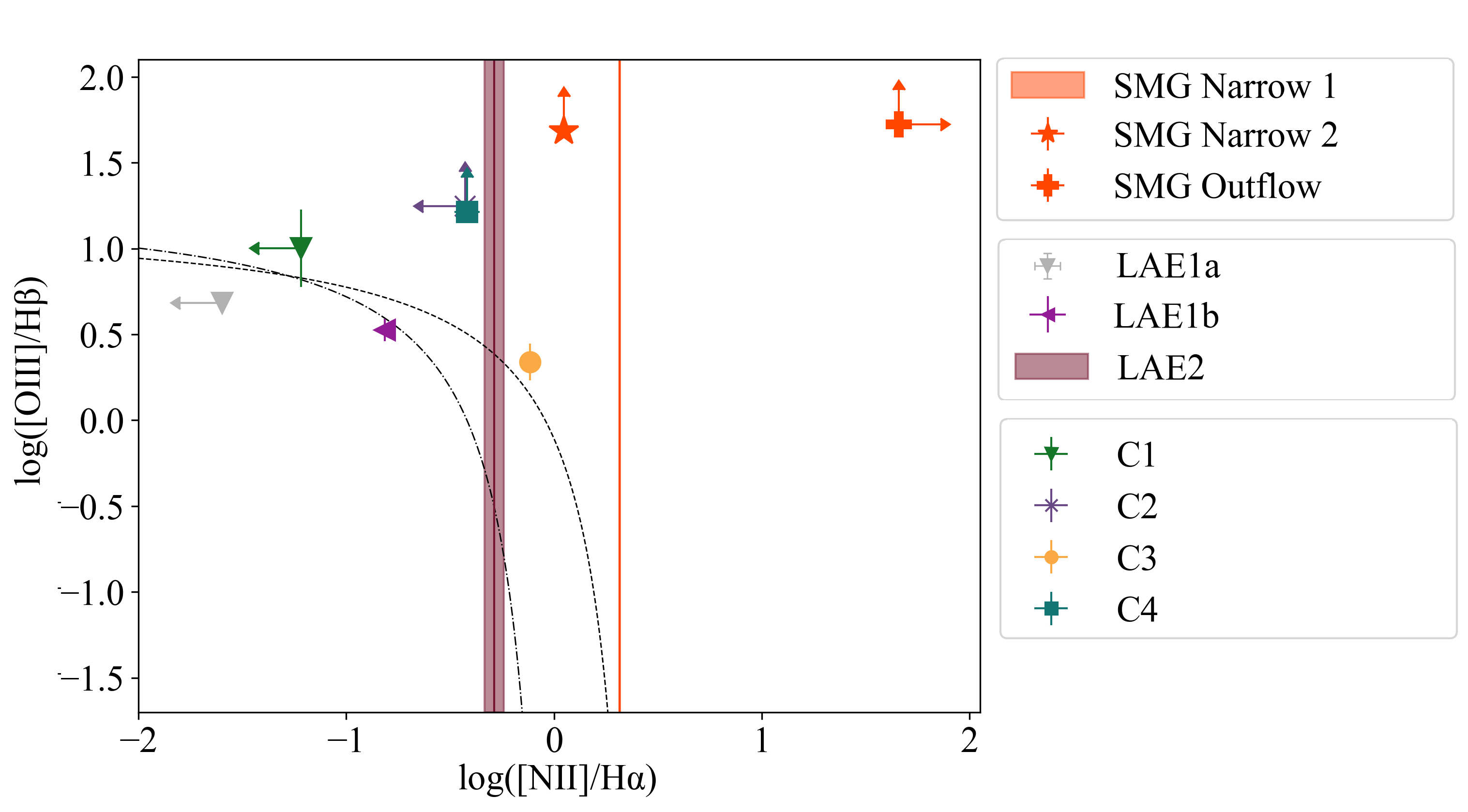}
\caption{The \oiii/H$\beta$ vs \nii/H$\alpha$ BPT diagnostic diagram for all objects analyzed. The demarcation between H\,\textsc{ii} and AGN-ionized regions \citep[][dash-dotted and dashed black lines, respectively]{kauffmann2003,kewley2001} are reported. The vertical lines indicate the \nii/\ha ratio for the two sources in which we have not values of \oiii and \hb emission lines, with the errors showed with shaded bands (the errors are not visible for the \textit{SMG Narrow 1} component).}
\label{fig:bpt}
\end{figure*}

To investigate the mechanism of ionization of the various sources in the BR1202-0725 system, we made use of the so-called BPT diagnostic diagram \citep{bpt}, which exploits the \oiii$\lambda$5007 \AA\ / \hb and \nii$\lambda$6584 \AA\ / H$\alpha$ line ratios to distinguish between star-forming regions and those dominated by AGN radiation, shocks, or non-thermal activity.
We report the BPT diagram for the galaxies analyzed in this study in Fig. \ref{fig:bpt}.

For the component associated with the gas at rest in the SMG (SMG Narrow 1, vertical orange line), there is no detection of the \oiii$\lambda$5007 \AA\ or H$\beta$ emission lines (see Fig. \ref{fig:sms_fit}), but both best-fit results of Scenario 1 and Scenario 2 return a log(\nii/H$\alpha$) ratio of 0.32 (Table \ref{tab:SMG}), not consistent with star formation processes.
The two other components of the SMG (\textit{SMG Narrow 2} and \textit{SMG Outflow} in Scenario 1) have extreme \oiii/\hb and \nii/\ha, fully consistent with the gas being AGN photoionized, rather than SF-ionized. 
Then, the location of the SMG in the BPT together with both scenarios proposed to reproduce the observed integrated spectrum (a BLR or just an outflow with velocity shift of $\sim$ --2500~\kms), indicate that the SMG hosts an accreting supermassive black hole. This supports a dual AGN scenario in this complex system at $z$ $\sim$ 4.7.

The LAE1a and LAE1b and the companion C1 exhibit \nii/H$\alpha$ ratios consistent with star formation processes if they were in the local Universe. However, at high $z$ the traditional low-$z$ BPT becomes insensitive to the mechanism of ionization in this part of the diagram, where both low-metallicity star-forming galaxies and AGN accumulate \citep[see e.g.,][]{feltre2016, uebler2023, maiolino2023, hirschmann2023, cameron2023, scholtz2023}.
Therefore, the nature of LAE1a, LAE1b, and C1 is uncertain.
For the LAE2 there is no detection of either \oiii $\lambda$5007 \AA\ or H$\beta$ emission lines, then it might be ionized by either AGN or star formation processes. The nature of the C3 companion is also uncertain because it lies close to the demarcation line between AGN- and star formation-dominated ionization in the BPT diagram. Finally, C2 and C4 are the companions which are mostly consistent with being ionized by the QSO photons.

We  calculated the ionization parameter ($U$) for the companion galaxies in the system using the \oii$\lambda \lambda$3727,29 \AA\ / \oiii$\lambda $5007 \AA\ ratio \citep{diaz2000}.
We obtained values ranging from --3.16 to --2.56 in the logarithmic scale (see Table \ref{tab:companions_fluxes}). 
%\textbf{and it is important to emphasize that these values represent only upper limits for the ionization parameter, as the spectra do not have sufficient S/N to estimate the extinction of these sources.}

Assuming that the ionization of the companions is due to star formation and not AGN photons (though this cannot be unambiguously assessed based on the BPT diagram; see Fig. \ref{fig:bpt}), we estimated their metallicities by using the following strong-line diagnostics:
R2 = $\rm \oii\lambda\lambda3727,29\AA / \hb$,
R3 = $\rm \oiii\lambda5007\AA / \hb$,
O3O2 = $\rm \oiii\lambda5007\AA / \oii\lambda\lambda3727,29 \AA$,
$\rm \hat{R}$ = $\rm 0.47 \, R2 + 0.88 \, R3$,
N2 = $\rm \nii\lambda 6584\AA / H\alpha$, 
O3N2 = $\rm (\oiii\lambda 5007 \AA /H\beta) / (\nii\lambda 6584\AA / H\alpha)$.
We adopted the recently introduced diagnostic ratio $\rm \hat{R}$ \citep{Laseter2024} in place of the traditional $\rm R23 = (\oii\lambda\lambda3727,29 \AA + \oiii\lambda\lambda4958,5007\AA) / \hb$, since the former is more suited for high-$z$ galaxies. 
For C1, the only source for which we found a non-zero extinction from \ha/hb ($A_V$ = 0.16 $\pm$ 0.08), we corrected the line fluxes for extinction, assuming a \cite{Calzetti2000} reddening curve with $R_V$ = 4.05.
To obtain the gas metallicity from the above ratios, we adopted the best-fit polynomial calibrations from \cite{Curti2017,Curti2020}, slightly revisited in \cite{Curti2023,Curti2024} to better probe the low-O/H regime at high $z$. 
Assuming a solar photospheric value 12+log(O/H)$_\odot$ = 8.69 \citep[][]{asplund2009}, most of the sources have values consistent with $Z/Z_\odot$ $\sim$ 0.5, while LAE1a and C1 have lower values, of $\sim$ 0.3 solar, and only C3 has a higher value with $Z/Z_\odot$ = 0.8 (see Table \ref{tab:companions_fluxes}).

Still under the assumption that gas excitation driving line emission is due to star formation and not to the QSO, we find that each of the companions has a SFR higher than $\sim$ 1 $M_\odot$~yr$^{-1}$ (using the \ha-to-SFR relation from \citealt{Kennicutt2012}).

We stress again, however, that these estimates should be taken with great caution, given that a contribution to the gas ionization from the bright QSO may affect the line emission fluxes and their ratios. This is especially the case of C2 and C4, whose line ratios are more clearly in the AGN region of the BPT (Fig.~\ref{fig:bpt}).

\subsection{X-ray detection of the SMG}\label{sec:chandra}
To further probe the AGN nature of the SMG, we investigated its X-ray properties with archival data by \textit{Chandra}, the only X-ray facility that can spatially resolve the QSO and the SMG of the system. The X-ray emission of the QSO is detected in both observations, while only \citet{iono2006} addressed the X-ray emission of the SMG, presenting a tentative detection (below the 3$\sigma$ level).

\begin{figure}
\centering
\includegraphics[width=0.45\textwidth]{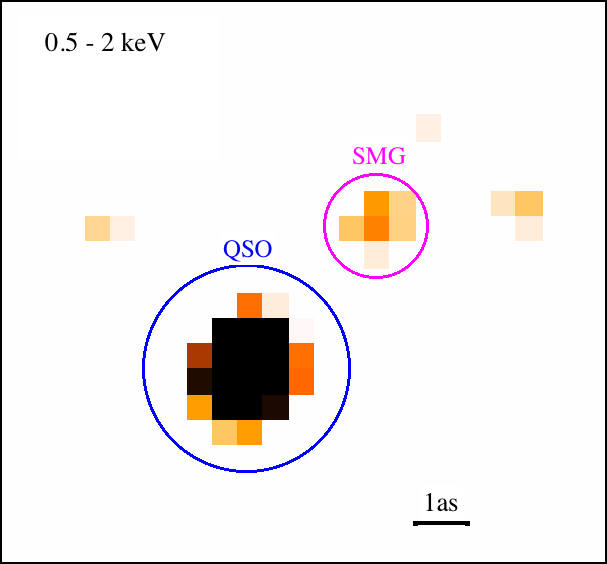}
\caption{Soft-band (0.5--2 keV) Chandra image of BR1202-0725 (10\arcsec$\times$10\arcsec\ box). Data were smoothed with a Gaussian kernel of 2-pixel radius. Circular regions mark the positions of the QSO (blue) and of the SMG (magenta).}
\label{fig:chandra}
\end{figure}

We reduced both archival observations (see Sect.~\ref{sec:observations}) with the {\tt chandra\_repro} tool of CIAO v.4.16 \citep{fruscione2006_ciao} and merged the two datasets with the {\tt merge\_obs} script of CIAO, to maximize the chances of detecting the X-ray emission from the SMG. 
A soft-band (0.5--2 keV) image (smoothed with a 2-pixel Gaussian kernel) from the resulting data is shown in Fig.~\ref{fig:chandra}.
With a standard circular extraction region of 1$\arcsec$ radius, we measured 3 counts at the position of the SMG in the soft band (0.5--2 keV observed-frame) and no counts in the hard band (2--7 keV observed-frame), which is not surprising given the drop of \textit{Chandra}'s effective area above 2 keV. With a background level of 0.08 counts, we measured 2.9$\pm$1.7 net counts in the soft band associated with the SMG, corresponding to the $\simeq$2.8--11.4 keV rest-frame energy range. As a comparison, we detected 33.9$\pm$5.8 net counts in the same energy band in a circular aperture of 1$\arcsec$ radius centered on the QSO. 
We follow the approach of \citet{vito2019} to test the significance of the measured X-ray emission at the position of the SMG (i.e., 3 total counts in the soft band). First, we estimated the probability of this being a spurious detection, that is, due to background fluctuations, applying Eq. A21 of \citet{weisskopf2007}. We found a false-detection probability of $P=8.9\times10^{-5}$, making this a detection at $\gtrsim4\sigma$. To double-check this result, we measured the probability of detecting 3 total counts in the soft band by chance by performing aperture photometry in regions of 1$\arcsec$ radius randomly distributed in the field over $10^{6}$ positions. Considering the full merged field, only 24 regions out of the $10^6$ considered present more than 3 total counts with a false-detection probability of $P=7.4\times10^{-5}$. This result could be due to the variation of exposure time from the central region, where the two observations overlap, to the outskirts of each separate field, combined to a larger \textit{Chandra} PSF moving off-axis. We thus repeated the exercise restricting the sampled area to the part of the field that was exposed in both observations, finding a false-detection probability of $P=1.5\times10^{-4}$ and thus a significance of the X-ray emission at the position of the SMG that is slightly less than 4$\sigma$. 

Given this tentative 4$\sigma$ detection, we measured the X-ray flux of the SMG in the soft band and estimated its intrinsic X-ray luminosity. Assuming a photon index $\Gamma=1.9$ (in agreement with typical values at these redshifts; \citealp[e.g.,][]{vignali2005,just2007}) and converting the measured count rate with the PIMMS tool v.4.12d\footnote{\url{https://cxc.harvard.edu/toolkit/pimms.jsp}.}, we found a soft-band observed flux of $f_{0.5-2\,\rm keV}\simeq7.8\times10^{-16} \rm~erg~s^{-1}~cm^{-2} $. Assuming the same photon index, we converted this flux to intrinsic 2--10 keV luminosity  in the two extreme cases of no absorption ($N_{\rm H}<10^{20}\rm~cm^{-2}$) and Compton-thick ($N_{\rm H}=
10^{24}\rm~cm^{-2}$) column densities, which return  $L_{\rm 2-10\,keV}=2.1-17\times10^{44}\rm~erg~s^{-1}$, respectively. The estimated X-ray luminosity is higher than $2\times10^{44}\rm~erg~s^{-1}$ at any absorption level and, as such, cannot be explained by SF processes only: previous works estimate a SFR in the range 1000--5000 $M_{\odot}$~yr$^{-1}$ \citep[see][]{carniani2013}, which corresponds to an X-ray luminosity of $~3\times10^{43}\rm ~erg~s^{-1}$ in the 0.5--8 keV rest-frame band, based on the relation of \citet{mineo2014}. This is less than half the 0.5--8 keV lower limit we get from the estimated X-ray luminosity ($L_{\rm 0.5-8\,keV}\simeq3\times10^{44}\rm~ erg~s^{-1}$, assuming no absorption). This tentative X-ray detection thus supports the scenario of the SMG harboring an AGN. 

\subsection{Black hole properties}\label{sec:bh_props}

We determined the properties of the black hole of the QSO and the SMG from the best-fit results of the spectral fitting. We report them in Table \ref{tab:AGNprop}. For the SMG, we  used the results from Scenario 2, where we  assumed that the broad component of the \ha emission line is associated with the BLR.

Initially, we estimated the impact of dust attenuation on the observed flux of the optical lines by measuring the Balmer decrement.   We  measured a (\ha/\hb)$_{\rm BLR}$ ratio of 2.36 ($\pm$ 0.14) for the BLR component of the QSO. The theoretical ratio in low-density gas is 2.87 assuming Case-B recombination and an electron temperature of $T_{\rm e} \sim 10^{4}$~K \citep{Osterbrock2006}, while it spans a range between 3 and 10 in the dense BLR gas where the optical line depths and collisional effects are different from those in H\,\textsc{ii} regions \citep{dong2008, baron2016}. The ratio measured in the QSO therefore suggests that a dust extinction correction is not necessary.

We used the \ha\ luminosity to determine the bolometric luminosity of the AGN hosted in the SMG by using the relation from \cite{stern2012}. This gives a $\log(L_{\rm bol, H\alpha}/[{\rm erg~s^{-1}}])=47.2\pm0.4$ for the QSO, which is consistent with the bolometric luminosity calculated from the monochromatic continuum in UV at $\lambda$1350 \AA, of $47.32 \pm 0.02$ \citep{jun2015} and $47.929 \pm0.029$ \citep{yu2021}. We also determined the bolometric luminosity from the monochromatic luminosity at the rest-frame wavelength of 5100~\AA\ ($\lambda L_{5100}$) following \cite{netzer2019}:
\begin{equation}
    \begin{cases}
    \log(L_{\rm bol, 5100}) = \log(\lambda L_{5100})+\log(k_{\rm bol})\\
    \log(k_{\rm bol}) = 10-0.2\times\log(\lambda L_{5100})
    \end{cases}\,
\end{equation}
We infer $\log(L_{\rm bol,5100}/[{\rm erg~s^{-1}}])=47.52\pm0.02$, in agreement with the other measurements.

The properties of the \ha\ emission line can also be used to estimate the BH mass ($M_{\rm BH}$) assuming the gas in the BLR is virialized. In the local Universe, \citet{reines2013} infer  

\begin{equation}
\begin{split}
\log\left(\frac{M_{\rm BH}}{M_\odot}\right) \approx 6.57+0.47\cdot &\log \left(\frac{L_{\rm H\alpha}}{\rm 10^{42}~erg~s^{-1}}\right)+\\
& 2.06\cdot \log\left(\frac{FWHM_{\rm H\alpha}}{\rm 10^{3}~km~s^{-1}}\right) ,
\end{split}
\end{equation}
where $L_{\rm H\alpha}$ and $FWHM_{\rm H\alpha}$ are the line luminosity and width, respectively, of the \ha emission line associated to the BLR \citep[see also][]{greene2005}. The calibration has an intrinsic scatter of 0.5 dex \citep{reines2013}. We thus measured a BH mass of the QSO of $\log(M_{\rm BH}/{M_\odot}) = 10.1\pm0.5$, which is consistent within 2$\sigma$ with the estimates derived from \mgii ($\log(M_{\rm BH}/{M_\odot}) = 9.1\pm0.3$; \citealt{carniani2013}) and \civ corrected from outflows effect ($\log(M_{\rm BH}/M_\odot)=9.65\pm0.01$; \citealt{yu2021}). It is important to emphasize that degeneracies in the fitting of the QSO spectrum (see Sec. \ref{sec:QSO}) do not affect the estimated black hole mass, since all the values obtained from the different fittings are consistent with each other within the uncertainties, which are dominated by the error on the calibration.

The excitation diagnostic diagrams and the tentative X-ray detection indicate that the SMG hosts an AGN at its center. We thus estimated its black hole mass and bolometric luminosity from the best-fit results of the BLR \ha component inferred in Scenario 2. We measured a lower limit on the Balmer decrement of  (\ha/\hb)$_{\rm BLR}>6$, which is within the range of values measured in low-redshift Seyferts and QSOs \citep{dong2008, baron2016}. However, the rest-frame UV emission is heavily obscured due to the high dust content in the interstellar medium, so the measured \ha luminosity (Table~\ref{tab:AGNprop}) is a lower limit. Therefore, the quantities derived for the BH of the SMG are lower limits as well. We thus inferred $\log(L_{\rm bol, H\alpha}/[{\rm erg~s^{-1}}])>44.9$ and  $\log(M_{\rm BH}/{M_\odot}) > 8.0$. 

If the mass of the black hole of the SMG is confirmed by future observations, BR1202-0725 would be one of the most distant systems with two BHs with similar masses ($M_{\rm BH}\ga10^8~{M_\odot}$) within a projected distance of 24 kpc \citep[see also][]{matsuoka2024,uebler2024}. Simulations of $z\sim6$ luminous QSOs predict that highly accreting supermassive BHs live in overdense fields hosting galaxies with similar stellar and BH masses \citep{barai2018, valentini2020, zana2022,dimascia2021}. For example, \cite{dimascia2021} show a simulated system similar to the BR1202-0725 one where a bright UV QSO lies at the center of the field and another luminous AGN is hosted in dusty galaxies that are obscured in the UV due to the interstellar medium dust. In these simulations, the most massive BHs of the system grow by accretion of the surrounding gas or by mergers with other BHs. Therefore, BR1202-0725 may represent the typical system where bright QSOs form and evolve, but luminous companion AGN can be obscured by the dust of their host galaxy.

\begin{table}[h]
\caption{BH properties of the QSO and SMG. The properties of the BLR of the SMG are derived from the \ha \textit{BLR} component from Scenario 2.}
\centering
\begin{tabular}{lcc }
\hline \hline
 & QSO   & SMG   \\ \hline \hline
$\log(L_{\rm H\alpha}/{\rm [erg~s^{-1}]})$ & 45.07 $\pm$ 0.02 & > 42.8 \\  
$\log(\lambda L_{\rm 5100}/{\rm [erg~s^{-1}]})$ & 46.903 $\pm$ 0.007 & -- \\  
$\log(L_{\rm bol,5100}/{\rm [erg~s^{-1}]})$ & 47.52 $\pm$ 0.02 & --\\  
$\log(L_{\rm bol,H\alpha}/{\rm [erg~s^{-1}]})$ & 47.2 $\pm$ 0.4 & > 44.9 \\  
$\log(M_{\rm BH,H\alpha}/{M_\odot})$ & 10.1 $\pm$ 0.5 & > 8.0 \\  
$\log(L_{\rm edd}/{\rm [erg~s^{-1}]})$& 48.2 $\pm$ 0.5 & > 46.1\\   
$\lambda_{\rm edd}$ & 0.1 & 0.1 \\  
\hline   
\end{tabular}
\label{tab:AGNprop}  
\end{table}

\subsection{Outflow properties in the SMG}

The spectra of the SMG show evidence for broad blueshifted component likely associated with galactic outflows (Figs. \ref{fig:sms_esp}, and \ref{fig:sms_fit}). We determined the properties of the outflowing gas from the \textit{Outflow} broad component of the \oiii line because the spectral fitting of the oxygen line is less affected by degeneracies than the H$\alpha$+\nii complex. We assumed a uniformly filled conical outflow for which the mass outflow rate of the ionized gas is defined as $\dot{M}_{\rm ion}= M_{\rm ion}v_{\rm out}/R_{\rm out}$, where $M_{\rm ion}$ is the outflow mass, $v_{\rm out}$ the outflow velocity, and  $R_{\rm out}$ the size of the outflow. 

Following \cite{cresci2015} and \cite{carniani2015}, we determined the mass of the ionized outflowing gas from the flux of the \textit{Outflow} broad component of the optical lines fixing the electron density to 500~${\rm cm^{-3}}$ and solar gas-phase metallicity to be consistent with other outflow measurements in AGN at lower redshift \citep{carniani2015, fiore2017}. We estimated the outflow velocity as $v_{\rm out}= |\Delta v_{Outflow}| + 2\sigma_{Outflow}$, where $\Delta{v_{Outflow}}$ is the velocity shift of the centroid of the \textit{Outflow} component relative to the systemic redshift of the galaxy and $\sigma_{Outflow}$ its velocity dispersion (see Table \ref{tab:SMG}). We found that the SMG is expelling more than $5\times10^6~{M_\odot}$ of ionized gas at a velocity $\sim5000$~\kms (see Table~\ref{tab:OUTprop}). Such high velocity is very high even for an extreme starburst galaxy \citep{heckman2016, perrotta2021}, which indicates that this outflow is driven by an AGN \citep[e.g.,][]{carniani2015, fiore2017, fs2019,cresci2023}. Similar velocities are observed only in AGN with bolometric luminosities higher than $10^{47}~{\rm erg~s^{-1}}$ \citep{carniani2015, zakamska2016, fiore2017, perrotta2019, perna2023b}. This further supports that the SMG hosts an accreting supermassive BH. 

We estimates the mass outflow rate assuming a size of 1 kpc for the outflow (corresponding to the spatial resolution of NIRSpec, being the outflows unresolved; \citealt{DEugenio2024a}), resulting in a mass outflow rate of $\sim25~{M_{\odot}~\rm yr^{-1}}$. The measured mass outflow rate is more than one order of magnitude lower than the SFRs inferred from the far-infrared continuum luminosity \citep[2600 $M_\odot$~yr$^{-1}$;][]{carniani2013}. Then, the mass loading factor ($\dot{M}_\mathrm{ion}$/SFR) of the wind assuming the SFR from far-infrared is $\sim$ 1\%.
This low mass loading factor suggests that the star formation process is currently dominating gas consumption.

However, in this work, we are only probing the ionized gas phase, and contributions from molecular and neutral phases is missing \citep[see][]{rupkeveilleux2013, herreracamus2019, robertsborsani2020, fluetsch2021, belli2024, baron2022, cresci2023, perna2019}. We also emphasize that the measured ionized mass outflow rate value might represent a lower limit, as it is not corrected for dust attenuation, meaning the intrinsic \oiii luminosity associated with the outflow may be higher than observed.
%However, we stress that the measured mass outflow rate values might be lower limits as they are not corrected for dust attenuation and the intrinsic \oiii luminosity associated with the outflows might be higher than that observed.

\begin{table}[h]
\caption{Outflow properties of the SMG. Outflow mass and velocity for both galaxies are inferred from the \oiii\ broad component and assuming $n_e=500$~cm$^{-3}$, solar metallicity, and outflow extension of 1 kpc.}
\centering
\begin{tabular}{lc}
\hline \hline
$\log(M_{\rm out}/{M_\odot)}$                  &  $6.67\pm0.11$ \\ 
$v_{\rm out}~{\rm [\kms]}$                     &  $5000\pm1000$    \\ 
$\log(\dot{M}_{\rm out}/{[M_\odot~\rm yr^{-1}]})$   & $1.4\pm0.2$  \\ 
$\log(\dot{E}_{\rm out}/{\rm [erg~s^{-1}])}$        & $44.5\pm0.3$ \\  
$\log(\dot{P}_{\rm out}/{\rm [g~cm~s^{-2}]})$     & $33.8\pm0.2$ \\ 

\hline \hline 
\end{tabular}
\label{tab:OUTprop}  
\end{table}

\section{Conclusions}\label{sec:conclusions}

In this study, we have presented new JWST NIRSpec IFS observations of BR1202-0725, one of the most extensively studied overdense fields at $z$ $\sim$ 4.7. The high resolution and sensitivity of these observations enabled us to investigate for the first time the ionized properties of this system in the rest-frame optical wavelength range. We analyzed the kinematic and physical properties of the main sources in the system, the QSO and the SMG, from their integrated spectra, and we searched for galaxy companions that are part of this system. The main results of this work are the following: 
\begin{itemize}

    \item The two companion galaxies identified from previous ground-based observations,  LAE1 and LAE2, are detected in the strongest optical emission lines ([O\,\textsc{iii}]+\hb and/or H$\alpha$+\nii). LAE1 is detected with a high level of significance in both NIRSpec filters, while LAE2 is detected only in \ha and \nii. The optical lines of LAE1 exhibit two spectrally and spatially isolated components (LAE1a and LAE1b), which suggest a merger scenario between two galaxies at a projected distance of $\sim$ 2.8 kpc. The line ratios measured in LAE1 indicate that the gas excitation is driven by OB stars.

    \item We robustly identify four new additional companion galaxies within 1 arcsec (i.e., $\sim$ 6.5 kpc) of the QSO. They are detected in both \oiii and H$\alpha$ emission, and their redshifts are consistent with the systemic velocity of the QSO within 1050 \kms.

    \item The spectrum of the QSO clearly reveals broad ($\sigma $ $\sim $ 4300 \kms) emission lines of \ha and \hb and \feii emission multiplets, both arising from the BLR, an accretion-disk continuum, and nebular emission lines.
    The broad \ha line emission implies a bolometric luminosity of $\log(L_{\rm bol, H\alpha}/[{\rm erg~s^{-1}}])$ = 47.2 $\pm$ 0.4, consistent with those calculated from the monochromatic continuum at 1350 \AA\ and 5100 \AA. For the QSO, we measure a black hole mass of $\log(M_{\rm BH}/{M_\odot}) = 10.1\pm0.5$, compatible within 2$\sigma$ with those estimated from \mgii and \civ.

    \item The \oiii profile of the SMG is surprisingly blueshifted by more than --700~\kms at its peak relative to the systemic redshift of the galaxy (determined from \cii) and includes two distinct components. The broadest component has a velocity and a velocity dispersion of $-2450\pm40$~\kms and $1300\pm500$~\kms, respectively, and it is likely associated with a fast galactic outflow. The other \oiii component has a velocity dispersion more consistent with the one measured from \cii and CO for the cold gas, but is shifted by $-760\pm90$~\kms. Similar spectroscopic features have also been observed in some local and intermediate-redshift QSOs dubbed ``blue outliers''. These QSOs show  \oiii emission lines blueshifted by more than 250~\kms at peak with respect to the systemic redshift of the galaxy. These profiles have been associated with spatially compact outflows in the narrow line region of AGN.  
    
    \item Based on the analysis of the broad \oiii component, we find that the outflow in the SMG is expelling $\sim 5\times10^6~{M_\odot}$ of ionized gas at a velocity $\sim5000$~\kms. Similar velocities have only been observed in AGN with high bolometric luminosity ($L_{\rm bol}>10^{46}~{\rm erg~s^{-1}}$). Despite the high velocity, the mass loading factor (ratio between the mass outflow rate and the host galaxy star formation rate) is of the order of 1 percent, suggesting that the star formation processes are currently dominating gas consumption, rather than the material ejection due to the outflow.

    \item The integrated spectrum of the SMG also reveals a complex  H$\alpha$+\nii profile with a prominent blueshifted wing that requires different components to be reproduced. On the one hand, a narrow component at the systemic redshift of the galaxy is needed, and its velocity dispersion ($\sigma$ $\sim$ 470 \kms) is consistent within the errors with that measured from the far-infrared lines. On the other hand, the broad blueshifted and asymmetric wing in the H$\alpha$+\nii profile does not yield a simple interpretation.  It can be explained by either the high-velocity outflow also observed in \oiii emission or a BLR component in \ha. In this last scenario, the broad \ha implies $\log(L_{\rm bol, H\alpha}/[{\rm erg~s^{-1}}])>44.9$ and  $\log(M_{\rm BH}/{M_\odot}) > 8.0$.
    
    \item Independently on the interpretation of the broad and asymmetric wing of the H$\alpha$+\nii emission line complex, all components have high \oiii/\hb ($>10$) and/or \nii/\ha ($>1$), which robustly indicate AGN excitation. The presence of an accreting black hole at the center of the SMG is also supported by the analysis of \textit{Chandra} X-ray observations, which yield an X-ray luminosity $>10^{44}\rm~erg~s^{-1}$ at the location of the SMG, and the extreme outflow velocities comparable to those observed only in luminous quasars. 
 \end{itemize}

     We conclude that the BR1202-0725 QSO+SMG system hosts two supermassive black holes within a projected separation of 24 kpc and represents one of the most distant dual AGN systems in the Universe. Current JWST observations also reveal that the system is composed by at least ten sources (the QSO, the SMG, four LAEs, and four galaxy companions). 
     This makes BR1202-0725 one of the most overdense fields known in the early Universe. The number of galaxies found within a few tens of kiloparsecs from the QSO of BR1202-0725 is the order of magnitude higher than the average number of companions detected around high-redshift quasars in a similar field of view.

\begin{acknowledgements}
We thank F. Vito for the useful discussion regarding the X-ray data analysis. 
%Part of this work was supported by ...
GV, SC and SZ acknowledge support from the European Union (ERC, WINGS, 101040227).

EB and GC acknowledge support of the INAF Large Grant 2022 ``The metal circle: a new sharp view of the baryon cycle up to Cosmic Dawn with the latest generation IFU facilities''.

SA, MP, and BRdP acknowledge grant PID2021-127718NB-I00 funded by the Spanish Ministry of Science and Innovation/State Agency of Research (MICIN/AEI/ 10.13039/501100011033).

AJB acknowledges funding from the ``FirstGalaxies'' Advanced Grant from the European Research Council (ERC) under the European Union’s Horizon 2020 research and innovation program (Grant agreement No. 789056).

FDE and GCJ acknowledge support by the Science and Technology Facilities Council (STFC), by the ERC through Advanced Grant 695671 ``QUENCH'', and by the UKRI Frontier Research grant RISEandFALL.

IL acknowledges support from grant PRIN-MUR 2020ACSP5K\_002 financed by European Union – Next Generation EU.

RM acknowledges support by the Science and Technology Facilities Council (STFC), from the ERC Advanced Grant 695671 ``QUENCH'', and funding from a research professorship from the Royal Society.

H\"U acknowledges support through the ERC Starting Grant 101164796 ``APEX''.

\end{acknowledgements}

\bibliographystyle{aa} % style aa.bst
\bibliography{Article}

\begin{thebibliography}{154}
\expandafter\ifx\csname natexlab\endcsname\relax\def\natexlab#1{#1}\fi

\bibitem[{{Arribas} {et~al.}(2024){Arribas}, {Perna}, {Rodr{\'\i}guez Del Pino}, {Lamperti}, {D'Eugenio}, {P{\'e}rez-Gonz{\'a}lez}, {Jones}, {Crespo G{\'o}mez}, {Curti}, {Lim}, {{\'A}lvarez-M{\'a}rquez}, {Bunker}, {Carniani}, {Charlot}, {Jakobsen}, {Maiolino}, {{\"U}bler}, {Willott}, {B{\"o}ker}, {Chevallard}, {Circosta}, {Cresci}, {Kumari}, {Parlanti}, {Scholtz}, {Venturi}, \& {Witstok}}]{arribas2024}
{Arribas}, S., {Perna}, M., {Rodr{\'\i}guez Del Pino}, B., {et~al.} 2024, \aap, 688, A146

\bibitem[{{Asplund} {et~al.}(2009){Asplund}, {Grevesse}, {Sauval}, \& {Scott}}]{asplund2009}
{Asplund}, M., {Grevesse}, N., {Sauval}, A.~J., \& {Scott}, P. 2009, \araa, 47, 481

\bibitem[{{Baldwin} {et~al.}(1981){Baldwin}, {Phillips}, \& {Terlevich}}]{bpt}
{Baldwin}, J.~A., {Phillips}, M.~M., \& {Terlevich}, R. 1981, \pasp, 93, 5

\bibitem[{{Balmaverde} {et~al.}(2016){Balmaverde}, {Marconi}, {Brusa}, {Carniani}, {Cresci}, {Lusso}, {Maiolino}, {Mannucci}, \& {Nagao}}]{balmaverde2016}
{Balmaverde}, B., {Marconi}, A., {Brusa}, M., {et~al.} 2016, \aap, 585, A148

\bibitem[{{Barai} {et~al.}(2018){Barai}, {Gallerani}, {Pallottini}, {Ferrara}, {Marconi}, {Cicone}, {Maiolino}, \& {Carniani}}]{barai2018}
{Barai}, P., {Gallerani}, S., {Pallottini}, A., {et~al.} 2018, \mnras, 473, 4003

\bibitem[{{Baron} {et~al.}(2022){Baron}, {Netzer}, {Lutz}, {Prochaska}, \& {Davies}}]{baron2022}
{Baron}, D., {Netzer}, H., {Lutz}, D., {Prochaska}, J.~X., \& {Davies}, R.~I. 2022, \mnras, 509, 4457

\bibitem[{{Baron} {et~al.}(2016){Baron}, {Stern}, {Poznanski}, \& {Netzer}}]{baron2016}
{Baron}, D., {Stern}, J., {Poznanski}, D., \& {Netzer}, H. 2016, \apj, 832, 8

\bibitem[{{Belli} {et~al.}(2024){Belli}, {Park}, {Davies}, {Mendel}, {Johnson}, {Conroy}, {Benton}, {Bugiani}, {Emami}, {Leja}, {Li}, {Maheson}, {Mathews}, {Naidu}, {Nelson}, {Tacchella}, {Terrazas}, \& {Weinberger}}]{belli2024}
{Belli}, S., {Park}, M., {Davies}, R.~L., {et~al.} 2024, \nat, 630, 54

\bibitem[{{Benford} {et~al.}(1999){Benford}, {Cox}, {Omont}, {Phillips}, \& {McMahon}}]{benford1999}
{Benford}, D.~J., {Cox}, P., {Omont}, A., {Phillips}, T.~G., \& {McMahon}, R.~G. 1999, \apjl, 518, L65

\bibitem[{{Bischetti} {et~al.}(2024){Bischetti}, {Choi}, {Fiore}, {Feruglio}, {Carniani}, {D'Odorico}, {Ba{\~n}ados}, {Chen}, {Decarli}, {Gallerani}, {Hlavacek-Larrondo}, {Lai}, {Leighly}, {Mazzucchelli}, {Perreault-Levasseur}, {Tripodi}, {Walter}, {Wang}, {Yang}, {Zanchettin}, \& {Zhu}}]{Bischetti:2024}
{Bischetti}, M., {Choi}, H., {Fiore}, F., {et~al.} 2024, \apj, 970, 9

\bibitem[{{B{\"o}ker} {et~al.}(2022){B{\"o}ker}, {Arribas}, {L{\"u}tzgendorf}, {Alves de Oliveira}, {Beck}, {Birkmann}, {Bunker}, {Charlot}, {de Marchi}, {Ferruit}, {Giardino}, {Jakobsen}, {Kumari}, {L{\'o}pez-Caniego}, {Maiolino}, {Manjavacas}, {Marston}, {Moseley}, {Muzerolle}, {Ogle}, {Pirzkal}, {Rauscher}, {Rawle}, {Rix}, {Sabbi}, {Sargent}, {Sirianni}, {te Plate}, {Valenti}, {Willott}, \& {Zeidler}}]{boker2022}
{B{\"o}ker}, T., {Arribas}, S., {L{\"u}tzgendorf}, N., {et~al.} 2022, \aap, 661, A82

\bibitem[{{B{\"o}ker} {et~al.}(2023){B{\"o}ker}, {Beck}, {Birkmann}, {Giardino}, {Keyes}, {Kumari}, {Muzerolle}, {Rawle}, {Zeidler}, {Abul-Huda}, {Alves de Oliveira}, {Arribas}, {Bechtold}, {Bhatawdekar}, {Bonaventura}, {Bunker}, {Cameron}, {Carniani}, {Charlot}, {Curti}, {Espinoza}, {Ferruit}, {Franx}, {Jakobsen}, {Karakla}, {L{\'o}pez-Caniego}, {L{\"u}tzgendorf}, {Maiolino}, {Manjavacas}, {Marston}, {Moseley}, {Ogle}, {Perna}, {Pe{\~n}a-Guerrero}, {Pirzkal}, {Plesha}, {Proffitt}, {Rauscher}, {Rix}, {Rodr{\'\i}guez del Pino}, {Rustamkulov}, {Sabbi}, {Sing}, {Sirianni}, {te Plate}, {{\'U}beda}, {Wahlgren}, {Wislowski}, {Wu}, \& {Willott}}]{boker2023}
{B{\"o}ker}, T., {Beck}, T.~L., {Birkmann}, S.~M., {et~al.} 2023, \pasp, 135, 038001

\bibitem[{{Boroson} \& {Green}(1992)}]{boroson1992}
{Boroson}, T.~A. \& {Green}, R.~F. 1992, \apjs, 80, 109

\bibitem[{{Bosman} {et~al.}(2020){Bosman}, {Kakiichi}, {Meyer}, {Gronke}, {Laporte}, \& {Ellis}}]{bosman2020}
{Bosman}, S. E.~I., {Kakiichi}, K., {Meyer}, R.~A., {et~al.} 2020, \apj, 896, 49

\bibitem[{{Brusa} {et~al.}(2016){Brusa}, {Perna}, {Cresci}, {Schramm}, {Delvecchio}, {Lanzuisi}, {Mainieri}, {Mignoli}, {Zamorani}, {Berta}, {Bongiorno}, {Comastri}, {Fiore}, {Kakkad}, {Marconi}, {Rosario}, {Contini}, \& {Lamareille}}]{brusa2016}
{Brusa}, M., {Perna}, M., {Cresci}, G., {et~al.} 2016, \aap, 588, A58

\bibitem[{{Calzetti} {et~al.}(2000){Calzetti}, {Armus}, {Bohlin}, {Kinney}, {Koornneef}, \& {Storchi-Bergmann}}]{Calzetti2000}
{Calzetti}, D., {Armus}, L., {Bohlin}, R.~C., {et~al.} 2000, \apj, 533, 682

\bibitem[{{Cameron} {et~al.}(2023){Cameron}, {Saxena}, {Bunker}, {D'Eugenio}, {Carniani}, {Maiolino}, {Curtis-Lake}, {Ferruit}, {Jakobsen}, {Arribas}, {Bonaventura}, {Charlot}, {Chevallard}, {Curti}, {Looser}, {Maseda}, {Rawle}, {Rodr{\'\i}guez Del Pino}, {Smit}, {{\"U}bler}, {Willott}, {Witstok}, {Egami}, {Eisenstein}, {Johnson}, {Hainline}, {Rieke}, {Robertson}, {Stark}, {Tacchella}, {Williams}, {Willmer}, {Bhatawdekar}, {Bowler}, {Boyett}, {Circosta}, {Helton}, {Jones}, {Kumari}, {Ji}, {Nelson}, {Parlanti}, {Sandles}, {Scholtz}, \& {Sun}}]{cameron2023}
{Cameron}, A.~J., {Saxena}, A., {Bunker}, A.~J., {et~al.} 2023, \aap, 677, A115

\bibitem[{{Carilli} {et~al.}(2002){Carilli}, {Kohno}, {Kawabe}, {Ohta}, {Henkel}, {Menten}, {Yun}, {Petric}, \& {Tutui}}]{carilli2002}
{Carilli}, C.~L., {Kohno}, K., {Kawabe}, R., {et~al.} 2002, \aj, 123, 1838

\bibitem[{{Carilli} {et~al.}(2013){Carilli}, {Riechers}, {Walter}, {Maiolino}, {Wagg}, {Lentati}, {McMahon}, \& {Wolfe}}]{carilli2013}
{Carilli}, C.~L., {Riechers}, D., {Walter}, F., {et~al.} 2013, \apj, 763, 120

\bibitem[{{Carniani} {et~al.}(2013){Carniani}, {Marconi}, {Biggs}, {Cresci}, {Cupani}, {D'Odorico}, {Humphreys}, {Maiolino}, {Mannucci}, {Molaro}, {Nagao}, {Testi}, \& {Zwaan}}]{carniani2013}
{Carniani}, S., {Marconi}, A., {Biggs}, A., {et~al.} 2013, \aap, 559, A29

\bibitem[{{Carniani} {et~al.}(2015){Carniani}, {Marconi}, {Maiolino}, {Balmaverde}, {Brusa}, {Cano-D{\'\i}az}, {Cicone}, {Comastri}, {Cresci}, {Fiore}, {Feruglio}, {La Franca}, {Mainieri}, {Mannucci}, {Nagao}, {Netzer}, {Piconcelli}, {Risaliti}, {Schneider}, \& {Shemmer}}]{carniani2015}
{Carniani}, S., {Marconi}, A., {Maiolino}, R., {et~al.} 2015, \aap, 580, A102

\bibitem[{{Costa} {et~al.}(2014){Costa}, {Sijacki}, {Trenti}, \& {Haehnelt}}]{costa2014}
{Costa}, T., {Sijacki}, D., {Trenti}, M., \& {Haehnelt}, M.~G. 2014, \mnras, 439, 2146

\bibitem[{{Couto} {et~al.}(2016){Couto}, {Colina}, {L{\'o}pez}, {Storchi-Bergmann}, \& {Arribas}}]{Couto:2016}
{Couto}, G.~S., {Colina}, L., {L{\'o}pez}, J.~P., {Storchi-Bergmann}, T., \& {Arribas}, S. 2016, \aap, 594, A74

\bibitem[{{Cracco} {et~al.}(2016){Cracco}, {Ciroi}, {Berton}, {Di Mille}, {Foschini}, {La Mura}, \& {Rafanelli}}]{cracco2016}
{Cracco}, V., {Ciroi}, S., {Berton}, M., {et~al.} 2016, \mnras, 462, 1256

\bibitem[{{Cresci} {et~al.}(2015){Cresci}, {Mainieri}, {Brusa}, {Marconi}, {Perna}, {Mannucci}, {Piconcelli}, {Maiolino}, {Feruglio}, {Fiore}, {Bongiorno}, {Lanzuisi}, {Merloni}, {Schramm}, {Silverman}, \& {Civano}}]{cresci2015}
{Cresci}, G., {Mainieri}, V., {Brusa}, M., {et~al.} 2015, \apj, 799, 82

\bibitem[{{Cresci} {et~al.}(2023){Cresci}, {Tozzi}, {Perna}, {Brusa}, {Marconcini}, {Marconi}, {Carniani}, {Brienza}, {Giroletti}, {Belfiore}, {Ginolfi}, {Mannucci}, {Ulivi}, {Scholtz}, {Venturi}, {Arribas}, {{\"U}bler}, {D'Eugenio}, {Mingozzi}, {Balmaverde}, {Capetti}, {Parlanti}, \& {Zana}}]{cresci2023}
{Cresci}, G., {Tozzi}, G., {Perna}, M., {et~al.} 2023, \aap, 672, A128

\bibitem[{{Curti} {et~al.}(2017){Curti}, {Cresci}, {Mannucci}, {Marconi}, {Maiolino}, \& {Esposito}}]{Curti2017}
{Curti}, M., {Cresci}, G., {Mannucci}, F., {et~al.} 2017, \mnras, 465, 1384

\bibitem[{{Curti} {et~al.}(2023){Curti}, {D'Eugenio}, {Carniani}, {Maiolino}, {Sandles}, {Witstok}, {Baker}, {Bennett}, {Piotrowska}, {Tacchella}, {Charlot}, {Nakajima}, {Maheson}, {Mannucci}, {Amiri}, {Arribas}, {Belfiore}, {Bonaventura}, {Bunker}, {Chevallard}, {Cresci}, {Curtis-Lake}, {Hayden-Pawson}, {Jones}, {Kumari}, {Laseter}, {Looser}, {Marconi}, {Maseda}, {Scholtz}, {Smit}, {{\"U}bler}, \& {Wallace}}]{Curti2023}
{Curti}, M., {D'Eugenio}, F., {Carniani}, S., {et~al.} 2023, \mnras, 518, 425

\bibitem[{{Curti} {et~al.}(2024){Curti}, {Maiolino}, {Curtis-Lake}, {Chevallard}, {Carniani}, {D'Eugenio}, {Looser}, {Scholtz}, {Charlot}, {Cameron}, {{\"U}bler}, {Witstok}, {Boyett}, {Laseter}, {Sandles}, {Arribas}, {Bunker}, {Giardino}, {Maseda}, {Rawle}, {Rodr{\'\i}guez Del Pino}, {Smit}, {Willott}, {Eisenstein}, {Hausen}, {Johnson}, {Rieke}, {Robertson}, {Tacchella}, {Williams}, {Willmer}, {Baker}, {Bhatawdekar}, {Egami}, {Helton}, {Ji}, {Kumari}, {Perna}, {Shivaei}, \& {Sun}}]{Curti2024}
{Curti}, M., {Maiolino}, R., {Curtis-Lake}, E., {et~al.} 2024, \aap, 684, A75

\bibitem[{{Curti} {et~al.}(2020){Curti}, {Mannucci}, {Cresci}, \& {Maiolino}}]{Curti2020}
{Curti}, M., {Mannucci}, F., {Cresci}, G., \& {Maiolino}, R. 2020, \mnras, 491, 944

\bibitem[{{De Rosa} {et~al.}(2019){De Rosa}, {Vignali}, {Bogdanovi{\'c}}, {Capelo}, {Charisi}, {Dotti}, {Husemann}, {Lusso}, {Mayer}, {Paragi}, {Runnoe}, {Sesana}, {Steinborn}, {Bianchi}, {Colpi}, {del Valle}, {Frey}, {Gab{\'a}nyi}, {Giustini}, {Guainazzi}, {Haiman}, {Herrera Ruiz}, {Herrero-Illana}, {Iwasawa}, {Komossa}, {Lena}, {Loiseau}, {Perez-Torres}, {Piconcelli}, \& {Volonteri}}]{derosa2019}
{De Rosa}, A., {Vignali}, C., {Bogdanovi{\'c}}, T., {et~al.} 2019, \nar, 86, 101525

\bibitem[{{Decarli} {et~al.}(2024){Decarli}, {Loiacono}, {Farina}, {Dotti}, {Lupi}, {Meyer}, {Mignoli}, {Pensabene}, {Strauss}, {Venemans}, {Yang}, {Walter}, {Wolf}, {Ba{\~n}ados}, {Blecha}, {Bosman}, {Carilli}, {Comastri}, {Connor}, {Costa}, {Eilers}, {Fan}, {Gilli}, {Jun}, {Liu}, {Marshall}, {Mazzucchelli}, {Neeleman}, {Onoue}, {Overzier}, {Pudoka}, {Riechers}, {Rix}, {Schindler}, {Trakhtenbrot}, {Trebitsch}, {Vestergaard}, {Volonteri}, {Wang}, {Zhang}, \& {Zou}}]{decarli2024}
{Decarli}, R., {Loiacono}, F., {Farina}, E.~P., {et~al.} 2024, arXiv e-prints, arXiv:2406.06697

\bibitem[{{Decarli} {et~al.}(2014){Decarli}, {Walter}, {Carilli}, {Bertoldi}, {Cox}, {Ferkinhoff}, {Groves}, {Maiolino}, {Neri}, {Riechers}, \& {Weiss}}]{decarli2014}
{Decarli}, R., {Walter}, F., {Carilli}, C., {et~al.} 2014, \apjl, 782, L17

\bibitem[{{Decarli} {et~al.}(2017){Decarli}, {Walter}, {Venemans}, {Ba{\~n}ados}, {Bertoldi}, {Carilli}, {Fan}, {Farina}, {Mazzucchelli}, {Riechers}, {Rix}, {Strauss}, {Wang}, \& {Yang}}]{decarli2017}
{Decarli}, R., {Walter}, F., {Venemans}, B.~P., {et~al.} 2017, \nat, 545, 457

\bibitem[{{Decarli} {et~al.}(2018){Decarli}, {Walter}, {Venemans}, {Ba{\~n}ados}, {Bertoldi}, {Carilli}, {Fan}, {Farina}, {Mazzucchelli}, {Riechers}, {Rix}, {Strauss}, {Wang}, \& {Yang}}]{Decarli:2018}
{Decarli}, R., {Walter}, F., {Venemans}, B.~P., {et~al.} 2018, \apj, 854, 97

\bibitem[{{D'Eugenio} {et~al.}(2024){D'Eugenio}, {P{\'e}rez-Gonz{\'a}lez}, {Maiolino}, {Scholtz}, {Perna}, {Circosta}, {{\"U}bler}, {Arribas}, {B{\"o}ker}, {Bunker}, {Carniani}, {Charlot}, {Chevallard}, {Cresci}, {Curtis-Lake}, {Jones}, {Kumari}, {Lamperti}, {Looser}, {Parlanti}, {Rix}, {Robertson}, {Rodr{\'\i}guez Del Pino}, {Tacchella}, {Venturi}, \& {Willott}}]{DEugenio2024a}
{D'Eugenio}, F., {P{\'e}rez-Gonz{\'a}lez}, P.~G., {Maiolino}, R., {et~al.} 2024, Nature Astronomy [\eprint[arXiv]{2308.06317}]

\bibitem[{{Di Mascia} {et~al.}(2021){Di Mascia}, {Gallerani}, {Behrens}, {Pallottini}, {Carniani}, {Ferrara}, {Barai}, {Vito}, \& {Zana}}]{dimascia2021}
{Di Mascia}, F., {Gallerani}, S., {Behrens}, C., {et~al.} 2021, \mnras, 503, 2349

\bibitem[{{Di Matteo} {et~al.}(2012){Di Matteo}, {Khandai}, {DeGraf}, {Feng}, {Croft}, {Lopez}, \& {Springel}}]{dimatteo2012}
{Di Matteo}, T., {Khandai}, N., {DeGraf}, C., {et~al.} 2012, \apjl, 745, L29

\bibitem[{{D{\'\i}az} \& {P{\'e}rez-Montero}(2000)}]{diaz2000}
{D{\'\i}az}, A.~I. \& {P{\'e}rez-Montero}, E. 2000, \mnras, 312, 130

\bibitem[{{Dong} {et~al.}(2008){Dong}, {Wang}, {Wang}, {Yuan}, {Zhou}, {Dai}, \& {Zhang}}]{dong2008}
{Dong}, X., {Wang}, T., {Wang}, J., {et~al.} 2008, \mnras, 383, 581

\bibitem[{{Drake} {et~al.}(2020){Drake}, {Walter}, {Novak}, {Farina}, {Neeleman}, {Riechers}, {Carilli}, {Decarli}, {Mazzucchelli}, \& {Onoue}}]{drake2020}
{Drake}, A.~B., {Walter}, F., {Novak}, M., {et~al.} 2020, \apj, 902, 37

\bibitem[{{Eracleous} {et~al.}(2012){Eracleous}, {Boroson}, {Halpern}, \& {Liu}}]{eracleous2012}
{Eracleous}, M., {Boroson}, T.~A., {Halpern}, J.~P., \& {Liu}, J. 2012, \apjs, 201, 23

\bibitem[{{Feltre} {et~al.}(2016){Feltre}, {Charlot}, \& {Gutkin}}]{feltre2016}
{Feltre}, A., {Charlot}, S., \& {Gutkin}, J. 2016, \mnras, 456, 3354

\bibitem[{{Fiore} {et~al.}(2017){Fiore}, {Feruglio}, {Shankar}, {Bischetti}, {Bongiorno}, {Brusa}, {Carniani}, {Cicone}, {Duras}, {Lamastra}, {Mainieri}, {Marconi}, {Menci}, {Maiolino}, {Piconcelli}, {Vietri}, \& {Zappacosta}}]{fiore2017}
{Fiore}, F., {Feruglio}, C., {Shankar}, F., {et~al.} 2017, \aap, 601, A143

\bibitem[{{Fluetsch} {et~al.}(2021){Fluetsch}, {Maiolino}, {Carniani}, {Arribas}, {Belfiore}, {Bellocchi}, {Cazzoli}, {Cicone}, {Cresci}, {Fabian}, {Gallagher}, {Ishibashi}, {Mannucci}, {Marconi}, {Perna}, {Sturm}, \& {Venturi}}]{fluetsch2021}
{Fluetsch}, A., {Maiolino}, R., {Carniani}, S., {et~al.} 2021, \mnras, 505, 5753

\bibitem[{{F{\"o}rster Schreiber} {et~al.}(2019){F{\"o}rster Schreiber}, {{\"U}bler}, {Davies}, {Genzel}, {Wisnioski}, {Belli}, {Shimizu}, {Lutz}, {Fossati}, {Herrera-Camus}, {Mendel}, {Tacconi}, {Wilman}, {Beifiori}, {Brammer}, {Burkert}, {Carollo}, {Davies}, {Eisenhauer}, {Fabricius}, {Lilly}, {Momcheva}, {Naab}, {Nelson}, {Price}, {Renzini}, {Saglia}, {Sternberg}, {van Dokkum}, \& {Wuyts}}]{fs2019}
{F{\"o}rster Schreiber}, N.~M., {{\"U}bler}, H., {Davies}, R.~L., {et~al.} 2019, \apj, 875, 21

\bibitem[{{Fruscione} {et~al.}(2006){Fruscione}, {McDowell}, {Allen}, {Brickhouse}, {Burke}, {Davis}, {Durham}, {Elvis}, {Galle}, {Harris}, {Huenemoerder}, {Houck}, {Ishibashi}, {Karovska}, {Nicastro}, {Noble}, {Nowak}, {Primini}, {Siemiginowska}, {Smith}, \& {Wise}}]{fruscione2006_ciao}
{Fruscione}, A., {McDowell}, J.~C., {Allen}, G.~E., {et~al.} 2006, in Society of Photo-Optical Instrumentation Engineers (SPIE) Conference Series, Vol. 6270, Society of Photo-Optical Instrumentation Engineers (SPIE) Conference Series, ed. D.~R. {Silva} \& R.~E. {Doxsey}, 62701V

\bibitem[{{Gaia Collaboration} {et~al.}(2023){Gaia Collaboration}, {Vallenari}, {Brown}, {Prusti}, {de Bruijne}, {Arenou}, {Babusiaux}, {Biermann}, {Creevey}, {Ducourant}, {Evans}, {Eyer}, {Guerra}, {Hutton}, {Jordi}, {Klioner}, {Lammers}, {Lindegren}, {Luri}, {Mignard}, {Panem}, {Pourbaix}, {Randich}, {Sartoretti}, {Soubiran}, {Tanga}, {Walton}, {Bailer-Jones}, {Bastian}, {Drimmel}, {Jansen}, {Katz}, {Lattanzi}, {van Leeuwen}, {Bakker}, {Cacciari}, {Casta{\~n}eda}, {De Angeli}, {Fabricius}, {Fouesneau}, {Fr{\'e}mat}, {Galluccio}, {Guerrier}, {Heiter}, {Masana}, {Messineo}, {Mowlavi}, {Nicolas}, {Nienartowicz}, {Pailler}, {Panuzzo}, {Riclet}, {Roux}, {Seabroke}, {Sordo}, {Th{\'e}venin}, {Gracia-Abril}, {Portell}, {Teyssier}, {Altmann}, {Andrae}, {Audard}, {Bellas-Velidis}, {Benson}, {Berthier}, {Blomme}, {Burgess}, {Busonero}, {Busso}, {C{\'a}novas}, {Carry}, {Cellino}, {Cheek}, {Clementini}, {Damerdji}, {Davidson}, {de Teodoro}, {Nu{\~n}ez Campos}, {Delchambre}, {Dell'Oro}, {Esquej},
  {Fern{\'a}ndez-Hern{\'a}ndez}, {Fraile}, {Garabato}, {Garc{\'\i}a-Lario}, {Gosset}, {Haigron}, {Halbwachs}, {Hambly}, {Harrison}, {Hern{\'a}ndez}, {Hestroffer}, {Hodgkin}, {Holl}, {Jan{\ss}en}, {Jevardat de Fombelle}, {Jordan}, {Krone-Martins}, {Lanzafame}, {L{\"o}ffler}, {Marchal}, {Marrese}, {Moitinho}, {Muinonen}, {Osborne}, {Pancino}, {Pauwels}, {Recio-Blanco}, {Reyl{\'e}}, {Riello}, {Rimoldini}, {Roegiers}, {Rybizki}, {Sarro}, {Siopis}, {Smith}, {Sozzetti}, {Utrilla}, {van Leeuwen}, {Abbas}, {{\'A}brah{\'a}m}, {Abreu Aramburu}, {Aerts}, {Aguado}, {Ajaj}, {Aldea-Montero}, {Altavilla}, {{\'A}lvarez}, {Alves}, {Anders}, {Anderson}, {Anglada Varela}, {Antoja}, {Baines}, {Baker}, {Balaguer-N{\'u}{\~n}ez}, {Balbinot}, {Balog}, {Barache}, {Barbato}, {Barros}, {Barstow}, {Bartolom{\'e}}, {Bassilana}, {Bauchet}, {Becciani}, {Bellazzini}, {Berihuete}, {Bernet}, {Bertone}, {Bianchi}, {Binnenfeld}, {Blanco-Cuaresma}, {Blazere}, {Boch}, {Bombrun}, {Bossini}, {Bouquillon}, {Bragaglia}, {Bramante}, {Breedt},
  {Bressan}, {Brouillet}, {Brugaletta}, {Bucciarelli}, {Burlacu}, {Butkevich}, {Buzzi}, {Caffau}, {Cancelliere}, {Cantat-Gaudin}, {Carballo}, {Carlucci}, {Carnerero}, {Carrasco}, {Casamiquela}, {Castellani}, {Castro-Ginard}, {Chaoul}, {Charlot}, {Chemin}, {Chiaramida}, {Chiavassa}, {Chornay}, {Comoretto}, {Contursi}, {Cooper}, {Cornez}, {Cowell}, {Crifo}, {Cropper}, {Crosta}, {Crowley}, {Dafonte}, {Dapergolas}, {David}, {David}, {de Laverny}, {De Luise}, {De March}, {De Ridder}, {de Souza}, {de Torres}, {del Peloso}, {del Pozo}, {Delbo}, {Delgado}, {Delisle}, {Demouchy}, {Dharmawardena}, {Di Matteo}, {Diakite}, {Diener}, {Distefano}, {Dolding}, {Edvardsson}, {Enke}, {Fabre}, {Fabrizio}, {Faigler}, {Fedorets}, {Fernique}, {Fienga}, {Figueras}, {Fournier}, {Fouron}, {Fragkoudi}, {Gai}, {Garcia-Gutierrez}, {Garcia-Reinaldos}, {Garc{\'\i}a-Torres}, {Garofalo}, {Gavel}, {Gavras}, {Gerlach}, {Geyer}, {Giacobbe}, {Gilmore}, {Girona}, {Giuffrida}, {Gomel}, {Gomez}, {Gonz{\'a}lez-N{\'u}{\~n}ez},
  {Gonz{\'a}lez-Santamar{\'\i}a}, {Gonz{\'a}lez-Vidal}, {Granvik}, {Guillout}, {Guiraud}, {Guti{\'e}rrez-S{\'a}nchez}, {Guy}, {Hatzidimitriou}, {Hauser}, {Haywood}, {Helmer}, {Helmi}, {Sarmiento}, {Hidalgo}, {Hilger}, {H{\l}adczuk}, {Hobbs}, {Holland}, {Huckle}, {Jardine}, {Jasniewicz}, {Jean-Antoine Piccolo}, {Jim{\'e}nez-Arranz}, {Jorissen}, {Juaristi Campillo}, {Julbe}, {Karbevska}, {Kervella}, {Khanna}, {Kontizas}, {Kordopatis}, {Korn}, {K{\'o}sp{\'a}l}, {Kostrzewa-Rutkowska}, {Kruszy{\'n}ska}, {Kun}, {Laizeau}, {Lambert}, {Lanza}, {Lasne}, {Le Campion}, {Lebreton}, {Lebzelter}, {Leccia}, {Leclerc}, {Lecoeur-Taibi}, {Liao}, {Licata}, {Lindstr{\o}m}, {Lister}, {Livanou}, {Lobel}, {Lorca}, {Loup}, {Madrero Pardo}, {Magdaleno Romeo}, {Managau}, {Mann}, {Manteiga}, {Marchant}, {Marconi}, {Marcos}, {Marcos Santos}, {Mar{\'\i}n Pina}, {Marinoni}, {Marocco}, {Marshall}, {Martin Polo}, {Mart{\'\i}n-Fleitas}, {Marton}, {Mary}, {Masip}, {Massari}, {Mastrobuono-Battisti}, {Mazeh}, {McMillan}, {Messina}, {Michalik},
  {Millar}, {Mints}, {Molina}, {Molinaro}, {Moln{\'a}r}, {Monari}, {Mongui{\'o}}, {Montegriffo}, {Montero}, {Mor}, {Mora}, {Morbidelli}, {Morel}, {Morris}, {Muraveva}, {Murphy}, {Musella}, {Nagy}, {Noval}, {Oca{\~n}a}, {Ogden}, {Ordenovic}, {Osinde}, {Pagani}, {Pagano}, {Palaversa}, {Palicio}, {Pallas-Quintela}, {Panahi}, {Payne-Wardenaar}, {Pe{\~n}alosa Esteller}, {Penttil{\"a}}, {Pichon}, {Piersimoni}, {Pineau}, {Plachy}, {Plum}, {Poggio}, {Pr{\v{s}}a}, {Pulone}, {Racero}, {Ragaini}, {Rainer}, {Raiteri}, {Rambaux}, {Ramos}, {Ramos-Lerate}, {Re Fiorentin}, {Regibo}, {Richards}, {Rios Diaz}, {Ripepi}, {Riva}, {Rix}, {Rixon}, {Robichon}, {Robin}, {Robin}, {Roelens}, {Rogues}, {Rohrbasser}, {Romero-G{\'o}mez}, {Rowell}, {Royer}, {Ruz Mieres}, {Rybicki}, {Sadowski}, {S{\'a}ez N{\'u}{\~n}ez}, {Sagrist{\`a} Sell{\'e}s}, {Sahlmann}, {Salguero}, {Samaras}, {Sanchez Gimenez}, {Sanna}, {Santove{\~n}a}, {Sarasso}, {Schultheis}, {Sciacca}, {Segol}, {Segovia}, {S{\'e}gransan}, {Semeux}, {Shahaf}, {Siddiqui}, {Siebert},
  {Siltala}, {Silvelo}, {Slezak}, {Slezak}, {Smart}, {Snaith}, {Solano}, {Solitro}, {Souami}, {Souchay}, {Spagna}, {Spina}, {Spoto}, {Steele}, {Steidelm{\"u}ller}, {Stephenson}, {S{\"u}veges}, {Surdej}, {Szabados}, {Szegedi-Elek}, {Taris}, {Taylor}, {Teixeira}, {Tolomei}, {Tonello}, {Torra}, {Torra}, {Torralba Elipe}, {Trabucchi}, {Tsounis}, {Turon}, {Ulla}, {Unger}, {Vaillant}, {van Dillen}, {van Reeven}, {Vanel}, {Vecchiato}, {Viala}, {Vicente}, {Voutsinas}, {Weiler}, {Wevers}, {Wyrzykowski}, {Yoldas}, {Yvard}, {Zhao}, {Zorec}, {Zucker}, \& {Zwitter}}]{gaia_collaboration2023}
{Gaia Collaboration}, {Vallenari}, A., {Brown}, A.~G.~A., {et~al.} 2023, \aap, 674, A1

\bibitem[{{Gaskell}(2010)}]{2010Natur.463E...1G}
{Gaskell}, C.~M. 2010, \nat, 463, E1

\bibitem[{{Greene} \& {Ho}(2005)}]{greene2005}
{Greene}, J.~E. \& {Ho}, L.~C. 2005, \apj, 630, 122

\bibitem[{{Habouzit} {et~al.}(2019){Habouzit}, {Genel}, {Somerville}, {Kocevski}, {Hirschmann}, {Dekel}, {Choi}, {Nelson}, {Pillepich}, {Torrey}, {Hernquist}, {Vogelsberger}, {Weinberger}, \& {Springel}}]{habouzit2019}
{Habouzit}, M., {Genel}, S., {Somerville}, R.~S., {et~al.} 2019, \mnras, 484, 4413

\bibitem[{{Harrison} {et~al.}(2016){Harrison}, {Alexander}, {Mullaney}, {Stott}, {Swinbank}, {Arumugam}, {Bauer}, {Bower}, {Bunker}, \& {Sharples}}]{harrison2016}
{Harrison}, C.~M., {Alexander}, D.~M., {Mullaney}, J.~R., {et~al.} 2016, \mnras, 456, 1195

\bibitem[{{Harrison} {et~al.}(2015){Harrison}, {Thomson}, {Alexander}, {Bauer}, {Edge}, {Hogan}, {Mullaney}, \& {Swinbank}}]{harrison2015}
{Harrison}, C.~M., {Thomson}, A.~P., {Alexander}, D.~M., {et~al.} 2015, \apj, 800, 45

\bibitem[{{Heckman} \& {Borthakur}(2016)}]{heckman2016}
{Heckman}, T.~M. \& {Borthakur}, S. 2016, \apj, 822, 9

\bibitem[{{Herrera-Camus} {et~al.}(2019){Herrera-Camus}, {Tacconi}, {Genzel}, {F{\"o}rster Schreiber}, {Lutz}, {Bolatto}, {Wuyts}, {Renzini}, {Lilly}, {Belli}, {{\"U}bler}, {Shimizu}, {Davies}, {Sturm}, {Combes}, {Freundlich}, {Garc{\'\i}a-Burillo}, {Cox}, {Burkert}, {Naab}, {Colina}, {Saintonge}, {Cooper}, {Feruglio}, \& {Weiss}}]{herreracamus2019}
{Herrera-Camus}, R., {Tacconi}, L., {Genzel}, R., {et~al.} 2019, \apj, 871, 37

\bibitem[{{Hirschmann} {et~al.}(2023){Hirschmann}, {Charlot}, {Feltre}, {Curtis-Lake}, {Somerville}, {Chevallard}, {Choi}, {Nelson}, {Morisset}, {Plat}, \& {Vidal-Garcia}}]{hirschmann2023}
{Hirschmann}, M., {Charlot}, S., {Feltre}, A., {et~al.} 2023, \mnras, 526, 3610

\bibitem[{{Hu} \& {McMahon}(1996)}]{hu1996}
{Hu}, E. \& {McMahon}, R.~G. 1996, \nat, 382, 281

\bibitem[{{Iono} {et~al.}(2006){Iono}, {Yun}, {Elvis}, {Peck}, {Ho}, {Wilner}, {Hunter}, {Matsushita}, \& {Muller}}]{iono2006}
{Iono}, D., {Yun}, M.~S., {Elvis}, M., {et~al.} 2006, \apjl, 645, L97

\bibitem[{{Irwin} {et~al.}(1991){Irwin}, {McMahon}, \& {Hazard}}]{irwin1991}
{Irwin}, M., {McMahon}, R.~G., \& {Hazard}, C. 1991, in Astronomical Society of the Pacific Conference Series, Vol.~21, The Space Distribution of Quasars, ed. D.~{Crampton}, 117--126

\bibitem[{{Jakobsen} {et~al.}(2022){Jakobsen}, {Ferruit}, {Alves de Oliveira}, {Arribas}, {Bagnasco}, {Barho}, {Beck}, {Birkmann}, {B{\"o}ker}, {Bunker}, {Charlot}, {de Jong}, {de Marchi}, {Ehrenwinkler}, {Falcolini}, {Fels}, {Franx}, {Franz}, {Funke}, {Giardino}, {Gnata}, {Holota}, {Honnen}, {Jensen}, {Jentsch}, {Johnson}, {Jollet}, {Karl}, {Kling}, {K{\"o}hler}, {Kolm}, {Kumari}, {Lander}, {Lemke}, {L{\'o}pez-Caniego}, {L{\"u}tzgendorf}, {Maiolino}, {Manjavacas}, {Marston}, {Maschmann}, {Maurer}, {Messerschmidt}, {Moseley}, {Mosner}, {Mott}, {Muzerolle}, {Pirzkal}, {Pittet}, {Plitzke}, {Posselt}, {Rapp}, {Rauscher}, {Rawle}, {Rix}, {R{\"o}del}, {Rumler}, {Sabbi}, {Salvignol}, {Schmid}, {Sirianni}, {Smith}, {Strada}, {te Plate}, {Valenti}, {Wettemann}, {Wiehe}, {Wiesmayer}, {Willott}, {Wright}, {Zeidler}, \& {Zincke}}]{jakobsen2022}
{Jakobsen}, P., {Ferruit}, P., {Alves de Oliveira}, C., {et~al.} 2022, \aap, 661, A80

\bibitem[{{Jones} {et~al.}(2024{\natexlab{a}}){Jones}, {Bunker}, {Telikova}, {Arribas}, {Carniani}, {Charlot}, {D'Eugenio}, {Maiolino}, {Perna}, {Rodriguez Del Pino}, {Ubler}, {Willott}, {Aravena}, {Boker}, {Cresci}, {Curti}, {Herrera-Camus}, {Lamperti}, {Parlanti}, {Perez-Gonzalez}, \& {Villanueva}}]{jones2024}
{Jones}, G.~C., {Bunker}, A.~J., {Telikova}, K., {et~al.} 2024{\natexlab{a}}, arXiv e-prints, arXiv:2405.12955

\bibitem[{{Jones} {et~al.}(2016){Jones}, {Carilli}, {Momjian}, {Wagg}, {Riechers}, {Walter}, {Decarli}, {Ota}, \& {McMahon}}]{jones2016}
{Jones}, G.~C., {Carilli}, C.~L., {Momjian}, E., {et~al.} 2016, \apj, 830, 63

\bibitem[{{Jones} {et~al.}(2024{\natexlab{b}}){Jones}, {{\"U}bler}, {Perna}, {Arribas}, {Bunker}, {Carniani}, {Charlot}, {Maiolino}, {Del Pino}, {Willott}, {Bowler}, {B{\"o}ker}, {Cameron}, {Chevallard}, {Cresci}, {Curti}, {D'Eugenio}, {Kumari}, {Saxena}, {Scholtz}, {Venturi}, \& {Witstok}}]{jones2024-2}
{Jones}, G.~C., {{\"U}bler}, H., {Perna}, M., {et~al.} 2024{\natexlab{b}}, \aap, 682, A122

\bibitem[{{Ju} {et~al.}(2013){Ju}, {Greene}, {Rafikov}, {Bickerton}, \& {Badenes}}]{2013ApJ...777...44J}
{Ju}, W., {Greene}, J.~E., {Rafikov}, R.~R., {Bickerton}, S.~J., \& {Badenes}, C. 2013, \apj, 777, 44

\bibitem[{{Jun} {et~al.}(2015){Jun}, {Im}, {Lee}, {Ohyama}, {Woo}, {Fan}, {Goto}, {Kim}, {Kim}, {Kim}, {Lee}, {Nakagawa}, {Pearson}, \& {Serjeant}}]{jun2015}
{Jun}, H.~D., {Im}, M., {Lee}, H.~M., {et~al.} 2015, \apj, 806, 109

\bibitem[{{Just} {et~al.}(2007){Just}, {Brandt}, {Shemmer}, {Steffen}, {Schneider}, {Chartas}, \& {Garmire}}]{just2007}
{Just}, D.~W., {Brandt}, W.~N., {Shemmer}, O., {et~al.} 2007, \apj, 665, 1004

\bibitem[{{Kashino} {et~al.}(2023){Kashino}, {Lilly}, {Matthee}, {Eilers}, {Mackenzie}, {Bordoloi}, \& {Simcoe}}]{kashino2023}
{Kashino}, D., {Lilly}, S.~J., {Matthee}, J., {et~al.} 2023, \apj, 950, 66

\bibitem[{{Kauffmann} {et~al.}(2003){Kauffmann}, {Heckman}, {Tremonti}, {Brinchmann}, {Charlot}, {White}, {Ridgway}, {Brinkmann}, {Fukugita}, {Hall}, {Ivezi{\'c}}, {Richards}, \& {Schneider}}]{kauffmann2003}
{Kauffmann}, G., {Heckman}, T.~M., {Tremonti}, C., {et~al.} 2003, \mnras, 346, 1055

\bibitem[{{Kennicutt} \& {Evans}(2012)}]{Kennicutt2012}
{Kennicutt}, R.~C. \& {Evans}, N.~J. 2012, \araa, 50, 531

\bibitem[{{Kewley} {et~al.}(2001){Kewley}, {Dopita}, {Sutherland}, {Heisler}, \& {Trevena}}]{kewley2001}
{Kewley}, L.~J., {Dopita}, M.~A., {Sutherland}, R.~S., {Heisler}, C.~A., \& {Trevena}, J. 2001, \apj, 556, 121

\bibitem[{{Komossa} \& {Merritt}(2008)}]{2008ApJ...683L..21K}
{Komossa}, S. \& {Merritt}, D. 2008, \apjl, 683, L21

\bibitem[{{Lamperti} {et~al.}(2024){Lamperti}, {Arribas}, {Perna}, {Rodr{\'\i}guez Del Pino}, {Circosta}, {P{\'e}rez-Gonz{\'a}lez}, {Bunker}, {Carniani}, {Charlot}, {D'Eugenio}, {Maiolino}, {{\"U}bler}, {Willott}, {Bertola}, {B{\"o}ker}, {Cresci}, {Curti}, {Jones}, {Kumari}, {Parlanti}, {Scholtz}, \& {Venturi}}]{lamperti2024}
{Lamperti}, I., {Arribas}, S., {Perna}, M., {et~al.} 2024, arXiv e-prints, arXiv:2406.10348

\bibitem[{{Lanzuisi} {et~al.}(2015){Lanzuisi}, {Ranalli}, {Georgantopoulos}, {Georgakakis}, {Delvecchio}, {Akylas}, {Berta}, {Bongiorno}, {Brusa}, {Cappelluti}, {Civano}, {Comastri}, {Gilli}, {Gruppioni}, {Hasinger}, {Iwasawa}, {Koekemoer}, {Lusso}, {Marchesi}, {Mainieri}, {Merloni}, {Mignoli}, {Piconcelli}, {Pozzi}, {Rosario}, {Salvato}, {Silverman}, {Trakhtenbrot}, {Vignali}, \& {Zamorani}}]{lanzuisi2015}
{Lanzuisi}, G., {Ranalli}, P., {Georgantopoulos}, I., {et~al.} 2015, \aap, 573, A137

\bibitem[{{Laseter} {et~al.}(2024){Laseter}, {Maseda}, {Curti}, {Maiolino}, {D'Eugenio}, {Cameron}, {Looser}, {Arribas}, {Baker}, {Bhatawdekar}, {Boyett}, {Bunker}, {Carniani}, {Charlot}, {Chevallard}, {Curtis-lake}, {Egami}, {Eisenstein}, {Hainline}, {Hausen}, {Ji}, {Kumari}, {Perna}, {Rawle}, {Rix}, {Robertson}, {Rodr{\'\i}guez Del Pino}, {Sandles}, {Scholtz}, {Smit}, {Tacchella}, {{\"U}bler}, {Williams}, {Willott}, \& {Witstok}}]{Laseter2024}
{Laseter}, I.~H., {Maseda}, M.~V., {Curti}, M., {et~al.} 2024, \aap, 681, A70

\bibitem[{{Lee} {et~al.}(2019){Lee}, {Nagao}, {De Breuck}, {Carniani}, {Cresci}, {Hatsukade}, {Kawabe}, {Kohno}, {Maiolino}, {Mannucci}, {Marconi}, {Nakanishi}, {Saito}, {Tamura}, {Troncoso}, {Umehata}, \& {Yun}}]{lee2019}
{Lee}, M.~M., {Nagao}, T., {De Breuck}, C., {et~al.} 2019, \apjl, 883, L29

\bibitem[{{Lehnert} {et~al.}(2020){Lehnert}, {Yang}, {Emonts}, {Omont}, {Falgarone}, {Cox}, \& {Guillard}}]{lehnert2020}
{Lehnert}, M.~D., {Yang}, C., {Emonts}, B.~H.~C., {et~al.} 2020, \aap, 641, A124

\bibitem[{{Leung} {et~al.}(2019){Leung}, {Coil}, {Aird}, {Azadi}, {Kriek}, {Mobasher}, {Reddy}, {Shapley}, {Siana}, {Fetherolf}, {Fornasini}, {Freeman}, {Price}, {Sanders}, {Shivaei}, \& {Zick}}]{leung2019}
{Leung}, G. C.~K., {Coil}, A.~L., {Aird}, J., {et~al.} 2019, \apj, 886, 11

\bibitem[{{Li} {et~al.}(2021){Li}, {Wang}, {Yang}, {Bregman}, {Fan}, \& {Zhang}}]{li2021_chandra2}
{Li}, J.-T., {Wang}, F., {Yang}, J., {et~al.} 2021, \mnras, 504, 2767

\bibitem[{{Li} {et~al.}(2022){Li}, {Feng}, {Liu}, {Bai}, {Li}, {Lu}, {Wang}, {Huang}, \& {Zhang}}]{li2022}
{Li}, S.-S., {Feng}, H.-C., {Liu}, H.~T., {et~al.} 2022, \apj, 936, 75

\bibitem[{{Loiacono} {et~al.}(2024){Loiacono}, {Decarli}, {Mignoli}, {Farina}, {Ba{\~n}ados}, {Bosman}, {Eilers}, {Schindler}, {Strauss}, {Vestergaard}, {Wang}, {Blecha}, {Carilli}, {Comastri}, {Connor}, {Costa}, {Dotti}, {Fan}, {Gilli}, {Jun}, {Liu}, {Lupi}, {Marshall}, {Mazzucchelli}, {Meyer}, {Neeleman}, {Overzier}, {Pensabene}, {Riechers}, {Trakhtenbrot}, {Trebitsch}, {Venemans}, {Walter}, \& {Yang}}]{loiacono2024}
{Loiacono}, F., {Decarli}, R., {Mignoli}, M., {et~al.} 2024, \aap, 685, A121

\bibitem[{{Lu} {et~al.}(2017){Lu}, {Zhao}, {D{\'\i}az-Santos}, {Xu}, {Charmandaris}, {Gao}, {van der Werf}, {Privon}, {Inami}, {Rigopoulou}, {Sanders}, \& {Zhu}}]{lu2017}
{Lu}, N., {Zhao}, Y., {D{\'\i}az-Santos}, T., {et~al.} 2017, \apjl, 842, L16

\bibitem[{{Maiolino} {et~al.}(2023){Maiolino}, {Scholtz}, {Curtis-Lake}, {Carniani}, {Baker}, {de Graaff}, {Tacchella}, {{\"U}bler}, {D'Eugenio}, {Witstok}, {Curti}, {Arribas}, {Bunker}, {Charlot}, {Chevallard}, {Eisenstein}, {Egami}, {Ji}, {Jones}, {Lyu}, {Rawle}, {Robertson}, {Rujopakarn}, {Perna}, {Sun}, {Venturi}, {Williams}, \& {Willott}}]{maiolino2023}
{Maiolino}, R., {Scholtz}, J., {Curtis-Lake}, E., {et~al.} 2023, arXiv e-prints, arXiv:2308.01230

\bibitem[{{Marshall} {et~al.}(2023){Marshall}, {Perna}, {Willott}, {Maiolino}, {Scholtz}, {{\"U}bler}, {Carniani}, {Arribas}, {L{\"u}tzgendorf}, {Bunker}, {Charlot}, {Ferruit}, {Jakobsen}, {Rix}, {Rodr{\'\i}guez Del Pino}, {B{\"o}ker}, {Cameron}, {Cresci}, {Curtis-Lake}, {Jones}, {Kumari}, {P{\'e}rez-Gonz{\'a}lez}, \& {Reed}}]{marshall2023}
{Marshall}, M.~A., {Perna}, M., {Willott}, C.~J., {et~al.} 2023, \aap, 678, A191

\bibitem[{{Marshall} {et~al.}(2024){Marshall}, {Yue}, {Eilers}, {Scholtz}, {Perna}, {Willott}, {Maiolino}, {{\"U}bler}, {Arribas}, {Bunker}, {Charlot}, {Rodr{\'\i}guez Del Pino}, {B{\"o}ker}, {Carniani}, {Cresci}, {D'Eugenio}, {Jones}, {Venturi}, {Bordoloi}, {Kashino}, {Mackenzie}, {Matthee}, {Naidu}, \& {Simcoe}}]{marshall2024}
{Marshall}, M.~A., {Yue}, M., {Eilers}, A.-C., {et~al.} 2024, arXiv e-prints, arXiv:2410.11035

\bibitem[{{Marziani} {et~al.}(2016){Marziani}, {Mart{\'\i}nez Carballo}, {Sulentic}, {Del Olmo}, {Stirpe}, \& {Dultzin}}]{marziani2016}
{Marziani}, P., {Mart{\'\i}nez Carballo}, M.~A., {Sulentic}, J.~W., {et~al.} 2016, \apss, 361, 29

\bibitem[{{Marziani} {et~al.}(2003){Marziani}, {Zamanov}, {Sulentic}, {Calvani}, \& {Dultzin-Hacyan}}]{marziani2003}
{Marziani}, P., {Zamanov}, R., {Sulentic}, J.~W., {Calvani}, M., \& {Dultzin-Hacyan}, D. 2003, \memsai, 74, 492

\bibitem[{{Matsuoka} {et~al.}(2024){Matsuoka}, {Izumi}, {Onoue}, {Strauss}, {Iwasawa}, {Kashikawa}, {Akiyama}, {Aoki}, {Arita}, {Imanishi}, {Ishimoto}, {Kawaguchi}, {Kohno}, {Lee}, {Nagao}, {Silverman}, \& {Toba}}]{matsuoka2024}
{Matsuoka}, Y., {Izumi}, T., {Onoue}, M., {et~al.} 2024, \apjl, 965, L4

\bibitem[{{Meyer} {et~al.}(2022){Meyer}, {Decarli}, {Walter}, {Li}, {Wang}, {Mazzucchelli}, {Ba{\~n}ados}, {Farina}, \& {Venemans}}]{meyer2022}
{Meyer}, R.~A., {Decarli}, R., {Walter}, F., {et~al.} 2022, \apj, 927, 141

\bibitem[{{Mignoli} {et~al.}(2020){Mignoli}, {Gilli}, {Decarli}, {Vanzella}, {Balmaverde}, {Cappelluti}, {Cassar{\`a}}, {Comastri}, {Cusano}, {Iwasawa}, {Marchesi}, {Prandoni}, {Vignali}, {Vito}, {Zamorani}, {Chiaberge}, \& {Norman}}]{mignoli2020}
{Mignoli}, M., {Gilli}, R., {Decarli}, R., {et~al.} 2020, \aap, 642, L1

\bibitem[{{Miller} \& {Mathews}(1972)}]{reddening}
{Miller}, J.~S. \& {Mathews}, W.~G. 1972, \apj, 172, 593

\bibitem[{{Mineo} {et~al.}(2014){Mineo}, {Gilfanov}, {Lehmer}, {Morrison}, \& {Sunyaev}}]{mineo2014}
{Mineo}, S., {Gilfanov}, M., {Lehmer}, B.~D., {Morrison}, G.~E., \& {Sunyaev}, R. 2014, \mnras, 437, 1698

\bibitem[{{Morselli} {et~al.}(2014){Morselli}, {Mignoli}, {Gilli}, {Vignali}, {Comastri}, {Sani}, {Cappelluti}, {Zamorani}, {Brusa}, {Gallozzi}, \& {Vanzella}}]{morselli2014}
{Morselli}, L., {Mignoli}, M., {Gilli}, R., {et~al.} 2014, \aap, 568, A1

\bibitem[{{Neeleman} {et~al.}(2019){Neeleman}, {Ba{\~n}ados}, {Walter}, {Decarli}, {Venemans}, {Carilli}, {Fan}, {Farina}, {Mazzucchelli}, {Novak}, {Riechers}, {Rix}, \& {Wang}}]{Neeleman:2019}
{Neeleman}, M., {Ba{\~n}ados}, E., {Walter}, F., {et~al.} 2019, \apj, 882, 10

\bibitem[{{Netzer}(2019)}]{netzer2019}
{Netzer}, H. 2019, \mnras, 488, 5185

\bibitem[{{Ni} {et~al.}(2020){Ni}, {Di Matteo}, {Gilli}, {Croft}, {Feng}, \& {Norman}}]{ni2020}
{Ni}, Y., {Di Matteo}, T., {Gilli}, R., {et~al.} 2020, \mnras, 495, 2135

\bibitem[{{Ohta} {et~al.}(1996){Ohta}, {Yamada}, {Nakanishi}, {Kohno}, {Akiyama}, \& {Kawabe}}]{ohta1996}
{Ohta}, K., {Yamada}, T., {Nakanishi}, K., {et~al.} 1996, \nat, 382, 426

\bibitem[{{Omont}(1996)}]{omont1996}
{Omont}, A. 1996, in Science with Large Millimetre Arrays, ed. P.~A. {Shaver}, 82

\bibitem[{{Omont} {et~al.}(1996){Omont}, {Petitjean}, {Guilloteau}, {McMahon}, {Solomon}, \& {P{\'e}contal}}]{omont1996b}
{Omont}, A., {Petitjean}, P., {Guilloteau}, S., {et~al.} 1996, \nat, 382, 428

\bibitem[{{Ormerod} {et~al.}(2024){Ormerod}, {Conselice}, {Adams}, {Harvey}, {Austin}, {Trussler}, {Ferreira}, {Caruana}, {Lucatelli}, {Li}, \& {Roper}}]{Ormerod2024}
{Ormerod}, K., {Conselice}, C.~J., {Adams}, N.~J., {et~al.} 2024, \mnras, 527, 6110

\bibitem[{{Osterbrock} \& {Ferland}(2006)}]{Osterbrock2006}
{Osterbrock}, D.~E. \& {Ferland}, G.~J. 2006, {Astrophysics of gaseous nebulae and active galactic nuclei}

\bibitem[{{Overzier}(2022)}]{overzier2022}
{Overzier}, R.~A. 2022, \apj, 926, 114

\bibitem[{{Park} {et~al.}(2022){Park}, {Kim}, {Kim}, \& {van Putten}}]{park2022}
{Park}, H.-J., {Kim}, S.-J., {Kim}, S., \& {van Putten}, M. H.~P.~M. 2022, \apj, 938, 69

\bibitem[{{Pavesi} {et~al.}(2016){Pavesi}, {Riechers}, {Capak}, {Carilli}, {Sharon}, {Stacey}, {Karim}, {Scoville}, \& {Smol{\v{c}}i{\'c}}}]{pavesi2016}
{Pavesi}, R., {Riechers}, D.~A., {Capak}, P.~L., {et~al.} 2016, \apj, 832, 151

\bibitem[{{Perna} {et~al.}(2023{\natexlab{a}}){Perna}, {Arribas}, {Lamperti}, {Circosta}, {Bertola}, {P{\'e}rez-Gonz{\'a}lez}, {D'Eugenio}, {{\"U}bler}, {Cresci}, {Maiolino}, {Rodr{\'\i}guez Del Pino}, {Bunker}, {Charlot}, {Willott}, {Carniani}, {B{\"o}ker}, {Chevallard}, {Curti}, {Jones}, {Kumari}, {Marshall}, {Saxena}, {Scholtz}, {Venturi}, \& {Witstok}}]{perna2023b}
{Perna}, M., {Arribas}, S., {Lamperti}, I., {et~al.} 2023{\natexlab{a}}, arXiv e-prints, arXiv:2310.03067

\bibitem[{{Perna} {et~al.}(2023{\natexlab{b}}){Perna}, {Arribas}, {Marshall}, {D'Eugenio}, {{\"U}bler}, {Bunker}, {Charlot}, {Carniani}, {Jakobsen}, {Maiolino}, {Rodr{\'\i}guez Del Pino}, {Willott}, {B{\"o}ker}, {Circosta}, {Cresci}, {Curti}, {Husemann}, {Kumari}, {Lamperti}, {P{\'e}rez-Gonz{\'a}lez}, \& {Scholtz}}]{perna2023}
{Perna}, M., {Arribas}, S., {Marshall}, M., {et~al.} 2023{\natexlab{b}}, \aap, 679, A89

\bibitem[{{Perna} {et~al.}(2021){Perna}, {Arribas}, {Pereira Santaella}, {Colina}, {Bellocchi}, {Catal{\'a}n-Torrecilla}, {Cazzoli}, {Crespo G{\'o}mez}, {Maiolino}, {Piqueras L{\'o}pez}, \& {Rodr{\'\i}guez del Pino}}]{perna2021}
{Perna}, M., {Arribas}, S., {Pereira Santaella}, M., {et~al.} 2021, \aap, 646, A101

\bibitem[{{Perna} {et~al.}(2019){Perna}, {Cresci}, {Brusa}, {Lanzuisi}, {Concas}, {Mainieri}, {Mannucci}, \& {Marconi}}]{perna2019}
{Perna}, M., {Cresci}, G., {Brusa}, M., {et~al.} 2019, \aap, 623, A171

\bibitem[{{Perna} {et~al.}(2017){Perna}, {Lanzuisi}, {Brusa}, {Cresci}, \& {Mignoli}}]{perna2017}
{Perna}, M., {Lanzuisi}, G., {Brusa}, M., {Cresci}, G., \& {Mignoli}, M. 2017, \aap, 606, A96

\bibitem[{{Perrotta} {et~al.}(2021){Perrotta}, {George}, {Coil}, {Tremonti}, {Rupke}, {Davis}, {Diamond-Stanic}, {Geach}, {Hickox}, {Moustakas}, {Petter}, {Rudnick}, {Sell}, {Swiggum}, \& {Whalen}}]{perrotta2021}
{Perrotta}, S., {George}, E.~R., {Coil}, A.~L., {et~al.} 2021, \apj, 923, 275

\bibitem[{{Perrotta} {et~al.}(2019){Perrotta}, {Hamann}, {Zakamska}, {Alexandroff}, {Rupke}, \& {Wylezalek}}]{perrotta2019}
{Perrotta}, S., {Hamann}, F., {Zakamska}, N.~L., {et~al.} 2019, \mnras, 488, 4126

\bibitem[{{Reines} {et~al.}(2013){Reines}, {Greene}, \& {Geha}}]{reines2013}
{Reines}, A.~E., {Greene}, J.~E., \& {Geha}, M. 2013, \apj, 775, 116

\bibitem[{{Riechers} {et~al.}(2006){Riechers}, {Walter}, {Carilli}, {Knudsen}, {Lo}, {Benford}, {Staguhn}, {Hunter}, {Bertoldi}, {Henkel}, {Menten}, {Weiss}, {Yun}, \& {Scoville}}]{riechers2006}
{Riechers}, D.~A., {Walter}, F., {Carilli}, C.~L., {et~al.} 2006, \apj, 650, 604

\bibitem[{{Roberts-Borsani} {et~al.}(2020){Roberts-Borsani}, {Saintonge}, {Masters}, \& {Stark}}]{robertsborsani2020}
{Roberts-Borsani}, G.~W., {Saintonge}, A., {Masters}, K.~L., \& {Stark}, D.~V. 2020, \mnras, 493, 3081

\bibitem[{{Rodr{\'\i}guez Del Pino} {et~al.}(2024){Rodr{\'\i}guez Del Pino}, {Perna}, {Arribas}, {D'Eugenio}, {Lamperti}, {P{\'e}rez-Gonz{\'a}lez}, {{\"U}bler}, {Bunker}, {Carniani}, {Charlot}, {Maiolino}, {Willott}, {B{\"o}ker}, {Chevallard}, {Cresci}, {Curti}, {Jones}, {Parlanti}, {Scholtz}, \& {Venturi}}]{delpino2024}
{Rodr{\'\i}guez Del Pino}, B., {Perna}, M., {Arribas}, S., {et~al.} 2024, \aap, 684, A187

\bibitem[{{Rupke} \& {Veilleux}(2013)}]{rupkeveilleux2013}
{Rupke}, D. S.~N. \& {Veilleux}, S. 2013, \apj, 768, 75

\bibitem[{{Salom{\'e}} {et~al.}(2012){Salom{\'e}}, {Gu{\'e}lin}, {Downes}, {Cox}, {Guilloteau}, {Omont}, {Gavazzi}, \& {Neri}}]{salome2012}
{Salom{\'e}}, P., {Gu{\'e}lin}, M., {Downes}, D., {et~al.} 2012, \aap, 545, A57

\bibitem[{{Scholtz} {et~al.}(2023){Scholtz}, {Maiolino}, {D'Eugenio}, {Curtis-Lake}, {Carniani}, {Charlot}, {Curti}, {Silcock}, {Arribas}, {Baker}, {Bhatawdekar}, {Boyett}, {Bunker}, {Chevallard}, {Circosta}, {Eisenstein}, {Hainline}, {Hausen}, {Ji}, {Ji}, {Johnson}, {Kumari}, {Looser}, {Lyu}, {Maseda}, {Parlanti}, {Perna}, {Rieke}, {Robertson}, {Rodr{\'\i}guez Del Pino}, {Sun}, {Tacchella}, {{\"U}bler}, {Venturi}, {Williams}, {Willmer}, {Willott}, \& {Witstok}}]{scholtz2023}
{Scholtz}, J., {Maiolino}, R., {D'Eugenio}, F., {et~al.} 2023, arXiv e-prints, arXiv:2311.18731

\bibitem[{{Shields} {et~al.}(2009){Shields}, {Rosario}, {Smith}, {Bonning}, {Salviander}, {Kalirai}, {Strickler}, {Ramirez-Ruiz}, {Dutton}, {Treu}, \& {Marshall}}]{shields2009}
{Shields}, G.~A., {Rosario}, D.~J., {Smith}, K.~L., {et~al.} 2009, \apj, 707, 936

\bibitem[{{Sijacki} {et~al.}(2009){Sijacki}, {Springel}, \& {Haehnelt}}]{sijacki2009}
{Sijacki}, D., {Springel}, V., \& {Haehnelt}, M.~G. 2009, \mnras, 400, 100

\bibitem[{{Simpson} {et~al.}(2014){Simpson}, {Mortlock}, {Warren}, {Cantalupo}, {Hewett}, {McLure}, {McMahon}, \& {Venemans}}]{simpson2014}
{Simpson}, C., {Mortlock}, D., {Warren}, S., {et~al.} 2014, \mnras, 442, 3454

\bibitem[{{Speranza} {et~al.}(2024){Speranza}, {Ramos Almeida}, {Acosta-Pulido}, {Audibert}, {Holden}, {Tadhunter}, {Lapi}, {Gonz{\'a}lez-Mart{\'\i}n}, {Brusa}, {L{\'o}pez}, {Musiimenta}, \& {Shankar}}]{speranza2024}
{Speranza}, G., {Ramos Almeida}, C., {Acosta-Pulido}, J.~A., {et~al.} 2024, \aap, 681, A63

\bibitem[{{Steinhardt} {et~al.}(2012){Steinhardt}, {Schramm}, {Silverman}, {Alexandroff}, {Capak}, {Civano}, {Elvis}, {Masters}, {Mobasher}, {Pattarakijwanich}, \& {Strauss}}]{steinhardt2012}
{Steinhardt}, C.~L., {Schramm}, M., {Silverman}, J.~D., {et~al.} 2012, \apj, 759, 24

\bibitem[{{Stern} \& {Laor}(2012)}]{stern2012}
{Stern}, J. \& {Laor}, A. 2012, \mnras, 423, 600

\bibitem[{{Tsuzuki} {et~al.}(2006){Tsuzuki}, {Kawara}, {Yoshii}, {Oyabu}, {Tanab{\'e}}, \& {Matsuoka}}]{tsuzuki2006}
{Tsuzuki}, Y., {Kawara}, K., {Yoshii}, Y., {et~al.} 2006, \apj, 650, 57

\bibitem[{{{\"U}bler} {et~al.}(2023){{\"U}bler}, {Maiolino}, {Curtis-Lake}, {P{\'e}rez-Gonz{\'a}lez}, {Curti}, {Perna}, {Arribas}, {Charlot}, {Marshall}, {D'Eugenio}, {Scholtz}, {Bunker}, {Carniani}, {Ferruit}, {Jakobsen}, {Rix}, {Rodr{\'\i}guez Del Pino}, {Willott}, {Boeker}, {Cresci}, {Jones}, {Kumari}, \& {Rawle}}]{uebler2023}
{{\"U}bler}, H., {Maiolino}, R., {Curtis-Lake}, E., {et~al.} 2023, \aap, 677, A145

\bibitem[{{{\"U}bler} {et~al.}(2024){{\"U}bler}, {Maiolino}, {P{\'e}rez-Gonz{\'a}lez}, {D'Eugenio}, {Perna}, {Curti}, {Arribas}, {Bunker}, {Carniani}, {Charlot}, {Rodr{\'\i}guez Del Pino}, {Baker}, {B{\"o}ker}, {Cresci}, {Dunlop}, {Grogin}, {Jones}, {Kumari}, {Lamperti}, {Laporte}, {Marshall}, {Mazzolari}, {Parlanti}, {Rawle}, {Scholtz}, {Venturi}, \& {Witstok}}]{uebler2024}
{{\"U}bler}, H., {Maiolino}, R., {P{\'e}rez-Gonz{\'a}lez}, P.~G., {et~al.} 2024, \mnras, 531, 355

\bibitem[{{Valentini} {et~al.}(2021){Valentini}, {Gallerani}, \& {Ferrara}}]{valentini2021}
{Valentini}, M., {Gallerani}, S., \& {Ferrara}, A. 2021, \mnras, 507, 1

\bibitem[{{Valentini} {et~al.}(2020){Valentini}, {Murante}, {Borgani}, {Granato}, {Monaco}, {Brighenti}, {Tornatore}, {Bressan}, \& {Lapi}}]{valentini2020}
{Valentini}, M., {Murante}, G., {Borgani}, S., {et~al.} 2020, \mnras, 491, 2779

\bibitem[{{van Dokkum}(2001)}]{Vandokkum:2001}
{van Dokkum}, P.~G. 2001, \pasp, 113, 1420

\bibitem[{{Vanden Berk} {et~al.}(2001){Vanden Berk}, {Richards}, {Bauer}, {Strauss}, {Schneider}, {Heckman}, {York}, {Hall}, {Fan}, {Knapp}, {Anderson}, {Annis}, {Bahcall}, {Bernardi}, {Briggs}, {Brinkmann}, {Brunner}, {Burles}, {Carey}, {Castander}, {Connolly}, {Crocker}, {Csabai}, {Doi}, {Finkbeiner}, {Friedman}, {Frieman}, {Fukugita}, {Gunn}, {Hennessy}, {Ivezi{\'c}}, {Kent}, {Kunszt}, {Lamb}, {Leger}, {Long}, {Loveday}, {Lupton}, {Meiksin}, {Merelli}, {Munn}, {Newberg}, {Newcomb}, {Nichol}, {Owen}, {Pier}, {Pope}, {Rockosi}, {Schlegel}, {Siegmund}, {Smee}, {Snir}, {Stoughton}, {Stubbs}, {SubbaRao}, {Szalay}, {Szokoly}, {Tremonti}, {Uomoto}, {Waddell}, {Yanny}, \& {Zheng}}]{Vanden2001}
{Vanden Berk}, D.~E., {Richards}, G.~T., {Bauer}, A., {et~al.} 2001, \aj, 122, 549

\bibitem[{{Vayner} {et~al.}(2024){Vayner}, {Zakamska}, {Ishikawa}, {Sankar}, {Wylezalek}, {Rupke}, {Veilleux}, {Bertemes}, {Barrera-Ballesteros}, {Chen}, {Diachenko}, {Goulding}, {Greene}, {Hainline}, {Hamann}, {Heckman}, {Johnson}, {Grace Lim}, {Liu}, {Lutz}, {L{\"u}tzgendorf}, {Mainieri}, {McCrory}, {Murphree}, {Nesvadba}, {Ogle}, {Sturm}, \& {Whitesell}}]{vayner2024}
{Vayner}, A., {Zakamska}, N.~L., {Ishikawa}, Y., {et~al.} 2024, \apj, 960, 126

\bibitem[{{Vayner} {et~al.}(2021){Vayner}, {Zakamska}, {Riffel}, {Alexandroff}, {Cosens}, {Hamann}, {Perrotta}, {Rupke}, {Bergmann}, {Veilleux}, {Walth}, {Wright}, \& {Wylezalek}}]{vayner2021}
{Vayner}, A., {Zakamska}, N.~L., {Riffel}, R.~A., {et~al.} 2021, \mnras, 504, 4445

\bibitem[{{Venemans} {et~al.}(2019){Venemans}, {Neeleman}, {Walter}, {Novak}, {Decarli}, {Hennawi}, \& {Rix}}]{Venemans:2019}
{Venemans}, B.~P., {Neeleman}, M., {Walter}, F., {et~al.} 2019, \apjl, 874, L30

\bibitem[{{Venemans} {et~al.}(2020){Venemans}, {Walter}, {Neeleman}, {Novak}, {Otter}, {Decarli}, {Ba{\~n}ados}, {Drake}, {Farina}, {Kaasinen}, {Mazzucchelli}, {Carilli}, {Fan}, {Rix}, \& {Wang}}]{Venemans:2020}
{Venemans}, B.~P., {Walter}, F., {Neeleman}, M., {et~al.} 2020, \apj, 904, 130

\bibitem[{{Venturi} {et~al.}(2023){Venturi}, {Treister}, {Finlez}, {D'Ago}, {Bauer}, {Harrison}, {Ramos Almeida}, {Revalski}, {Ricci}, {Sartori}, {Girdhar}, {Keel}, \& {Tub{\'\i}n}}]{venturi2023}
{Venturi}, G., {Treister}, E., {Finlez}, C., {et~al.} 2023, \aap, 678, A127

\bibitem[{{V{\'e}ron-Cetty} {et~al.}(2004){V{\'e}ron-Cetty}, {Joly}, \& {V{\'e}ron}}]{veron-cetty2004}
{V{\'e}ron-Cetty}, M.~P., {Joly}, M., \& {V{\'e}ron}, P. 2004, \aap, 417, 515

\bibitem[{{Vignali} {et~al.}(2005){Vignali}, {Brandt}, {Schneider}, \& {Kaspi}}]{vignali2005}
{Vignali}, C., {Brandt}, W.~N., {Schneider}, D.~P., \& {Kaspi}, S. 2005, \aj, 129, 2519

\bibitem[{{Villforth}(2023)}]{villforth2023}
{Villforth}, C. 2023, The Open Journal of Astrophysics, 6, 34

\bibitem[{{Vito} {et~al.}(2019){Vito}, {Brandt}, {Bauer}, {Gilli}, {Luo}, {Zamorani}, {Calura}, {Comastri}, {Mazzucchelli}, {Mignoli}, {Nanni}, {Shemmer}, {Vignali}, {Brusa}, {Cappelluti}, {Civano}, \& {Volonteri}}]{vito2019}
{Vito}, F., {Brandt}, W.~N., {Bauer}, F.~E., {et~al.} 2019, \aap, 628, L6

\bibitem[{{Volonteri} {et~al.}(2022){Volonteri}, {Pfister}, {Beckmann}, {Dotti}, {Dubois}, {Massonneau}, {Musoke}, \& {Tremmel}}]{volonteri2022}
{Volonteri}, M., {Pfister}, H., {Beckmann}, R., {et~al.} 2022, \mnras, 514, 640

\bibitem[{{Wagg} {et~al.}(2012){Wagg}, {Wiklind}, {Carilli}, {Espada}, {Peck}, {Riechers}, {Walter}, {Wootten}, {Aravena}, {Barkats}, {Cortes}, {Hills}, {Hodge}, {Impellizzeri}, {Iono}, {Leroy}, {Mart{\'\i}n}, {Rawlings}, {Maiolino}, {McMahon}, {Scott}, {Villard}, \& {Vlahakis}}]{wagg2012}
{Wagg}, J., {Wiklind}, T., {Carilli}, C.~L., {et~al.} 2012, \apjl, 752, L30

\bibitem[{{Wang} {et~al.}(2023){Wang}, {Yang}, {Hennawi}, {Fan}, {Sun}, {Champagne}, {Costa}, {Habouzit}, {Endsley}, {Li}, {Lin}, {Meyer}, {Schindler}, {Wu}, {Ba{\~n}ados}, {Barth}, {Bhowmick}, {Bieri}, {Blecha}, {Bosman}, {Cai}, {Colina}, {Connor}, {Davies}, {Decarli}, {De Rosa}, {Drake}, {Egami}, {Eilers}, {Evans}, {Farina}, {Haiman}, {Jiang}, {Jin}, {Jun}, {Kakiichi}, {Khusanova}, {Kulkarni}, {Li}, {Liu}, {Loiacono}, {Lupi}, {Mazzucchelli}, {Onoue}, {Pudoka}, {Rojas-Ruiz}, {Shen}, {Strauss}, {Tee}, {Trakhtenbrot}, {Trebitsch}, {Venemans}, {Volonteri}, {Walter}, {Xie}, {Yue}, {Zhang}, {Zhang}, \& {Zou}}]{wang2023}
{Wang}, F., {Yang}, J., {Hennawi}, J.~F., {et~al.} 2023, \apjl, 951, L4

\bibitem[{{Weinberger} {et~al.}(2018){Weinberger}, {Springel}, {Pakmor}, {Nelson}, {Genel}, {Pillepich}, {Vogelsberger}, {Marinacci}, {Naiman}, {Torrey}, \& {Hernquist}}]{weinberger2018}
{Weinberger}, R., {Springel}, V., {Pakmor}, R., {et~al.} 2018, \mnras, 479, 4056

\bibitem[{{Weisskopf} {et~al.}(2007){Weisskopf}, {Wu}, {Trimble}, {O'Dell}, {Elsner}, {Zavlin}, \& {Kouveliotou}}]{weisskopf2007}
{Weisskopf}, M.~C., {Wu}, K., {Trimble}, V., {et~al.} 2007, \apj, 657, 1026

\bibitem[{{Williams} {et~al.}(2014){Williams}, {Wagg}, {Maiolino}, {Foster}, {Aravena}, {Wiklind}, {Carilli}, {McMahon}, {Riechers}, \& {Walter}}]{williams2014}
{Williams}, R.~J., {Wagg}, J., {Maiolino}, R., {et~al.} 2014, \mnras, 439, 2096

\bibitem[{{Willott} {et~al.}(2017){Willott}, {Bergeron}, \& {Omont}}]{Willott:2017}
{Willott}, C.~J., {Bergeron}, J., \& {Omont}, A. 2017, \apj, 850, 108

\bibitem[{{Willott} {et~al.}(2005){Willott}, {Percival}, {McLure}, {Crampton}, {Hutchings}, {Jarvis}, {Sawicki}, \& {Simard}}]{willott2005}
{Willott}, C.~J., {Percival}, W.~J., {McLure}, R.~J., {et~al.} 2005, \apj, 626, 657

\bibitem[{{Yu} {et~al.}(2021){Yu}, {Li}, {Qu}, {Roederer}, {Bregman}, {Fan}, {Fang}, {Johnson}, {Wang}, \& {Yang}}]{yu2021}
{Yu}, X., {Li}, J.-T., {Qu}, Z., {et~al.} 2021, \mnras, 505, 4444

\bibitem[{{Zakamska} {et~al.}(2016){Zakamska}, {Hamann}, {P{\^a}ris}, {Brandt}, {Greene}, {Strauss}, {Villforth}, {Wylezalek}, {Alexandroff}, \& {Ross}}]{zakamska2016}
{Zakamska}, N.~L., {Hamann}, F., {P{\^a}ris}, I., {et~al.} 2016, \mnras, 459, 3144

\bibitem[{{Zamanov} {et~al.}(2002){Zamanov}, {Marziani}, {Sulentic}, {Calvani}, {Dultzin-Hacyan}, \& {Bachev}}]{zamanov2002}
{Zamanov}, R., {Marziani}, P., {Sulentic}, J.~W., {et~al.} 2002, \apjl, 576, L9

\bibitem[{{Zana} {et~al.}(2023){Zana}, {Carniani}, {Prelogovi{\'c}}, {Vito}, {Allevato}, {Ferrara}, {Gallerani}, \& {Parlanti}}]{Zana:2023}
{Zana}, T., {Carniani}, S., {Prelogovi{\'c}}, D., {et~al.} 2023, \aap, 679, A77

\bibitem[{{Zana} {et~al.}(2022){Zana}, {Gallerani}, {Carniani}, {Vito}, {Ferrara}, {Lupi}, {Di Mascia}, \& {Barai}}]{zana2022}
{Zana}, T., {Gallerani}, S., {Carniani}, S., {et~al.} 2022, \mnras, 513, 2118

\bibitem[{{Zhang}(2024)}]{zhang2024}
{Zhang}, X.-G. 2024, \mnras, 529, 41

\bibitem[{{Zheng} {et~al.}(2006){Zheng}, {Overzier}, {Bouwens}, {White}, {Ford}, {Ben{\'\i}tez}, {Blakeslee}, {Bradley}, {Jee}, {Martel}, {Mei}, {Zirm}, {Illingworth}, {Clampin}, {Hartig}, {Ardila}, {Bartko}, {Broadhurst}, {Brown}, {Burrows}, {Cheng}, {Cross}, {Demarco}, {Feldman}, {Franx}, {Golimowski}, {Goto}, {Gronwall}, {Holden}, {Homeier}, {Infante}, {Kimble}, {Krist}, {Lesser}, {Menanteau}, {Meurer}, {Miley}, {Motta}, {Postman}, {Rosati}, {Sirianni}, {Sparks}, {Tran}, \& {Tsvetanov}}]{zheng2006}
{Zheng}, W., {Overzier}, R.~A., {Bouwens}, R.~J., {et~al.} 2006, \apj, 640, 574

\end{thebibliography}

\appendix
\onecolumn
\section{PSF modeling} \label{ap:psf_mod}
\begin{figure*}[h]
\centering
\includegraphics[width=0.8\textwidth]{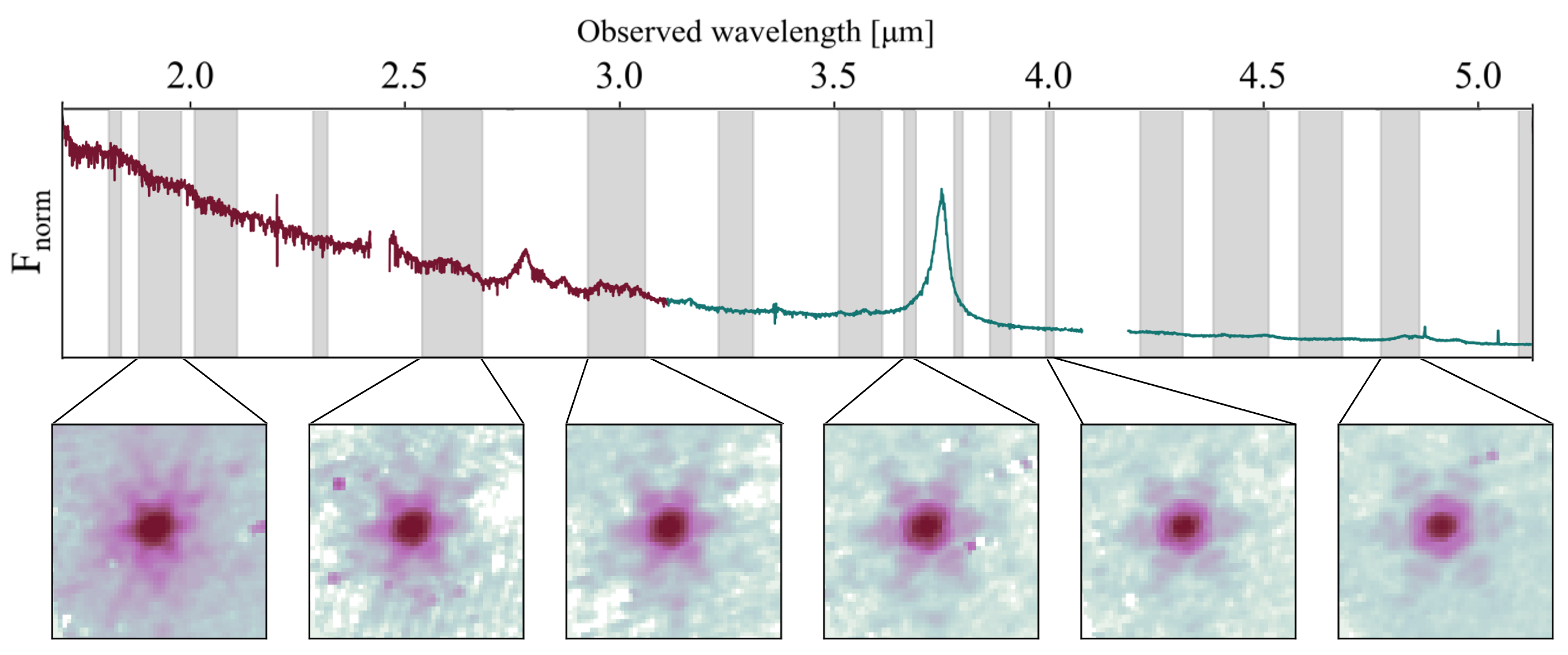}
\caption{QSO spectrum (top panel, G235H/F170LP and G395H/F290LP gratings/filters in red and blue, respectively) with examples of the empirical PSF models adopted in our analysis (bottom panels). The wavelength ranges used to compute the PSF are highlighted in gray.}
\label{fig:psf1}
\end{figure*}

\begin{figure*}[h]
\centering
\includegraphics[width=\textwidth]{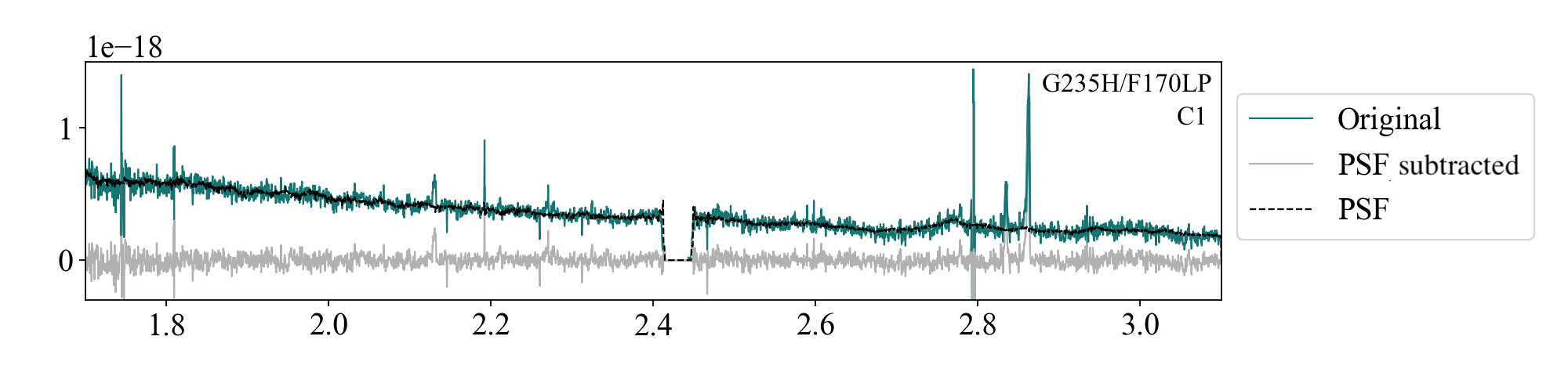}
\includegraphics[width=0.99\textwidth]{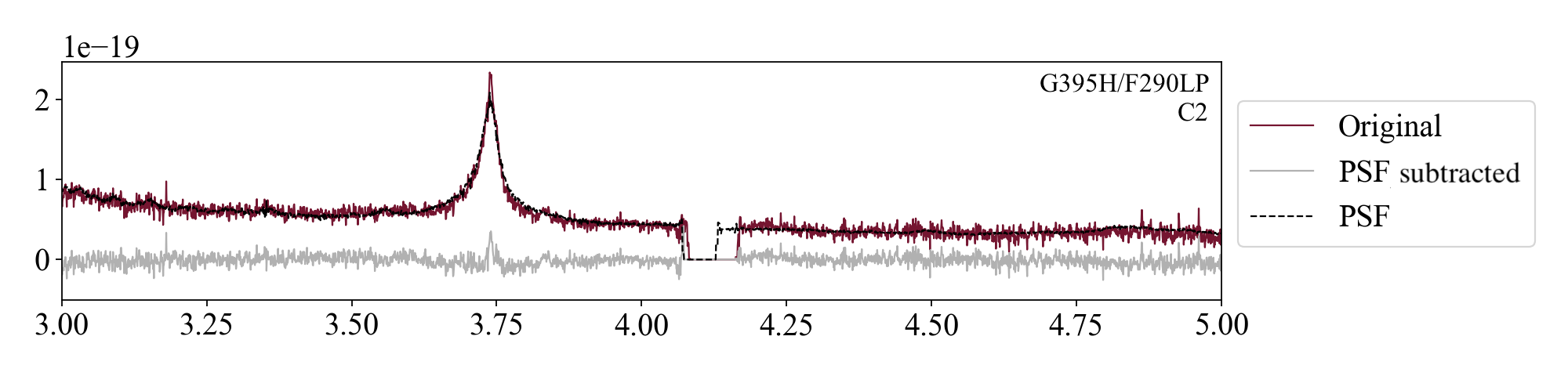}
\caption{Example of two PSF-subtracted spectra in the two different gratings for two different sources (G235H/F170LP for C1 and G395H/F290LP for C2, upper and lower panels, respectively). The modeled PSFs in the same aperture of the integrated spectra (see Sect. \ref{sec:psf_mod}) and the PSF-subtracted spectra are shown with dashed black and solid gray lines, respectively.}
\label{fig:psf-C1}
\end{figure*}

\newpage

\twocolumn
\section{Aperture correction results} \label{ap:cor_aper}
The aperture correction can be calculated using point source observations from the same instrument and configuration (grating/filter) of our observation. In Sect. \ref{sec:psf_mod}, we used the QSO emission itself to model the PSF from our data. Then, we used this modeling to calculate the wavelength-dependent correction adopting four different apertures of radius 0.2, 0.3, 0.4, and 0.5 arcsec centered on the QSO.

\begin{figure}[h!]
\includegraphics[width=\linewidth]{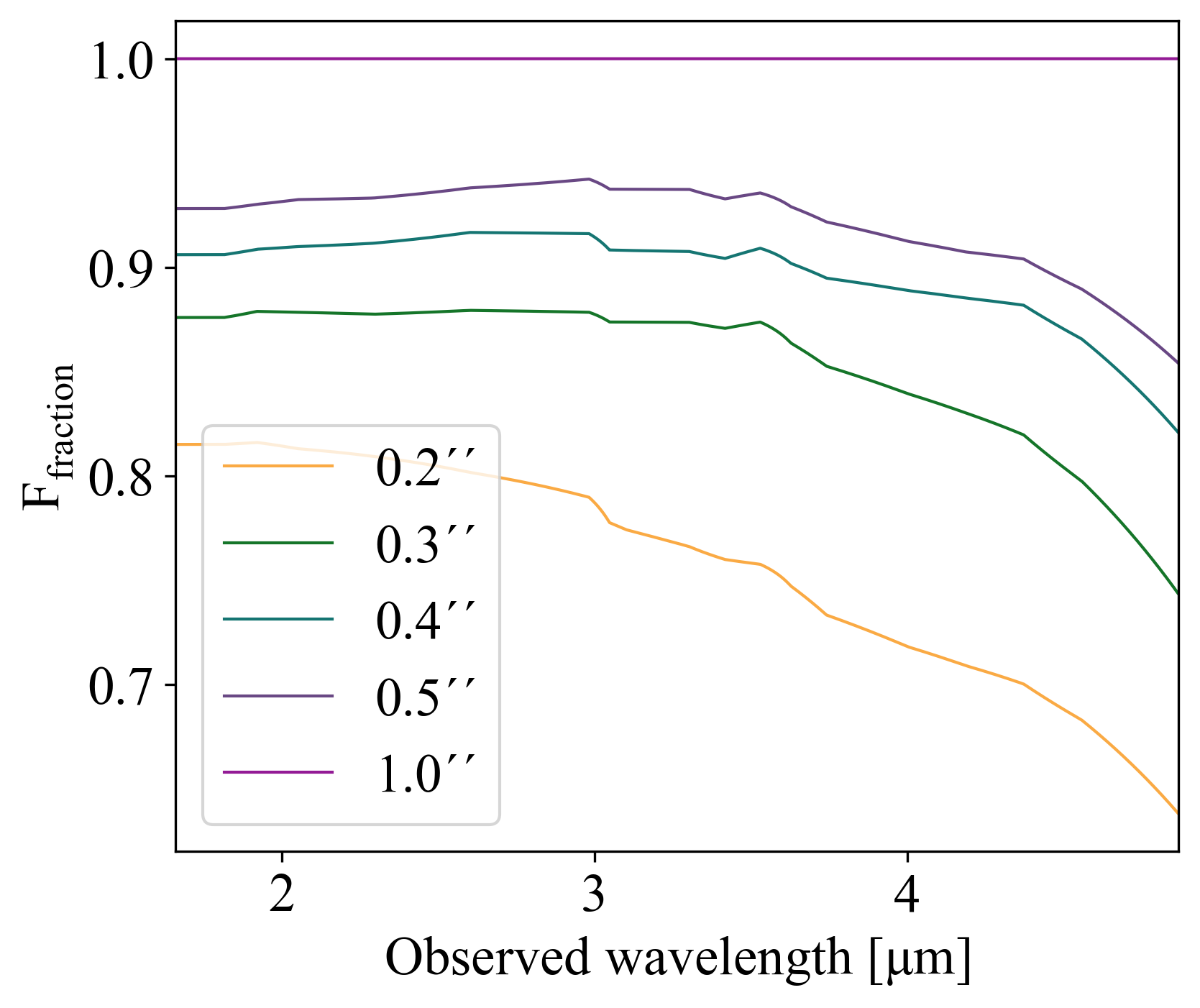}
\caption{Fraction of the total flux included in four different apertures ($R$ = 0.2\arcsec, 0.3\arcsec, 0.4\arcsec and 0.5\arcsec) 
as a function of the observed wavelengths, normalized to the value obtained in the aperture of $R$ = 1\arcsec.}
\label{fig:ap_cor}
\end{figure}

Figure \ref{fig:ap_cor} and Table \ref{tab:ap_cor} show the fraction of the flux at different observed wavelengths normalized to the value obtained in the aperture of 1 arcsec radius, which includes almost the entire emission of the PSF. This fraction decreases when the wavelength increases due to the larger PSF. For an aperture of radius = 0.5 arcsec, the flux lost at the H$\beta$ and H$\alpha$ wavelengths at redshift $z$ = 4.7 is around 6 \% and 8 \%, respectively, this value being larger for smaller apertures.

We verified that these results obtained from the continuum also work for the BLR broad Balmer lines. We thus extracted the integrated spectrum of the QSO in two different apertures: $R$ = 0.5\arcsec\ and 1.0\arcsec and modeled the broad \ha with a Gaussian with a velocity dispersion spanning a range  1000 \kms > $\sigma$ > 5000 \kms, and we also quantified the flux lost in the spectral domain. We  found that the integrated flux of the BLR component in both apertures is very similar, with a ratio $F_{\rm BLR}$($R$ = 0.5\arcsec) / $F_{\rm BLR}$($R$ = 1\arcsec) = 0.95 (we are losing less than a 5\% of the BLR flux in the smallest one). This value is similar to the aperture corrected flux reported in Table \ref{tab:ap_cor} with a value $F_{\rm fraction}$ ($\lambda$ = 3.74, $R$ = 0.5\arcsec) = 0.92. Then, the values calculated in Table \ref{tab:ap_cor} can be used to correct the integrated spectrum. 

\newpage 

\begin{table}[h]
\centering
\caption{Fraction of total flux included in four different apertures at
different observed wavelengths normalized to the value obtained in the aperture of $R$ = 1\arcsec.}\label{tab:ap_cor}
\begin{tabular}{ccccc}
\hline \hline
& \multicolumn{4}{c}{$F_{\rm fraction}$($\lambda$, R)} \\
$\lambda$ [$\mu$m] & $R$ = 0.2\arcsec & $R$ = 0.3\arcsec & $R$ = 0.4\arcsec & $R$ = 0.5\arcsec\\ 
\hline \hline
1.7 & 0.82 & 0.88 & 0.91 & 0.93 \\
1.8 & 0.82 & 0.88 & 0.91 & 0.93 \\
1.9 & 0.82 & 0.88 & 0.91 & 0.93 \\
2.0 & 0.81 & 0.88 & 0.91 & 0.93 \\
2.1 & 0.81 & 0.88 & 0.91 & 0.93 \\
2.2 & 0.81 & 0.88 & 0.91 & 0.93 \\
2.3 & 0.81 & 0.88 & 0.91 & 0.93 \\
2.4 & 0.81 & 0.88 & 0.91 & 0.93 \\
2.5 & 0.80 & 0.88 & 0.91 & 0.94 \\
2.6 & 0.80 & 0.88 & 0.92 & 0.94 \\
2.7 & 0.80 & 0.88 & 0.92 & 0.94 \\
2.8 & 0.80 & 0.88 & 0.92 & 0.94 \\
2.9 & 0.79 & 0.88 & 0.92 & 0.94 \\
3.0 & 0.79 & 0.88 & 0.91 & 0.94 \\
3.1 & 0.77 & 0.87 & 0.91 & 0.94 \\
3.2 & 0.77 & 0.87 & 0.91 & 0.94 \\
3.3 & 0.77 & 0.87 & 0.91 & 0.94 \\
3.4 & 0.76 & 0.87 & 0.90 & 0.93 \\
3.5 & 0.76 & 0.87 & 0.91 & 0.93 \\
3.6 & 0.75 & 0.87 & 0.91 & 0.93 \\
3.7 & 0.74 & 0.86 & 0.90 & 0.92 \\
3.8 & 0.73 & 0.85 & 0.89 & 0.92 \\
3.9 & 0.72 & 0.85 & 0.89 & 0.92 \\
4.0 & 0.72 & 0.84 & 0.89 & 0.91 \\
4.1 & 0.71 & 0.83 & 0.89 & 0.91 \\
4.2 & 0.71 & 0.83 & 0.89 & 0.91 \\
4.3 & 0.70 & 0.82 & 0.88 & 0.91 \\
4.4 & 0.70 & 0.82 & 0.88 & 0.90 \\
4.5 & 0.69 & 0.80 & 0.87 & 0.89 \\
4.6 & 0.68 & 0.79 & 0.86 & 0.89 \\
4.7 & 0.67 & 0.78 & 0.85 & 0.88 \\
4.8 & 0.65 & 0.76 & 0.83 & 0.86 \\
4.9 & 0.64 & 0.74 & 0.82 & 0.85 \\
5.0 & 0.64 & 0.74 & 0.82 & 0.85 \\
5.1 & 0.64 & 0.74 & 0.82 & 0.85 \\
5.2 & 0.64 & 0.74 & 0.82 & 0.85 \\
\hline
\end{tabular}
\end{table}

\section{NIRSpec errors}\label{ap:err}

The errors on the integrated spectrum of a source can be calculated (i) from the error extension present in the data cube in the same spaxels of the selected region propagating them in quadrature; or (ii) considering the spatial noise of our data calculating the root mean square (RMS) at each wavelength in apertures without source emission. 

We  calculated the errors using these two methods in three different random apertures (RA$_1$ = 181.3455 deg, Dec$_1$ = --7.7087; RA$_2$ = 181.3460 deg, Dec$_2$ = --7.7079; RA$_3$ = 181.3458 deg, Dec$_3$ = --7.7092). For the G235H grating, the mean error obtained for each method is 0.0065 and 0.0067 $\times$ 10$^{-20}$ erg s$^{-1}$ cm$^{-2}$ \AA$^{-1}$, respectively. The mean errors in the G395H grating are larger due to the low S/N of the data, 0.025 and  0.022 $\times$ 10$^{-20}$ erg s$^{-1}$ cm$^{-2}$ \AA$^{-1}$ with method (i) and (ii), respectively. The results of both methods are totally compatible between them according to the mean dispersion of the results in different apertures, then we  used the first method to determine the errors in the integrated spectra of the sources within the system.

\newpage

% \onecolumn

\section{LAEs and companions} \label{ap:comp}

\begin{figure}[h]
\centering
\includegraphics[width=\linewidth]{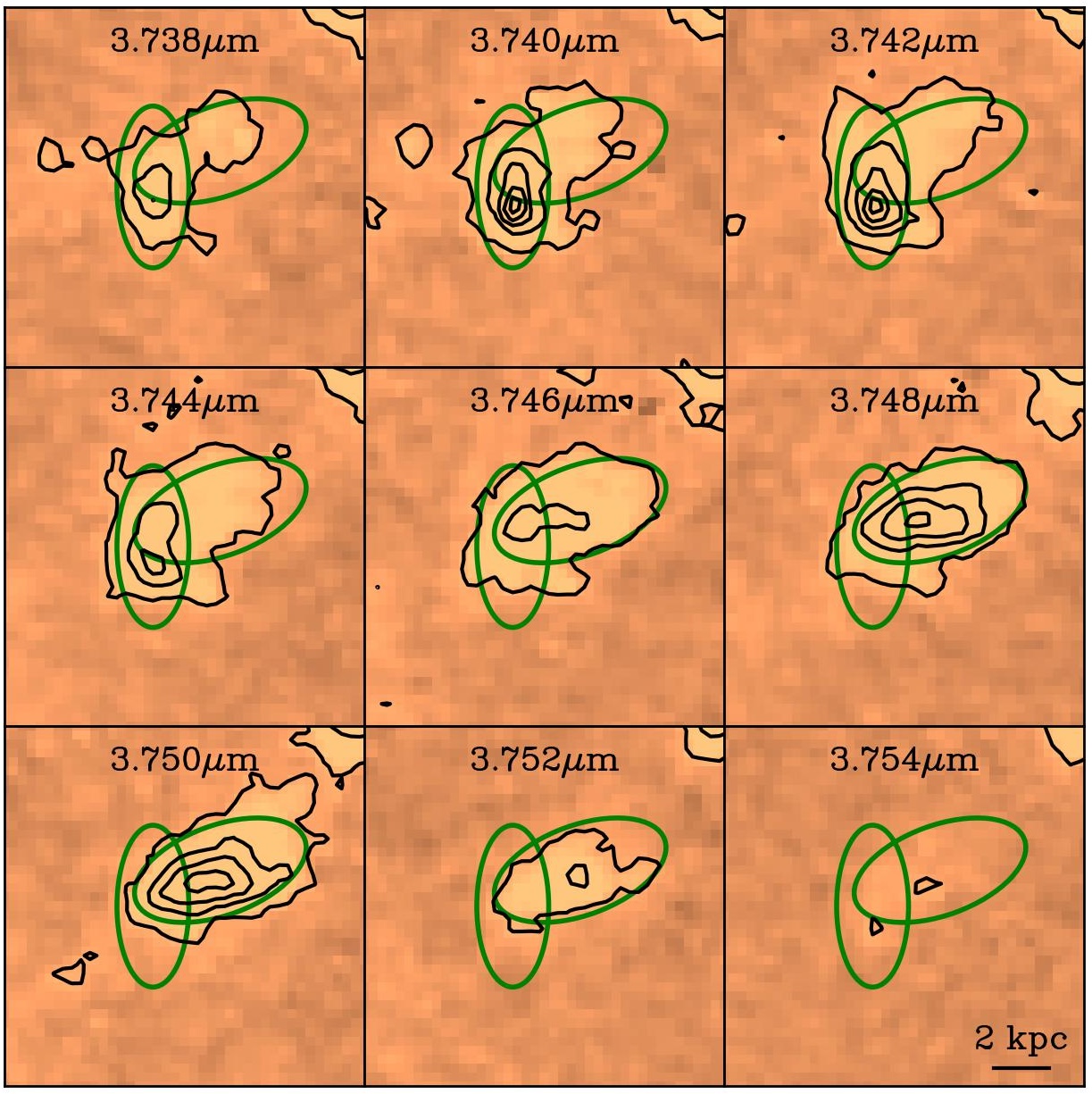}
\caption{
Channel maps of \ha emission from LAE1. Contours begin at 3$\sigma$ and are in steps of 5$\sigma$. Each channel map covers a FoV of $1^{\prime\prime}\times1^{\prime\prime}$ and spans 0.002 $\mu$m, which corresponds to 160~\kms. The green ellipses indicate the location and extension of the two merging galaxies: LAE1a and LAE1b. The emission around these sources is not aligned in the same direction, which discards a rotation scenario.}
\label{fig:fmapLae1}
\end{figure}

\newpage

\section{Additional candidate galaxies} \label{ap:comp_spec}

\begin{figure}[h]
\centering
\includegraphics[width=0.3\textwidth]{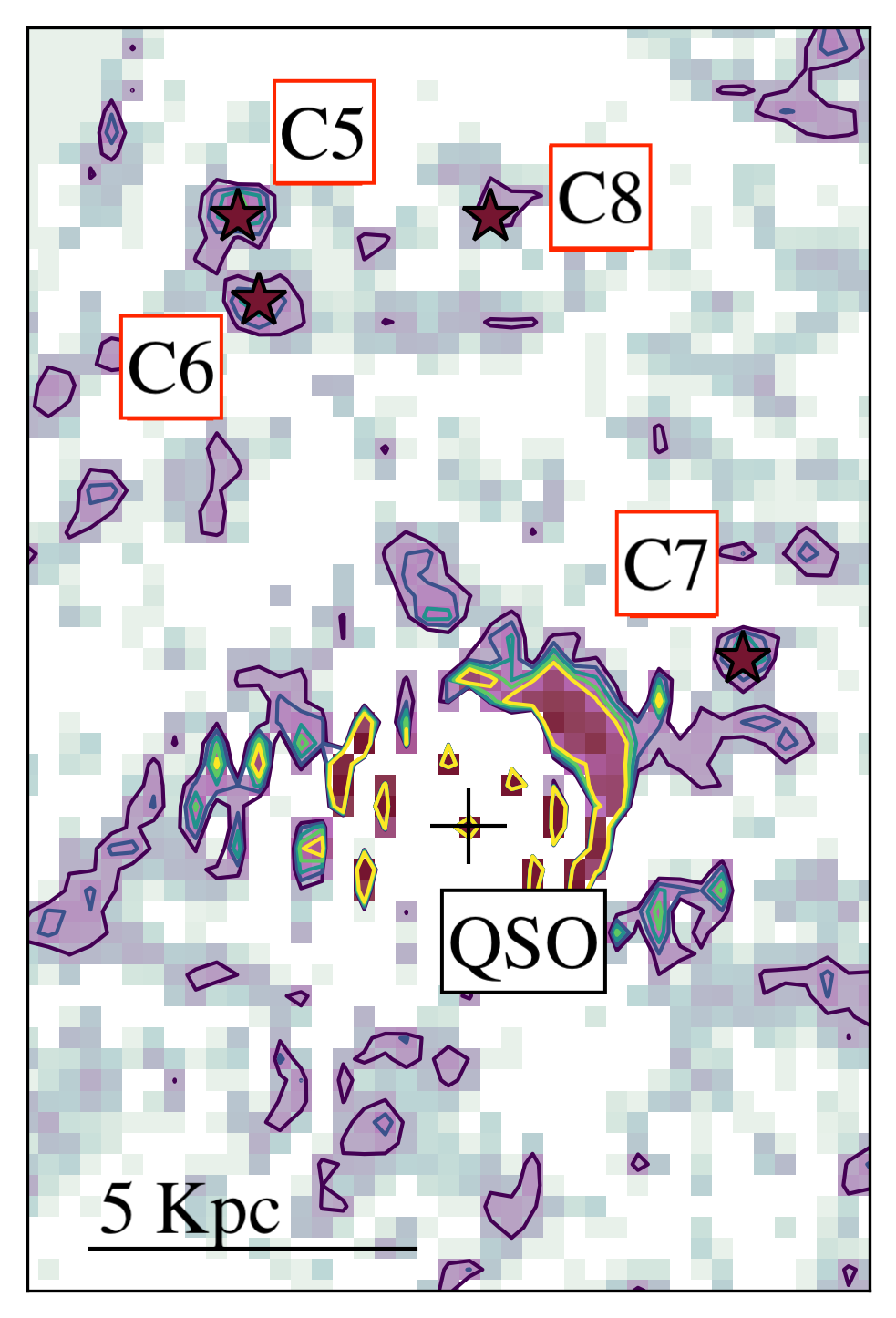}
\caption{JWST NIRSpec PSF-subtracted map integrated over the wavelength range 3.635 $\mu$m < $\lambda $ < 3.846 $\mu$m, encompassing H$\alpha$+\nii emission lines. The QSO position is marked with a plus symbol. The candidate galaxies found around the QSO are highlighted with stars.}
\label{fig:map_candidates}
\end{figure}

\begin{table}[h]
\caption{Coordinates of the additional candidate companion galaxies in BR1202-0725.}
\centering
\begin{tabular}{ccc}
\hline \hline
Name & RA [deg]$^a$    & Dec [deg]$^a$  \\ \hline \hline
C5   &  181.34649464 & --7.70866054 \\ 
C6   &  181.34648063 & --7.70872999 \\ 
C7   &  181.34615827 & --7.70895221 \\ 
C8   &  181.34632646 & --7.70866054 \\ 
 \hline   
\end{tabular}
\end{table}
\label{tab:comp}

\begin{figure*}[h!]
\centering
\includegraphics[width=\textwidth]{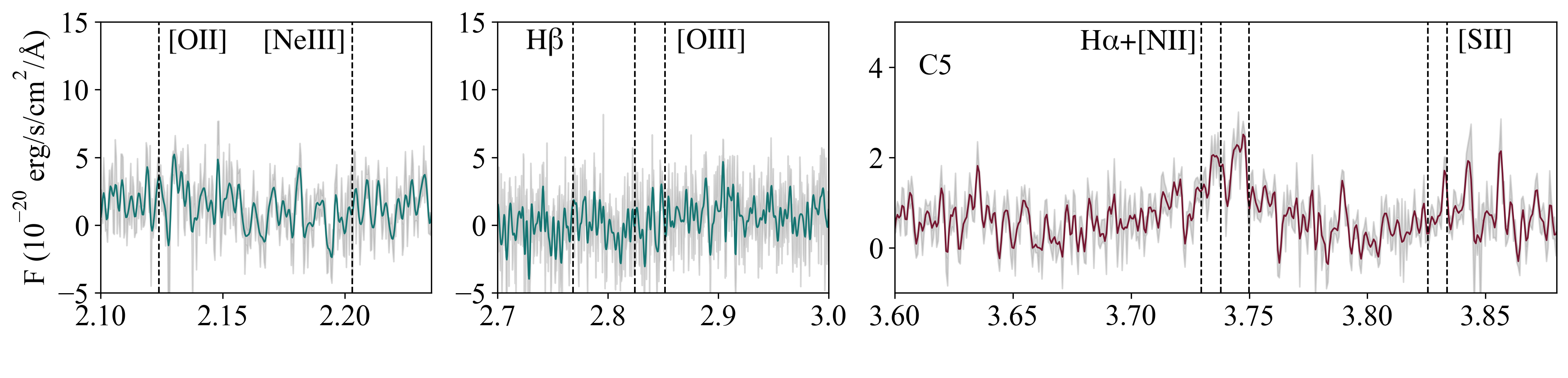}
\includegraphics[width=\textwidth]{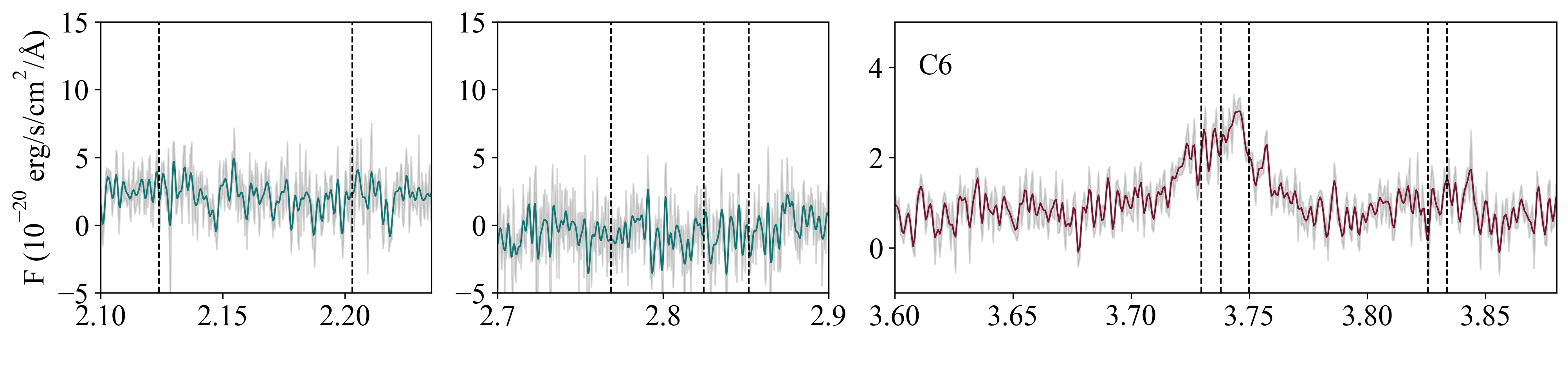}
\includegraphics[width=\textwidth]{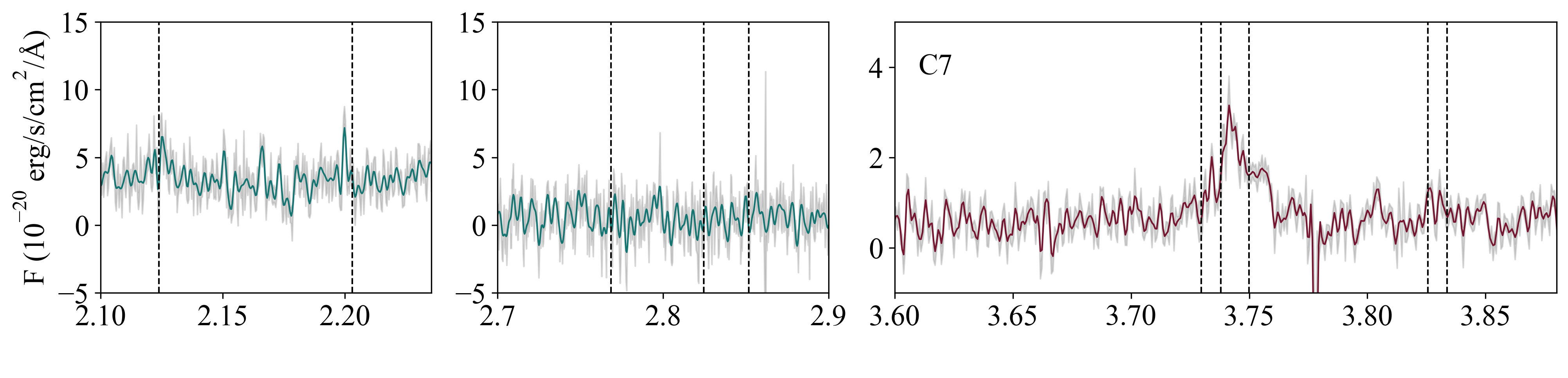}
\includegraphics[width=\textwidth]
{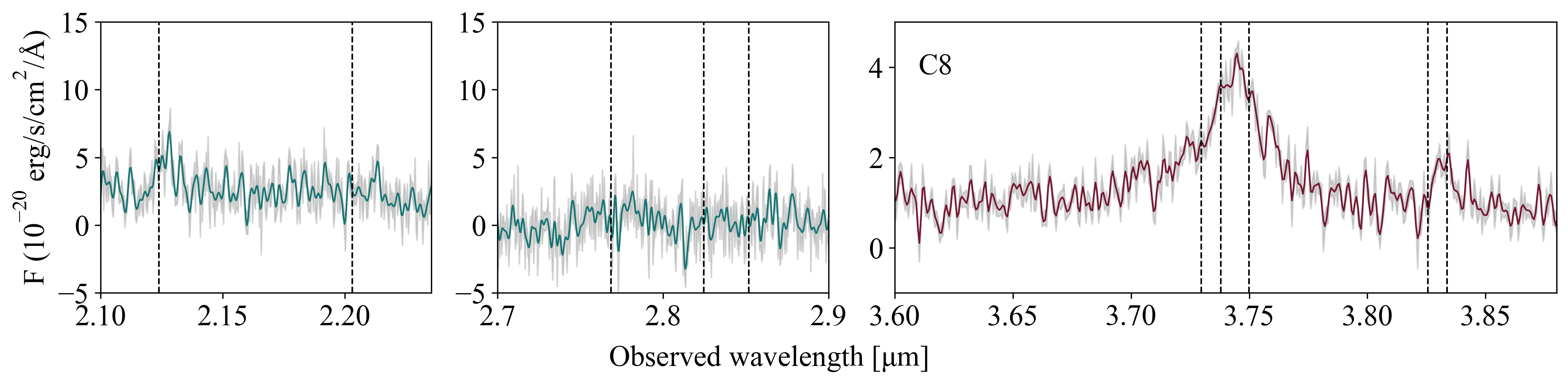}
\caption{JWST NIRSpec rest-frame optical spectra (integrated in circular apertures with R = 0.1 arcsec) of the candidate galaxies found around the QSO in the two gratings/filters used (G235H/F170LP and G395H/F290LP; blue and red lines, respectively). The errors in the spectra are shown in light gray. Vertical dashed lines mark the position of the main emission lines in the spectrum at the systemic redshift of the QSO. 
%(\oii$\lambda$3727,29 \AA, \neiii$\lambda$3869 \AA, H$\beta$, \oiii$\lambda$4959,5007 \AA, \nii$\lambda$6548 \AA, H$\alpha$, \nii$\lambda$6584, and \sii$\lambda$6716,31 \AA, from left to right, respectively).
}
\label{fig:comp}
\end{figure*}

\end{document}